\documentclass[a4paper,showpacs,preprintnumbers,floatfix,twocolumn,aps,pra]{revtex4-2}

\usepackage{ORCIDinREVTeX}
\usepackage{epsfig}
\usepackage{color}
\usepackage{amsmath}
\usepackage{gensymb}
\usepackage{hyperref} 
\usepackage[version=4]{mhchem}
\usepackage{cleveref}
\usepackage{caption}
\usepackage{subcaption}

\usepackage{graphicx}

\usepackage[utf8]{inputenc}

\def\markthis{}

\definecolor{red}{rgb}{1.0,0.0,0.0}\def\red{\color{red}}
\definecolor{RED}{rgb}{1.0,0.0,0.0}
\definecolor{blue}{rgb}{0.0,0.0,1.0}

\definecolor{green}{rgb}{0,0.5,0}
\definecolor{red}{rgb}{1.0,0,0}
\definecolor{blue}{rgb}{0,0,1.0}
\def\red{\color{red}}

\def\red{}

\newcommand {\citeref}[1]{Ref.~\citenum{#1}}
\newcommand {\citerefs}[1]{Refs.~\citenum{#1}}

\begin{document}

%

\title{Repulsive interatomic potentials calculated at three  levels of theory}

\author{Kai Nordlund}
\orcid{0000-0001-6244-1942}
\email[Corresponding author: ]{kai.nordlund@helsinki.fi}
\affiliation{
Department of Physics, P. O. Box 43,
FIN-00014 University of Helsinki, Finland
}

\author{Susi Lehtola}
\orcid{0000-0001-6296-8103}
\affiliation{
Department of Chemistry, P. O. Box 55,
FIN-00014 University of Helsinki, Finland
}

\author{Gerhard Hobler}
\orcid{0000-0002-2140-6101}
\affiliation{
Institute of Solid-State Electronics, TU Wien, 
Gu{\ss}hausstra{\ss}e 25-25a, A-1040 Wien, Austria
}

\date{\today}

\begin{abstract}
The high-energy repulsive interaction between nuclei at distances much smaller than the equilibrium bond length is the key quantity  determining the nuclear stopping power and atom scattering in keV and MeV radiation events.
This interaction is traditionally modeled within orbital-free density functional theory with frozen atomic electron densities, following the Ziegler--Biersack--Littmark (ZBL) model.
In this work, we calculate atom pair specific repulsive interatomic potentials with the ZBL model, and compare them to two kinds of quantum chemical calculations---second-order M\o{}ller--Plesset perturbation theory in flexible Gaussian basis sets as well as density functional theory with numerical atomic orbital basis sets---which go well beyond the limitations in the ZBL model, allowing the density to relax in the calculations.
We show that the repulsive interatomic potentials predicted by the two quantum chemical models agree within $\sim$ 1\% for potential energies above 30 eV,
while the ZBL pair-specific potentials and universal ZBL potentials differ much more from either of these calculations. 
We provide new pair-specific fits of the screening functions in terms of 3 exponentials to the calculations for all pairs $Z_1$-$Z_2$ for $1 \leq Z_i \leq 92$, and show that they agree within $\sim 2$\% with the raw data. We use the new potentials to simulate ion implantation depth profiles in single crystalline Si and show very good agreement with experiment. However, we also show that under channeling conditions, the attractive part of the potential can affect the depth profiles.
The full data sets of all the calculated interatomic potentials as well as analytic fits to the data are shared as open access.
\end{abstract}

\preprint{Accepted for publication in Physical Review A.}

\maketitle

\section{Introduction \label{sec:intro}}

Radiation effects occur widely in nature, both here on Earth and in space, and are important in numerous fields of human technology.
Technologies where radiation effects need to be considered include nuclear reactors and particle accelerators \cite{Was, Bernas}. Moreover, radiation is also widely used in material science: for example, ion implantation has become an indispensable process in the manufacture of integrated circuits \cite{Williams1998_MSEA_8}, and irradiation is therefore one of the key technologies behind the information revolution that began in the late $20^{\rm th}$ century.

Even though radiation events can be induced by many kinds of particles (e.g., natural fission fragments, neutrons in a reactor, cosmic muons, accelerated atoms or electrons, and X-ray and gamma photons),  the final damage is determined by the collision cascade that the energetic nucleus causes with the atoms in the sample  \cite{See62, LSS}.
In other words, the final material modification is induced by nuclei in the material that have received a recoil energy from the initially energetic particle \cite{Was, Nor18, Nor18b}. The interaction of such high-energy nuclei with each other is dominated by the highly repulsive part of the potential energy $V(\{{\bf r}_i\})$ \cite{Boh48}, where ${\bf r}_i$ are nuclear coordinates.
Because the potential energy increases rapidly in decreasing internuclear distance, the high-energy region of the potential is associated with atoms that are close to one another.
In the case of secondary radiation effects, $V(\{{\bf r}_i\})$ thus mainly depends on the interatomic separations $r_{ij}$, and not also on e.g. the angles between the various chemical bonds, or the number of neighbours that the atoms have, as equilibrium potentials often do \cite{Leach}.
As a result, it is commonly considered that the irradiation process can be modeled as a set of independent binary collisions \cite{LSS,BCA}.
The kinetics of these binary collisions (energy transfer, scattering angle) that controls the nuclear stopping power is then fully determined by the repulsive interatomic potential.

Several types of repulsive interatomic potentials have been developed in this general formalism since the 1940s \cite{Boh48, Mol47, Eck91a, ZBL, Zin15}. However, there have been no published reports on using first principles methods to examine systematically all interatomic potentials at first principles levels of theory. 
In this work we carry out \emph{ab initio} calculations of interatomic potentials at three different levels of theory.
First, we present calculations performed with second-order M\o{}ller--Plesset (MP2) perturbation theory with flexible Gaussian-type orbital (GTO) basis sets for all atom pairs for which the sum of the atomic numbers $Z_1 + Z_2 \leq 36$, for which relativistic effects are expected to be insignificant.
Second, we present density-functional calculations carried out with numerical atomic orbital (NAO) basis sets using the DMol code that includes relativistic effects for all pairs up to $Z_1,Z_2 \leq 92$.
Third, we present calculations with the orbital-free density functional (OF-DFT) model that underlies the widely used Ziegler--Biersack--Littmark (ZBL) universal potential used in the Stopping and Range of Ions in Matter (SRIM) program for all pairs of stable atoms up to $Z_1,Z_2 \leq 92$.
In contrast to the first two models, the OF-DFT calculations employ fixed atomic electron densities, that is, the electron density has not been optimized in the OF-DFT calculations.

The layout of this work is as follows.
In \cref{sec:quantumchem}, we discuss the general theoretical basis of calculating repulsive interatomic potentials with quantum chemical methods, and then in \cref{sec:methods}
the specific three approaches we use in this work for calculating repulsive potentials.
In \cref{sec:results}, we systematically compare the potentials computed with these approaches for all atom pairs.
We study the differences of our potentials to the ZBL potentials for select systems in \cref{sec:selectedpotentials}.
We contrast our quantum chemical calculations to our OF-DFT calculations in \cref{sec:solidstateresults}, and find significant differences between the quantum chemical calculations and the approximate OF-DFT calculations carried out following ZBL.
We proceed with a systematic analysis of the differences in \cref{sec:systematic}, finding that our quantum mechanical data indeed agree well with each other.
We then use our best data set to fit a new analytical potentials for all atom pairs in \cref{sec:analyticalpotential}. 
Finally, to explore how sensitive experimentally measurable quantities are to the choice of potential, we study differences of range profiles computed with various potentials and compare them with experiments in \cref{sec:range}.
The paper ends with a summary and conclusions in \cref{sec:summary}.

\section{Quantum chemistry of repulsive interatomic potentials \label{sec:quantumchem}}

\subsection{\red Screening function formalism for repulsive potentials}
\label{subsec:screening}

{\red To understand the scientific basis of the repulsive potential calculations, we first discuss the approaches used including recent developments that lay the basis of the results of the paper.}
The interaction of bare nuclei without any electrons (in the absence of nuclear resonances, which are beyond the scope of this work), is given by the Coulomb potential, expressed in {red international} standard {\red (SI)} units as
\begin{equation}\label{eq:Coulomb}
    V_{\rm Coul} (r) = {1 \over 4 \pi \varepsilon_0} {Z_1 e Z_2 e \over r}.
\end{equation}
{\red 
Here $\varepsilon_0$ is the dielectric constant $= 8.854187817\times10^{-12}$ F/m, $Z_1$ and $Z_2$ are the atomic numbers of the two colliding atoms, $e$ the elemental charge $= 1.602176634\times10^{-19}$ C, and $r$ the distance between the two atoms. 
}

For nuclei with electrons, the interaction can be described within the Born--Oppenheimer (BO) approximation \cite{Born1927_APB_457}: the motion of the nuclei is decoupled from that of the electrons. The coupled motion, including all kinds of excitation and ionization effects, is included in electronic stopping power \cite{LSS,Fer48, Kis62, Ech81, Sch11}.
The interatomic potential $V(r)$ can be computed quantum mechanically for binary collisions in the BO approach as the difference between the total energy $E^{\rm tot}_{1+2}(r)$ of a diatomic system with nuclei $Z_1$ and $Z_2$ separated by an internuclear distance $r$, and the total energies of the two atoms at dissociation, $E^{\rm tot}_{1}$ and $E^{\rm tot}_{2}$, respectively, as
\begin{equation}
    \label{eq:V_Etot}
    V (r) = E^{\rm tot}_{1+2} (r) - E^{\rm tot}_1 - E^{\rm tot}_2.
\end{equation}
This expression can be rearranged into
\begin{equation}
    \label{eq:V}
    V (r) = V_{\rm Coul} (r) + E^{\rm el}_{1+2}(r) - E^{\rm el}_1 - E^{\rm el}_2
\end{equation}
where $E^{\rm el}_{1+2}(r)$, $E^{\rm el}_1$ and $E^{\rm el}_2$ are the \emph{electronic energies} of the diatomic molecule and the two atoms, respectively.
The total energies $E^{\rm tot}_{1+2}(r)$, $E^{\rm tot}_{1}$ and $E^{\rm tot}_{2}$---or the electronic energies $E^{\rm el}_{1+2}(r)$, $E^{\rm el}_1$, and $E^{\rm el}_2$---can then be computed with established electronic structure methods (cf. \cref{sec:methods}).

As the Coulomb term $V_{\rm Coul}(r)\propto r^{-1}$ diverges in the limit $r \to 0$, while the electronic energy tends to a finite value --- the electronic energy $E_{A+B}^{\rm el}$ of the united atom A+B --- the interatomic potential is dominated by the Coulomb potential at small internuclear distances ($r \ll 0.1$ Å).
Therefore, repulsive potentials are commonly described in the form 
\begin{equation}\label{eq:screenedCoulomb}
    V (r) = V_{\rm Coul}(r) \phi(r)
\end{equation}
where the screening function $\phi(r)$, whose determination is the main focus of our work, is defined by
\begin{equation}
    \phi(r) = V(r)/V_{\rm Coul}(r).
    \label{eq:screen}
\end{equation}
The dominance of the Coulomb term at close range yields the property $\phi(r)\to 1$ when $r \to 0$.

In contrast, at large internuclear distances, the electrons surrounding the atoms screen the bare Coulomb interaction between the two nuclei, resulting in an interaction potential $V(r)$ that is weaker than the pure Coulomb potential $ V_{\rm Coul} (r) \propto r^{-1}$; for example, it is well known that $V(r)\propto -r^{-6}$ in the van der Waals limit \cite{Kittel}. 
Since $V(r)$ approaches zero faster than $V_{\rm Coul}(r)$ does, one therefore observes that $\phi(r) \to 0$ when $r \to \infty$.

The two features discussed above make the screening function attractive for use in simulations: unlike the bare interatomic potential $V(r)$, $\phi(r)$ has a limited numerical range and varies smoothly across all internuclear distances, making it easy to interpolate it accurately from a moderate number of tabulated points, for example.
It is worth noting that also chemical interactions can be considered in this formalism, since the mapping $r \to V_{\rm Coul}(r)$ is invertible, and it is easy to see that the description of a bonded diatomic molecule leads to $\phi(r) < 0$ at chemically bound interaction distances $r$ \cite{Eck91a}.
{\red Recently,  one of us has shown that also} antiproton-atom interactions can be modeled in this formalism \cite{Nor17}, even though the Coulomb interaction between the nuclei is then attractive instead of repulsive.

The simplest analytic expression for $\phi(r)$ of a repulsive potential that satisfies the criteria $\lim_{r \to 0} \phi(r) = 1$ and $\lim_{r \to \infty} \phi(r) = 0$ is an exponential, as proposed by \citet{Boh48}. 
However, as a single exponential cannot adequately describe interatomic interactions over the relevant range of internuclear distances $r$ that spans several orders of magnitude,
a sum of exponentials is commonly used instead \cite{Mol47, Wilson77, ZBL}. 
Other functions suitable for expansions of $\phi(r)$ have been proposed, as well \cite{Jen32, Som32, Bie82, Nak88, Zin11, Zin15}.

The most widely used repulsive interatomic potential to date is the ``universal'' ZBL potential, named after Ziegler, Biersack, and Littmark, who are the authors of the book, \citeref{ZBL}. 
ZBL carried out simplified quantum mechanical calculations for a large number of atom pairs $Z_1$--$Z_2$, and then fitted the results to an analytic universal form assumed to be valid for all $Z_1$--$Z_2$ pairs.
The quantum calculations of ZBL were based on the assumptions that (i) the electron densities of the two atoms don't change as the atoms approach each other, and that (ii) the change in energy due to overlapping electron densities can be modeled within the local density approximation (LDA) \cite{Thomas1927_MPCPS_542, Fermi1927_RdNdL_602, Bloch1929_ZfuP_545, Dirac1930_MPCPS_376} of orbital-free density functional theory (OF-DFT) \cite{Mi2023_CR_12039} by applying the expressions for the homogeneous electron gas to the inhomogeneous system of electrons moving in the field of the two nuclei.

\subsection{\red United atom approach for repulsive potentials}
\label{subsec:uniteatom}

Although the superposition of atomic densities assumed in the ZBL scheme is a reasonable approximation at distances beyond the equilibrium distance, $r \gtrsim r_{\rm equi}$, \cite{Almloef1982_JCC_385, VanLenthe2006_JCC_32, Lehtola2019_JCTC_1593} it does not take into account that when the two atoms approach one another, the exact electronic wave function is strongly modified from the ground state of the atoms at dissociation \cite{Sabelli1979_PRA_677}.

To understand this, let us consider a simple thought experiment: when $r\to 0$, the electron density of the system of two atoms $Z_1$ and $Z_2$ must in fact approach the electron density of the united atom with atomic number  $Z_1 + Z_2$.
This case is achieved in practice already when the internuclear distance is smaller than the $1s$ orbital of the united atom;
we denote this distance $r_{\rm ua}$, where ``$\rm ua$'' stands for united atom.
{\red This united atom approach has been recently introduced for repulsive potentials \citeref{Lehtola2020_PRA_32504}, and in the current work we apply and analyze it systematically for all atom pairs in the non-relativistic limit.}

If superposition could be applied when $Z_1 = Z_2 = Z$, the electron density of the united atom with atomic number $2 Z$ would be simply twice that of $Z$.
Yet, already the analysis of the hydrogenic atom shows that this is not true: the radius of the $1s$ orbital of the united atom is actually half that of the separate atoms.
As screening effects are negligible for the $1s$ orbital, this simple argument shows that the superposition of atomic densities leads to a completely incorrect form for the density at the united atom limit.

\begin{figure}
\begin{center} 
\includegraphics[width=0.9 \columnwidth]{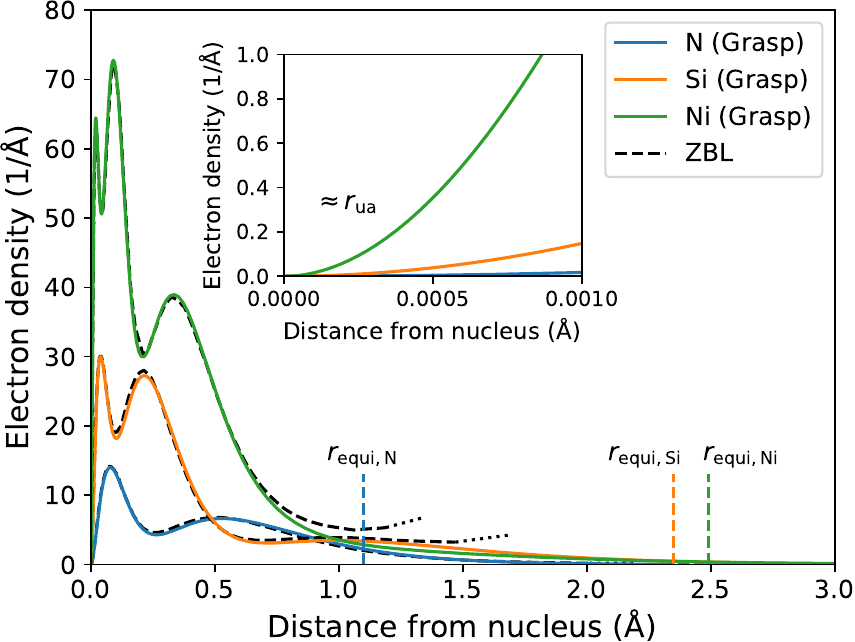} 
\end{center}
\caption{
Electron densities of the gas-phase N, Si and Ni atoms from the Grasp code \cite{Fro19, Sch22}.
Also indicated are the equilibrium nearest neighbour interatomic separations $r_{\rm equi}$ in the $N_2$ dimer and in solid crystalline Si and Ni.
The inset shows the distances $r \lesssim r_{\rm ua}$ where the united atom approximation may be used. 
Also shown are the electron densities used in the ZBL model (dashed lines).
}
\label{fig:eldens}
\end{figure}

The correct electronic structure is easy to determine at the united atom limit, and the resulting electron density is indeed found to differ from the superposed electron densities of the two atoms assumed by the ZBL model. 
\Cref{fig:eldens} illustrates that the electron density of $^{28}$Ni is not twice the electron density of $^{14}$Si, which is not twice the electron density of $^7$N, as the radial structures of the electron densities of the three atoms are pronouncedly different.
One can also note that while the real electronic wave function satisfies the Pauli exclusion principle, the superposition of atomic densities breaks this in a striking fashion by placing the electrons from both atoms in the same orbital at the united atom limit.

With a correct description of the united atom limit, the interatomic potential can be reliably computed at small $r$, $r \lesssim r_{\rm ua}$, using perturbation theory \cite{Pathak1987_JCP_2186}, for example.
However, capturing the electronic structure at intermediate separations, $r_{\rm ua} < r \ll r_{\rm equi}$, where the inner electronic shells of the two atoms partly overlap while the outermost shells resemble those of the united atom \cite{Sabelli1978_JCP_2767}, is quite challenging.
Excluding the case of hydrides, where the united atom model yields a qualitatively correct model of the electronic structure of the diatomic molecule \cite{Mulliken1932_RMP_1, Hirao2011_WCMS_337}, in the general case, the orbitals of the diatomic molecule undergo significant relaxation effects from those of the free atoms, and the chemistry of the high-energy region remains relatively unexplored.

In interesting cases, the occupied orbitals undergo thorough changes in character at intermediate separations: for example, in the \ce{NeCa <=> Zn}, \ce{MgAr <=> Zn} and \ce{Ar2 <=> Kr} barriers studied in \citeref{Lehtola2020_PRA_32504}, one goes from $s$ and $p$ electrons of the incoming atoms at separation to $s$, $p$, and $d$ electrons in the united atom limit.
These shell reorganizations occur somewhere in the intermediate region, and are associated with huge changes in the total energy: for example, the relaxation induced by allowing $d$ orbital occupation in the Zn atom is around 600 eV, and in the Kr atom around 2000 eV.

The ground spin state can also change along the internuclear distance.
For example, the collision of two Mg atoms---each of which has a singlet ground state---will produce a Cr atom at $r \to 0$, whose ground state is a septet.
It therefore appears that a systematic study of this system should consider the whole manifold of spin states from the singlet to the septet, as the transition from the singlet state at separation to the septet for the united atom may also involve the triplet and quintet states along specific internuclear distances.

Only around the chemical equilibrium distance $r_{\rm equi}$ is there broad knowledge of the electronic structure of diatomic molecules from conventional quantum chemistry.
Yet, the electronic structure of many diatomic molecules around the equilibrium distance remains unknown to this day, especially for systems involving open $d$ or $f$ shells, that is, molecules containing transition metals or lanthanides.
As an example, the ground states of many carbides and chlorides of first-row transition metals are still unknown \cite{Hait2019_JCTC_5370}, and molecules with e.g. two transition metal atoms are even more challenging.

Previous work in \citeref{Nor96c} has shown that the error in $\phi(r)$ is almost totally governed by the quality of the one-particle basis set, with the differences between various levels of theory (e.g. Hartree--Fock vs coupled-cluster theory vs density functional theory) being negligible with respect to the pursued level of accuracy.
Similarly, the role of the employed charge state is expected to have a negligible effect on the repulsive barrier, as differences of the order of 1 eV, that would be unacceptable for modeling the chemical equilibrium, become negligible in the highly repulsive part of the interatomic potential dominated by $V_{\rm Coul}(r)$.

As most electronic structure methods have been developed for the region $r\approx r_{\rm equi}$, a major problem in the present effort is to find a stable numerical representation that can adapt to the varying electronic configurations found in the regime $r_{\rm ua} < r < r_{\rm equi}$.
A {\red  major methodological advance } was recently achieved {\red by one of us} \citeref{Lehtola2020_PRA_32504}, in showing that the strongly repulsive region can be faithfully modeled with established atomic basis set approaches, if the united atom limit is included in the construction of the basis set and if the significant linear dependencies that arise in the basis set at small internuclear distances are properly taken care of.
{\red In the current work, we employ} this method to verify the accuracy of the calculations carried out with the two other methods discussed below.

\section{Methods \label{sec:methods}}

\subsection{Gaussian-type orbital calculations {\red ("MP2")}}
\label{subsec:MP2}

Preceding work in \citeref{Nor96c} found that the repulsive potential can be accurately reproduced with Gaussian-type orbital (GTO) calculations, if additional basis functions are added on the bond between the nuclei.
In this work, we follow \citeref{Lehtola2020_PRA_32504} and go beyond the use of bond functions by ensuring that the united atom limit is correctly reproduced by the calculation.
With this aim, we introduce ghost basis functions \cite{Boys1970_MP_553} corresponding to the united atom at the center of charge at $z(r)=Z_B r / (Z_A+Z_B)$, when atoms $A$ and $B$ are placed at the origin and at $(0,0,r)$, respectively.
State-of-the-art polarization consistent (pc-$n$) GTO basis sets by \cite{Jensen2001_JCP_9113} are employed in all calculations, with the basis set for the ghost atom also including additional polarization functions.
To allow even the deep core orbitals to hybridize, i.e., to form bonding and antibonding orbitals, all the GTO basis sets are employed in fully uncontracted form.
After testing, the final results are shown for the triple-$\zeta$ pc-2 basis set, which is expected to afford screening functions of high quality.

The resulting molecular basis set has pathological linear dependencies for vanishingly small interatomic distance, $r\to 0$.
A pivoted Cholesky decomposition procedure was employed to remove linearly dependent basis functions \cite{Lehtola2019_JCP_241102}; this method was shown in \citeref{Lehtola2020_PRA_32504} to afford excellent accuracy in Hartree--Fock (HF) calculations compared to ones performed with the HelFEM program \cite{Lehtola2019_IJQC_25968, Lehtola2019_IJQC_25944} that employs numerical basis functions following the finite element method, reproducing values directly at the complete basis set limit.

A threshold of $10^{-6}$ was employed in the pivoted Cholesky procedure to choose a linearly independent set of basis functions from the whole basis set, while a threshold of $10^{-5}$ was employed to form orthonormal linear combinations thereof to serve as the expansion basis for the HF molecular orbitals.
The PySCF program package was employed for the GTO calculations \cite{Sun2020_JCP_24109}; note that PySCF evaluates all the necessary integrals analytically.

The calculations were run separately for all atom pairs and each basis set.
The calculations for each such combination were carried out as spin-unrestricted.
Based on the findings of \citeref{Nor96c}, we assume that the chosen spin state or configuration will have negligible effect on the strongly repulsive barrier; correspondingly, the calculations were carried out for the lowest spin multiplicity $M$ possible: $M=1$ or $M=2$ for even and odd numbers of electrons, respectively.

The self-consistent field calculations were carried out with a Newton--Raphson solver in combination with stability analysis \cite{Seeger1977_JCP_3045} to allow the spin symmetry to break.
Note that in reference to the above discussion, this choice of methodology should also allow the calculations to converge on a higher spin state, which may be useful for systems such as the Mg-Mg case discussed in \cref{sec:intro}: the use of spin-unrestricted orbitals with stability analysis in principle allows switching the spin state along the internuclear coordinate by the introduction of ``spin contamination'', even though differences in the screening functions for various spin states are expected to be negligible at small $r$.

The calculation begins at the largest internuclear distance, for which the density is initialized with a superposition of atomic densities \cite{Almloef1982_JCC_385}, while the calculations for all successive points were initialized with the electron density of the previous point.
The atomic total energies $E_A$ and $E_B$ were determined in separate calculations.

When the HF calculation converged, the electron correlation effects neglected by HF were estimated with second-order M\o{}ller--Plesset perturbation \cite{Moller1934_PR_618} (MP2) theory, yielding the final energy employed in the GTO calculations of this work.


\subsection{Numerical atomic orbital calculations {\red ("DMol")}}
\label{subsec:dmol}

As a followup of the work presented in \citeref{Nor96c}, we used the DMol97 code to calculate the interatomic repulsive potential for all atom pairs $Z_1,Z_2 \leq 92$ soon after \citeref{Nor96c} was published. 
These results have since been used in numerous publications of radiation effects; see, for instance, \citerefs{Tar98b,Pel03d,Hen05a,Kot05b,Jus07,Nor16}. 
The novelty in the current work is to cross-check the DMol97 results against the new data obtained with the approach of \cref{subsec:MP2}, which should be more accurate from many points of view. 
Another contribution of this work is the publication of all the pair-specific DMol97 potentials as auxiliary data sets and a systematic comparison of the new models with the ZBL.

The approach to use DMol \cite{Del90, DMol, DMolManual} to calculate pair potentials within all-electron density functional theory (DFT) \cite{Hohenberg1964_PR_864, Kohn1965_PR_1133} was introduced in \citeref{Kei94}. 
The code has the advantage that it uses numerical atomic orbitals that can be of any form, and the program can be run in all-electron mode.
We found in \citeref{Nor96c} that it was crucial to augment the default DMol numerical atomic orbital basis sets with hydrogenic orbitals \cite{Del90, DMolManual} to obtain a good description of the repulsive potential also at high energies.
Moreover, the basis sets for each element with atomic number $Z$ were further augmented with the hydrogenic basis sets for element $Z-1$, which was found to further improve the convergence of the calculations.
For elements with $Z<40$, the hydrogenic basis sets for shells $n=1,2,3,4$ were used for both $Z$ and $Z-1$. 
For elements with $41 < Z < 55$, hydrogenic basis sets for shells $n=1,2,3,4,5$ were used for $Z$, and $n=1,2,3$ for $Z-1$.
For elements with $Z>54$, hydrogenic basis sets where used for shells $n=1,2,3$. 
Since the heaviest elements had a large number of regular numerical atomic orbitals, a limit of 19 on the number of orbitals used in the DMol97 code  prevented us from using a larger number of hydrogenic orbitals for the heaviest elements.

The DMol97 calculations were carried out including scalar relativity in the "vpsr" pseudopotential \cite{DMolManual}, employing LDA exchange \cite{Bloch1929_ZfuP_545, Dirac1930_MPCPS_376} and the Vosko--Wilk--Nusair (VWN) correlation functional \cite{Vosko1980_CJP_1200}.

The DMol calculations were carried out for 74 internuclear distances from
0.002 Å upwards.
The largest considered internuclear distances were intentionally set to 100 Å and 1000 Å, as the two atoms essentially do not interact at such large distances; hence, the energy at 1000 Å was used to determine the reference energy 
$E^{\rm tot}_1 + E^{\rm tot}_2$. 

Note that this approach can lead to an energy that differs from the sum of the energies of two free atoms, as a different spin state may be required to properly dissociate the atoms i.e. reproduce the latter energy exactly.
However, the difference in total energy is only of the order of $\sim 1$ eV, which is very small compared to the energies in the repulsive region $\gg 30$ eV, and the forces derived from the potential $F={\rm d}E/{\rm d}r$ that are used in MD simulations are not affected by the choice of the reference level, since the derivative of a constant vanishes. 

\subsection{Orbital-free density functional calculations {\red ("ZBL pair-specific")}}
\label{subsec:zbl}

While pair-specific ZBL interatomic potentials have occasionally been used in literature without reference to their source \cite{Park91, Hob93, Hob95a}, the code or data used by ZBL are no longer available, except for the electron densities and the obtained ``universal'' fit. 
We have reimplemented the OF-DFT method following \citet{Wed67}, which was likewise used by ZBL.
In this approach, the electronic terms in \cref{eq:V} are reorganized into the sum of an electronic Coulomb term $V_c(r)$ and quantum-mechanical terms that describe the electrons' kinetic and exchange energy, $V_k(r)$ and $V_x(r)$, respectively, as
\begin{equation} \label{eq:ofdft}
    V(r) = V_{\rm Coul} (r) + V_c(r) + V_k(r) + V_x(r) .
\end{equation}
As discussed in \cref{sec:intro}, the terms are evaluated assuming that the electron densities of the two atoms remain unchanged and add up linearly.
As \cref{eq:V} consisted of differences of energies between the diatomic system and the two atoms at separation, the lattermost three terms in \cref{eq:ofdft} represent quantities that are commonly referred to as ``excess energies'', as they compare the energy of the diatomic system to that of its constituent atoms at separation. 
The repulsive nuclear Coulomb term $V_{\rm Coul}(r)$ already appeared in \cref{eq:V}, and its expression was given in \cref{eq:Coulomb}.

The electronic Coulomb term $V_c(r)$ contains the classical nucleus-electron and electron-electron interactions, whose exact expressions are well-known.
The excess kinetic energy $V_k$ is approximated with the orbital-free Thomas--Fermi (TF) LDA functional \cite{Thomas1927_MPCPS_542, Fermi1927_RdNdL_602}. 
Denoting the electron densities of atoms 1 and 2  as $\rho_1$ and $\rho_2$, respectively, the TF expression for the excess kinetic energy reads
\begin{equation}
    \label{eq:tf}
    V_k = \kappa_k \int [ (\rho_1({\bf r})+\rho_2({\bf r}))^{5/3} 
          - \rho_1^{5/3}({\bf r}) - \rho_2^{5/3}({\bf r})] \, {\rm d}^3r
\end{equation}
where the constant is
\begin{equation}
    \label{eq:kappa-k}
\kappa_{k} = 
        \frac{3}{5} \, \frac{\hbar^2 \pi^2}{2 m_{e}} \left( \frac{3}{\pi} \right) ^{2/3} \approx 21.879\mathrm{\, eV\AA^2}. 
\end{equation} 
The excess exchange energy $V_x$ can be computed analogously by the LDA \cite{Bloch1929_ZfuP_545, Dirac1930_MPCPS_376} with a similar expression, where the power 5/3 is replaced by 4/3 and $\kappa_k$ is replaced by \cite{ZBL, Wed67}.
\begin{equation}
    \label{eq:kappa-x}
\kappa_{x} = 
        \frac{3}{4} \, 
        \frac{e^2}{4\pi\varepsilon_0} 
        \left( \frac{3}{\pi} \right) ^{1/3} \approx 10.635\mathrm{\, eV\AA}.
\end{equation} 

Our numerical approach to the evaluation of the integrals appearing in $V_c(r)$, $V_k(r)$, and $V_x(r)$ differs from that of \citerefs{ZBL} and \citenum{Wed67} and is presented in detail in \cref{sec:orbital-free-appendix}.
The  listings supplied by ZBL in \citeref{ZBL} were used for the electron densities $\rho_1(\bf r)$ and $\rho_2(\bf r)$, comprising all stable atoms plus Bi and U.
The densities had been obtained by ZBL by spherically and spin averaging models of solid-state electron densities $\rho({\bf r})$ constructed by superposing atomic electron densities from exchange-only LDA calculations, except that for some systems, they extracted averaged atomic electron densities from true solid-state LDA calculations.
We compared our results for  $V_c$, $V_k$, and $V_x$ for the B-B and Au-Au systems against data given in \citeref{ZBL}, and found our data to be in excellent agreement with the calculations of ZBL.

In addition, we also applied our reimplementation to electron densities of atoms in the gas phase, which were calculated using a recently published module \cite{Sch22} for the Grasp2018 program \cite{Fro19}. Grasp2018 is based on the fully relativistic multiconfiguration Dirac--Hartree--Fock (MCDHF) method. 
A comparison of the resulting spherically averaged solid-state and atomic electron densities for N, Si, and Ni is given in \cref{fig:eldens}.


%

\section{Results \label{sec:results}}

\subsection{Comparison with earlier model calculations \label{sec:earlier}}

As discussed in \cref{sec:intro}, it is by now well known that the quality of the screening function $\phi(r)$ in the repulsive region is primarily determined by the quality of the one-electron basis set.
The pioneering study in \citeref{Nor96c} examined the importance of the level of theory, and showed that the differences between screening functions reproduced by Hartree--Fock, DFT, MP2, and coupled-cluster theory are negligible for the \ce{Si\bond{-}Si} system.
\citet{Nor96c} also showed by comparison to numerically exact Hartree--Fock calculations (which are free of basis set truncation error) that the DMol approach, also employed in this work, can reproduce screening functions for \ce{H\bond{-}Si}, \ce{N\bond{-}Si}, \ce{Si\bond{-}Si}, \ce{C\bond{-}C} with an accuracy of the order of 1\% when a rich enough basis set is used.

The issue of the basis set truncation error was revisited in the recent study of \citet{Lehtola2020_PRA_32504}, who showed that an enriched basis similar to the one employed in this work (see \cref{subsec:MP2}) provides an excellent level of agreement with fully numerical calculations at the Hartree--Fock level of theory. 
In the following, we therefore assume that the MP2 calculations carried out in the fully uncontracted triple-$\zeta$ pc-2 basis set enriched with united-atom basis functions at the center of nuclear charge (see \cref{subsec:MP2}) provide a good estimate of the screening functions.

\begin{figure}
\begin{center} 
\includegraphics[width=0.9\columnwidth]{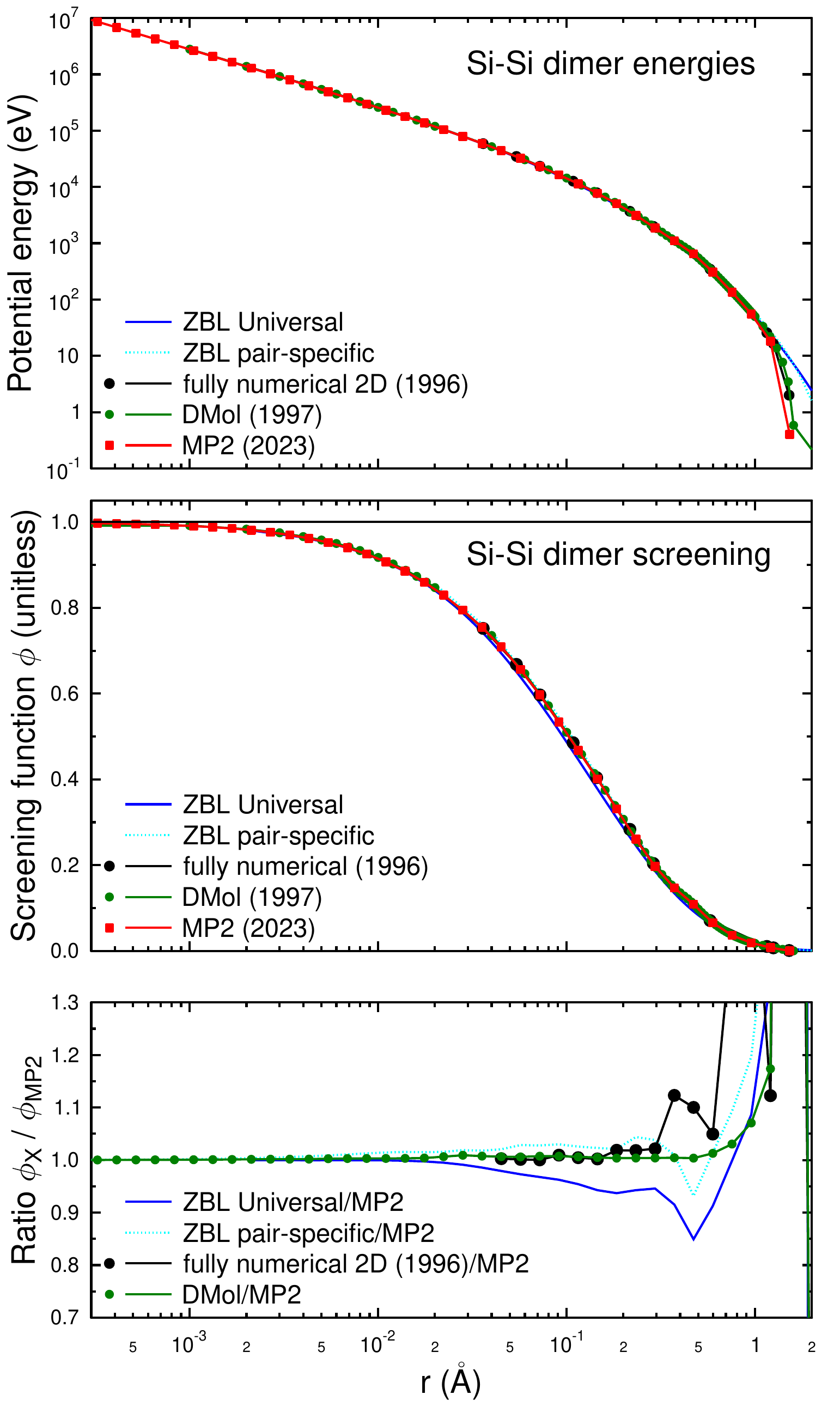} 
\end{center}
\caption{
Comparison of repulsive interatomic potentials calculated with four
different approaches: The universal ZBL approach, the fully numerical Hartree--Fock "2D" code
and DMol calculations from \citeref{Nor96c} as well as the new
MP2 data of this work.
}
\label{fig:sisi}
\end{figure}

We begin our analysis by comparing the new MP2 results with the two versions of the ZBL potential, and with the earlier results of \citet{Nor96c}, both their fully numerical Hartree--Fock calculation and their original DMol calculations.
Results for the Si-Si system are illustrated in \cref{fig:sisi}, where the top part shows the potential energy curve $V(r)$, the middle part shows the screening function $\phi(r)$, and the bottom part shows the ratio of the screening functions.
Since the differences between the methods are nearly undiscernable in the top and middle plots, we also plot the ratios of the resulting potential energies in the bottom part of the plot to make the differences better visible.

The results in \cref{fig:sisi} show that the universal potential of ZBL starts to deviate from the pair-specific quantum chemical approaches already above 0.02 Å, whereas the quantum chemical approaches agree well with each other up to a distance of 0.3 Å.
In terms of potential energy, these distances correspond to about 100 keV and 2 keV, respectively.
It is remarkable that even though the purely repulsive ZBL potential is specifically meant for the high-energy regions, it still starts deviating from the quantum chemical data already at {\red quite small distances and high energies. E.g. at the distance of $r=0.1$ Å where the ZBL potential has a value of 13.7 keV, it already deviates by 4 \% from the MP2 potential}.

The potential energy $V$ and screening function $\phi$ have the same information, as they are connected by a mathematically exact reversible mapping (multiplication or division by the internuclear Coulomb potential, see \cref{eq:screen}). 
In the remainder of the paper we illustrate the results only with the screening functions, and remind the reader that ratios of screening functions coincide with ratios of interatomic potentials.

\subsection{Comparison of selected potentials \label{sec:selectedpotentials}}

\begin{figure}
\begin{center} 
 \begin{subfigure}[b]{0.9\columnwidth}
    \centering
\includegraphics[width=0.9\columnwidth]{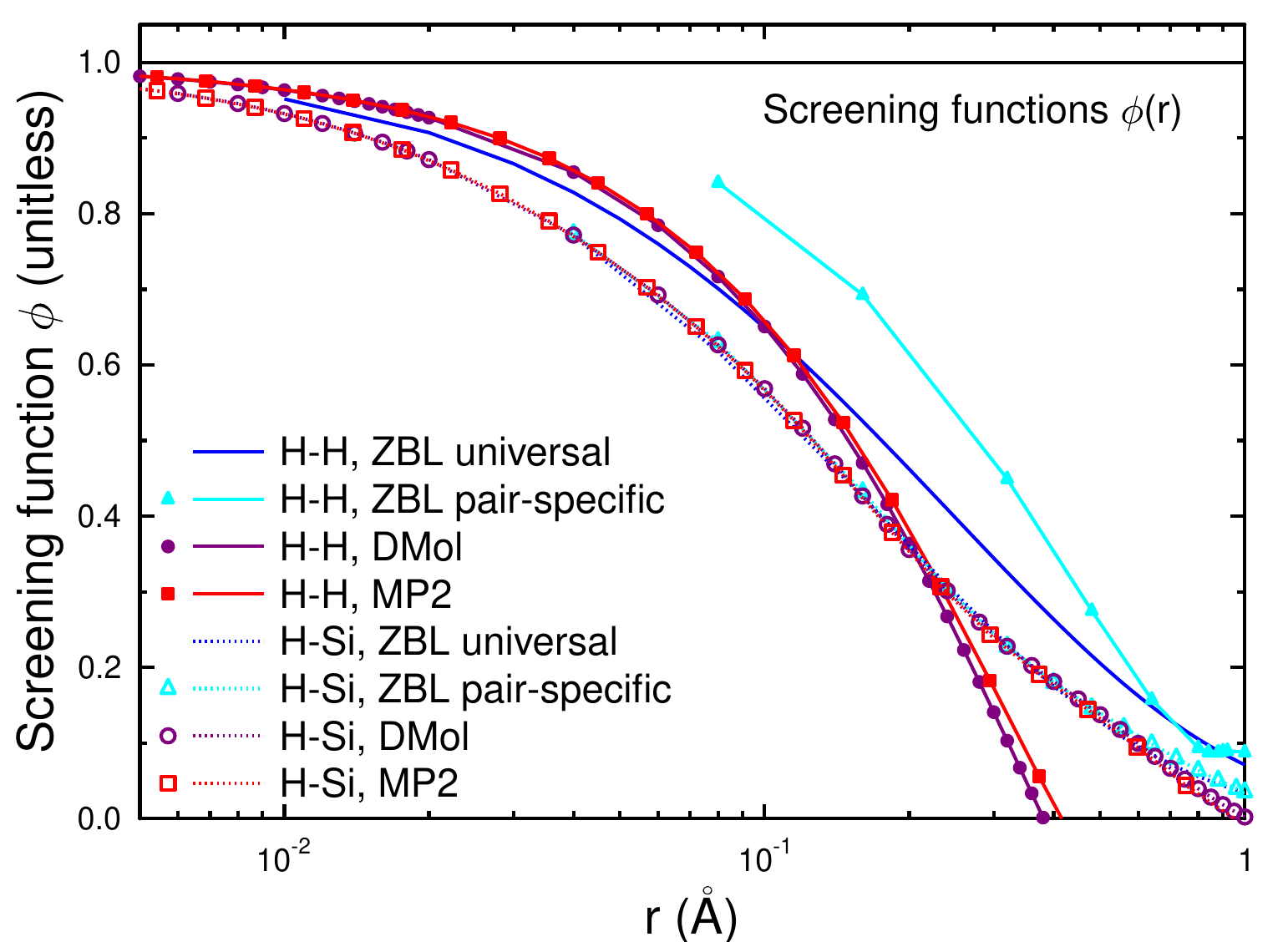} 
\caption{Screening functions for H-Si and Si-Si.}
\end{subfigure}
\begin{subfigure}[b]{0.9\columnwidth}
    \centering
\includegraphics[width=0.9\columnwidth]{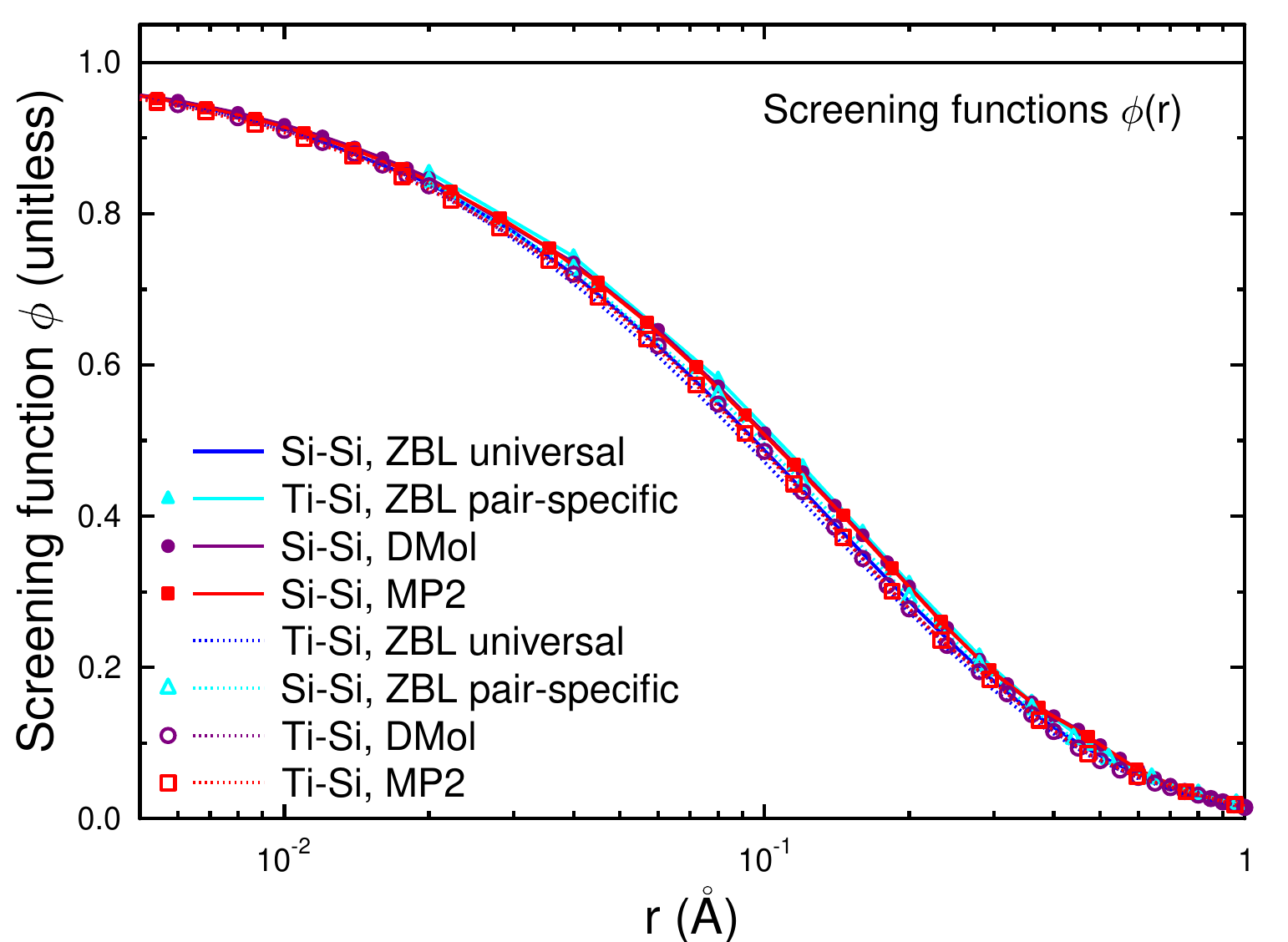} 
\caption{Screening functions for Si-Si and Ti-Si.}
\end{subfigure}
\end{center}
\caption{
Screening functions of repulsive interatomic potentials calculated with three approaches: the universal ZBL approach, DFT calculations with DMol, as well as the MP2 calculations of this work.
The black line shows $\phi(r)=1$ corresponding to unscreened Coulomb repulsion, for reference.
}
\label{fig:manyscreens}
\end{figure}

\Cref{fig:manyscreens} compares screening functions for four atom pairs: H-H, H-Si, Si-Si and Ti-Si.
In the case of H-H, the ZBL potential clearly deviates from the quantum chemical calculations (DMol and MP2). 
The quantum chemical methods agree well with each other within the plotting accuracy, in line with earlier findings of \citerefs{Lehtola2020_PRA_32504} and \citenum{Nor96c}.

\begin{figure}

\begin{center} 
\includegraphics[width=0.95\columnwidth]{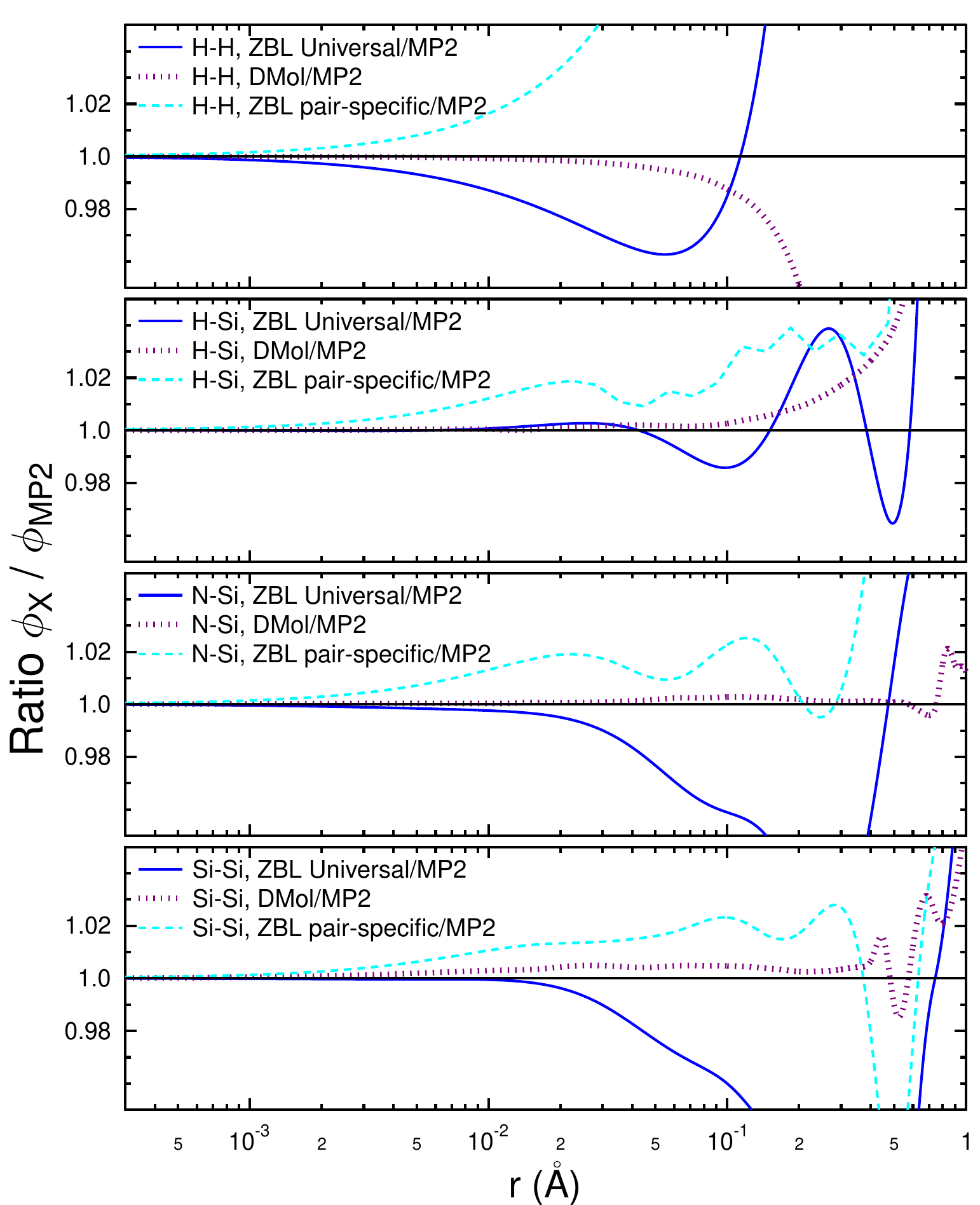} 
\end{center}
\caption{
Comparison of screening functions of repulsive interatomic potentials calculated with three different approaches.
The plots show the screening function produced by the universal ZBL approach, the DMol calculations, and the pair-specific ZBL calculations divided by the MP2 screening function calculated in this work for the H-H, H-Si, N-Si, and Si-Si atom pairs.
}
\label{fig:manyscreenratios}
\end{figure}

\begin{figure}
\begin{center} 
\includegraphics[width=0.95\columnwidth]{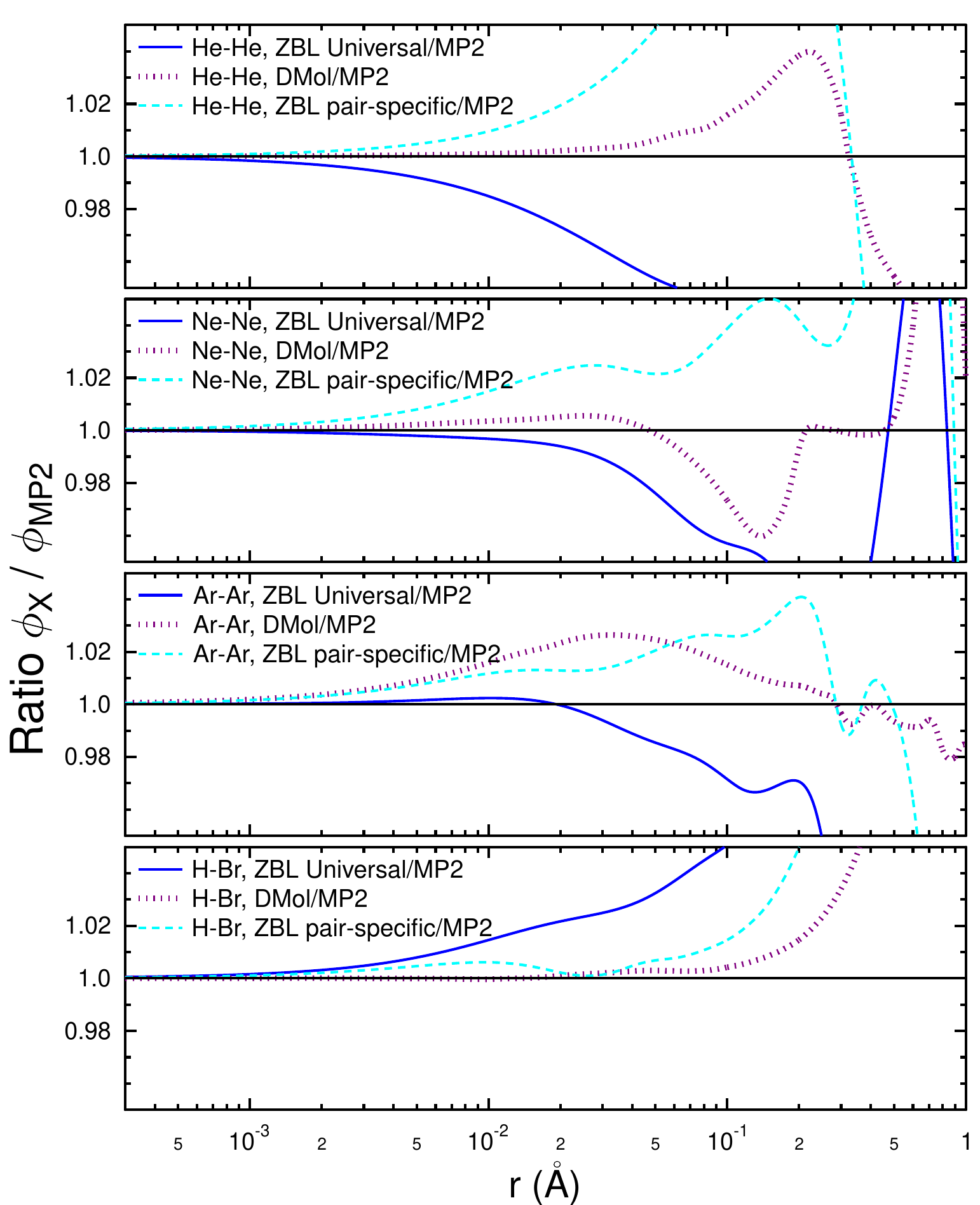} 
\end{center}
\caption{
As \cref{fig:manyscreenratios}, but for the He-He, Ne-Ne, Ar-Ar and H-Br
element pairs.
}
\label{fig:manyscreenratios2}
\end{figure}

\begin{figure}
\begin{center} 
\includegraphics[width=0.95\columnwidth]{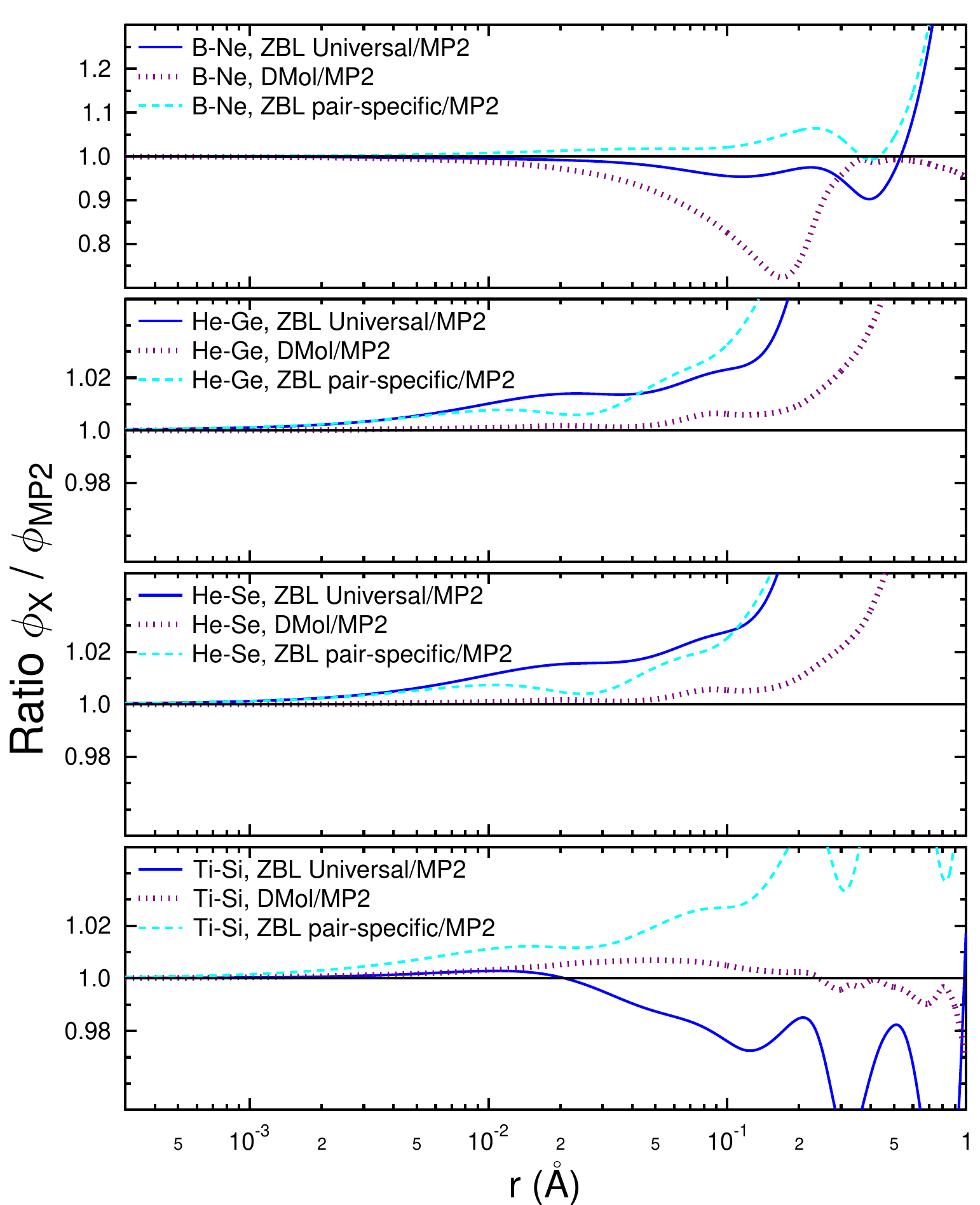} 
\end{center}
\caption{
As \cref{fig:manyscreenratios}, but for B-Ne, He-Ge, He-Se, and Ti-Si
element pairs.
Note that the case of B-Ne is plotted on a different ordinate scale than the He-Ge, He-Se and Ti-Si curves, as well as all the data in \cref{fig:manyscreenratios,fig:manyscreenratios2}.
}
\label{fig:manyscreenratios3}
\end{figure}

To better distinguish the potentials, we show the corresponding ratios of the ZBL universal, DMol, and pair-specific potentials to the MP2 potential in \cref{fig:manyscreenratios} for the same element pairs that were examined in \cref{fig:manyscreens}. \cref{fig:manyscreens} shows Ti-Si, while \cref{fig:manyscreenratios} shows N-Si.
Further data for other element pairs are shown in \cref{fig:manyscreenratios2} (He-He, Ne-Ne, Ar-Ar, and H-Br), and in \cref{fig:manyscreenratios3} (B-Ne, He-Ge, He-Se, and Ti-Si). 
As the ZBL universal potential is an averaging fit, it is not surprising that it differs strongly from the MP2 data. 
However, it is somewhat surprising that also the pair-specific ZBL potentials differ from the MP2 data, and that several of these pair-specific potentials exhibit similar-magnitude differences from MP2 as the universal ZBL potential.

Comparison of the DMol and MP2 data shows that these quantum chemical potentials are in excellent agreement with each other: the agreement is within 2\% at almost all distances considered in the repulsive region. 
The only exception is the case of B-Ne, where the difference is up to 20\% (\cref{fig:manyscreenratios3}). 
This indicates that the DMol approach is not suitable for calculation of repulsive potentials for the B-Ne case, and we recommend using the MP2 data instead for this system.

\subsection{Comparison to orbital-free calculations \label{sec:solidstateresults}}

\begin{figure}
\begin{center} 
\includegraphics[width=0.9\columnwidth]{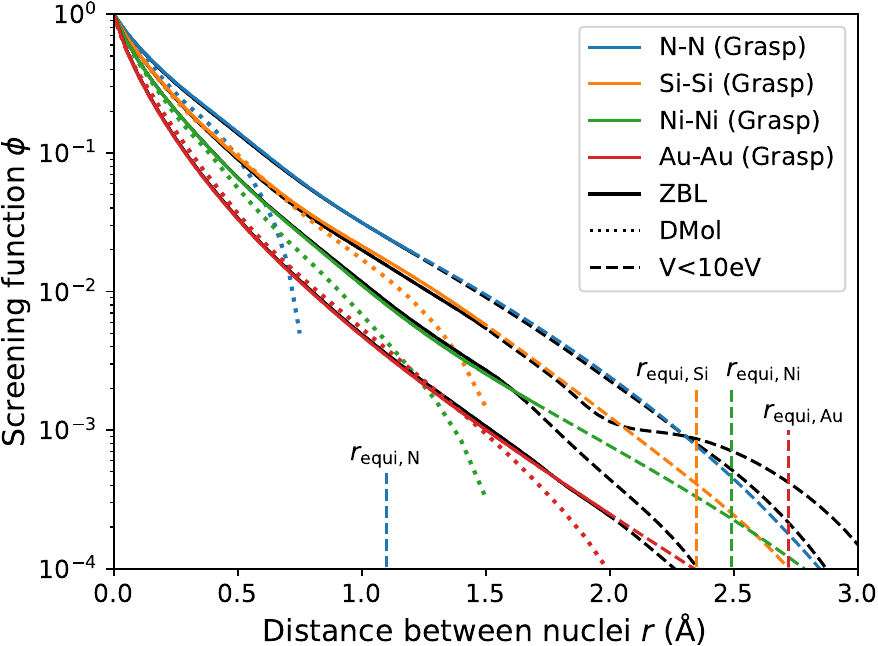}
\end{center}
\caption{
Screening functions obtained from OF-DFT calculations using electron densities from Grasp (colored solid lines with dashed extensions) or ZBL solid-state electron densities (black solid lines with dashed extensions).
Screening functions from DMol are shown as reference (dotted colored lines); the OF-DFT calculations (Grasp, ZBL) are much closer to each other than to the DMol calculations.
The dashed lines indicate distances for which the interatomic potential is less than 10~eV.
}
\label{fig:screen-grasp-zbl-dmol}
\end{figure}

The screening functions  for four homonuclear systems (N-N, Si-Si, Ni-Ni, and Au-Au) obtained within OF-DFT are compared to those obtained with the self-consistent DMol calculations in \cref{fig:screen-grasp-zbl-dmol}.
OF-DFT data is presented for both calculations based on electron densities of gas-phase atoms (Grasp), as well as solid-state electron densities (ZBL).
The results of the two OF-DFT calculations are rather close to each other, indicating that the choice of the electron density between atoms in the solid or gas phase is of minor relevance; this is particularly true for interaction energies larger than 10~eV (solid lines in \cref{fig:screen-grasp-zbl-dmol}).

\begin{figure}
\begin{center} 
\includegraphics[width=0.9\columnwidth]{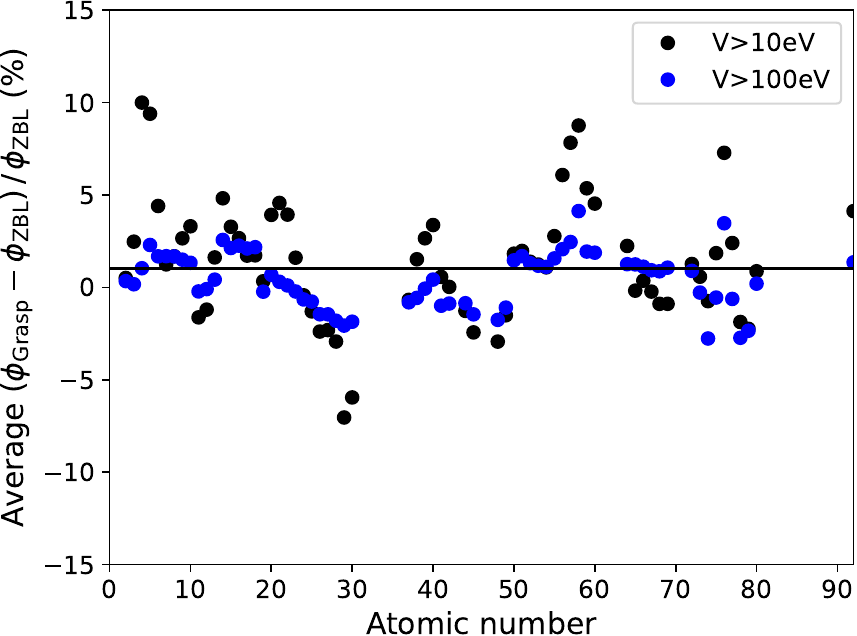}
\caption{
Average relative difference between gas phase and solid-phase potentials in the LDA to OF-DFT for homonuclear interactions where valid data have been obtained for both cases.
}
\label{fig:error-screen-grasp}
\end{center}
\end{figure}

A thorough analysis of the average relative differences between gas-phase (Grasp) and solid-state (ZBL) screening functions
\begin{equation}\label{eq:zblerror}
    \Delta_{Z_1,Z_2} = {1 \over N} \sum_{i}^N {\phi^{Z_1,Z_2}_{\rm Grasp}(r_i) - \phi^{Z_1,Z_2}_{\rm ZBL}(r_i) \over \phi^{Z_1,Z_2}_{\rm ZBL}(r_i)} 
\end{equation}
is shown in \cref{fig:error-screen-grasp} for the studied set of homoatomic systems ($Z_1=Z_2=Z$ in \cref{eq:zblerror}). 
As the near-equilibrium attractive part of the potential is not of interest in this work, the average in \cref{eq:zblerror} was taken over the subset of points for which the potential is either larger than 10~eV or 100~eV; this also ensures that $\phi(r_i) > 0$ in \cref{eq:zblerror}.
The resulting mean absolute errors for the two cutoffs are 2.7\% and 1.3\%, respectively, confirming that the OF-DFT model yields similar screening functions regardless of whether the atomic electron density is taken from gas-phase or solid-state calculations.
{\red In Appendix A, a systematic approach to join diatomic pair potentials to equilibrium potentials based on solid state DFT data is presented. 
This joining procedure should, at least in principle, also correct for the lack of solid-state effects in the diatomic potentials.}

In contrast, the results of the OF-DFT calculations deviate strongly from the DMol data, consistently with the findings above in  \cref{sec:selectedpotentials}.
The differences are particularly large at large internuclear distance, see \cref{fig:screen-grasp-zbl-dmol}: the DMol potential drops roughly down to zero in a region where the ZBL potential is still around 10 eV (Si, Ni, and Au), or even earlier (N).
This likely reflects the failure of the OF-DFT calculation to adequately describe bond formation, which is a well-known failure of Thomas--Fermi theory \cite{Teller1962_RMP_627}. Also, these calculations employed frozen electron densities, introducing another significant source of error at small internuclear distances where the electron density is strongly modified.

\subsection{Quantitative analysis of differences \label{sec:systematic}}

\subsubsection{MP2 reference \label{sec:mp2ref}}

To quantitatively assess the average difference between the potentials, we summed up all the differences between the ZBL universal, ZBL pair-specific, and DMol screening functions and the MP2 reference value with a metric similar to \cref{eq:zblerror}: \begin{equation}\label{eq:error}
    \Delta_{Z_1,Z_2} = {1 \over N} \sum_{i}^N {\phi^{Z_1,Z_2}_{\rm X}(r_i) \over \phi^{Z_1,Z_2}_{\rm MP2}(r_i)} - 1
\end{equation}
where the sum runs over the $N$ data points in the MP2 potential. 
Analogously to \cref{eq:zblerror}, the average in \cref{eq:error} was restricted to points either above 10, 30 or 100 eV in the MP2 value.

The comparison of potentials is slightly complicated by the different choices of $r$ grids at which the potentials were tabulated. 
The choice of points may also weigh the comparison towards small distances. 
To make a systematic comparison regardless of the choice of tabulations, all the screening functions were interpolated to an equidistant grid with a 0.01 Å interval starting from the origin, while the analytical ZBL potential was evaluated exactly.
To ensure high accuracy between tabulation points, cubic spline interpolation was used \cite{NumericalRecipes}; the same interpolation scheme is also used in the MDRANGE code for read-in potentials.
We tested that using an interval of 0.001 Å gave essentially the same average deviations.

\begin{figure}
  \centering
  \begin{subfigure}[b]{0.9\columnwidth}
    \centering
    \includegraphics[width=\textwidth]{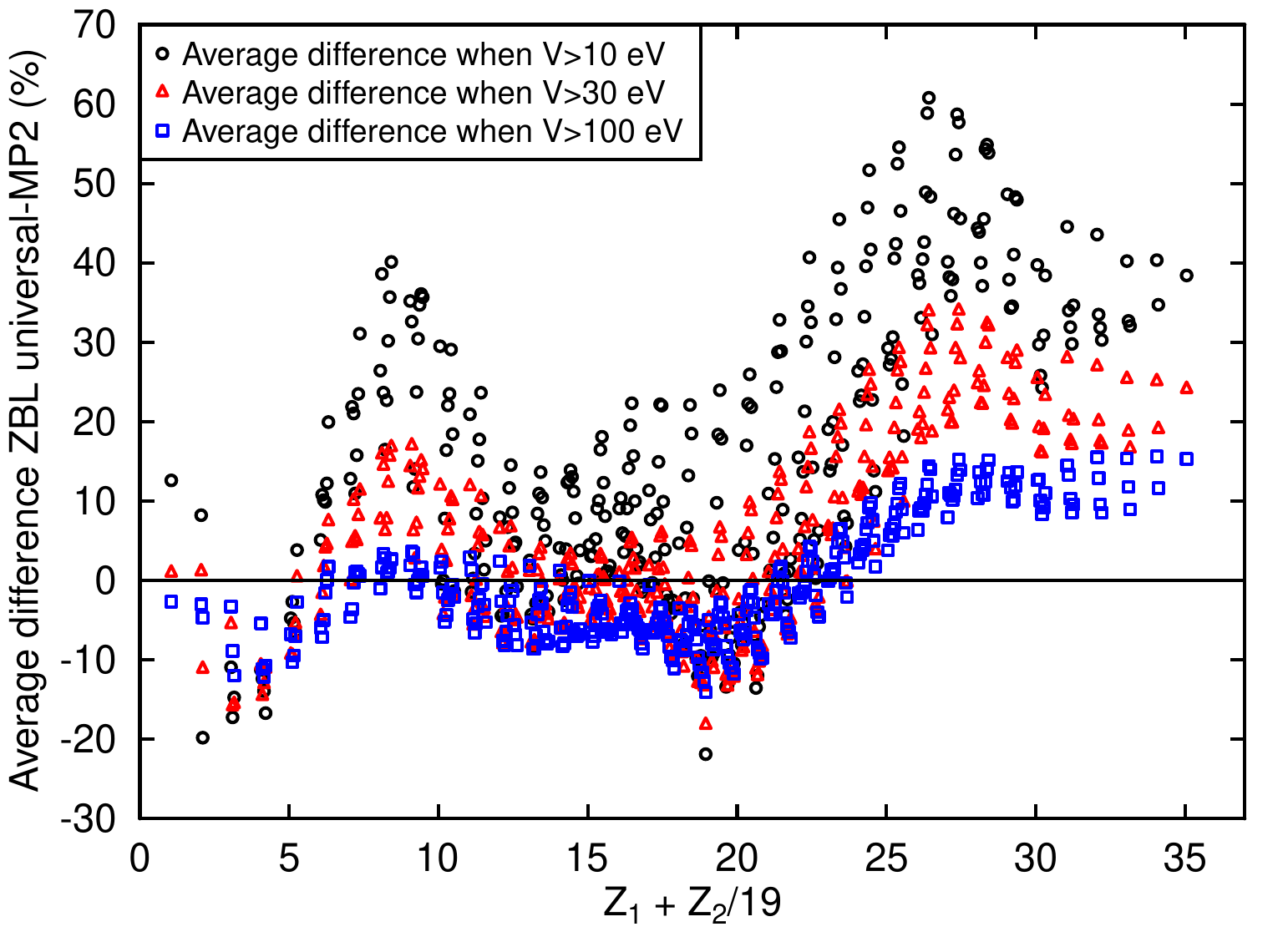}
    \label{fig:allaverage-zbl-mp2}
    \caption{ZBL universal compared to MP2}
  \end{subfigure}

  \begin{subfigure}[b]{0.9\columnwidth}
    \centering
    \includegraphics[width=\textwidth]{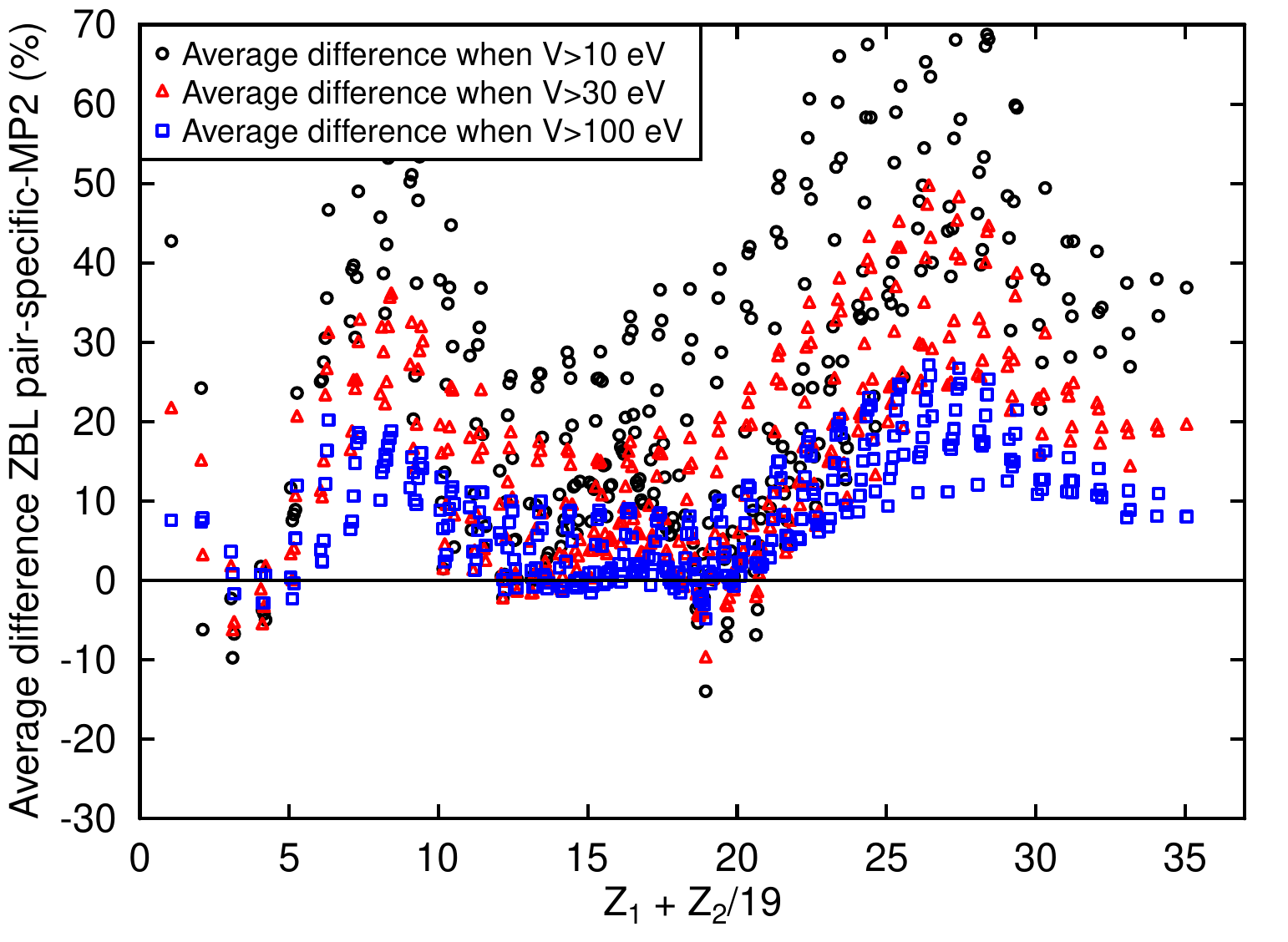}
    \label{fig:allaverage-zblpair-mp2}
       \caption{ZBL pair-specific compared to MP2}
  \end{subfigure}

  \begin{subfigure}[b]{0.9\columnwidth}
    \centering
    \includegraphics[width=\textwidth]{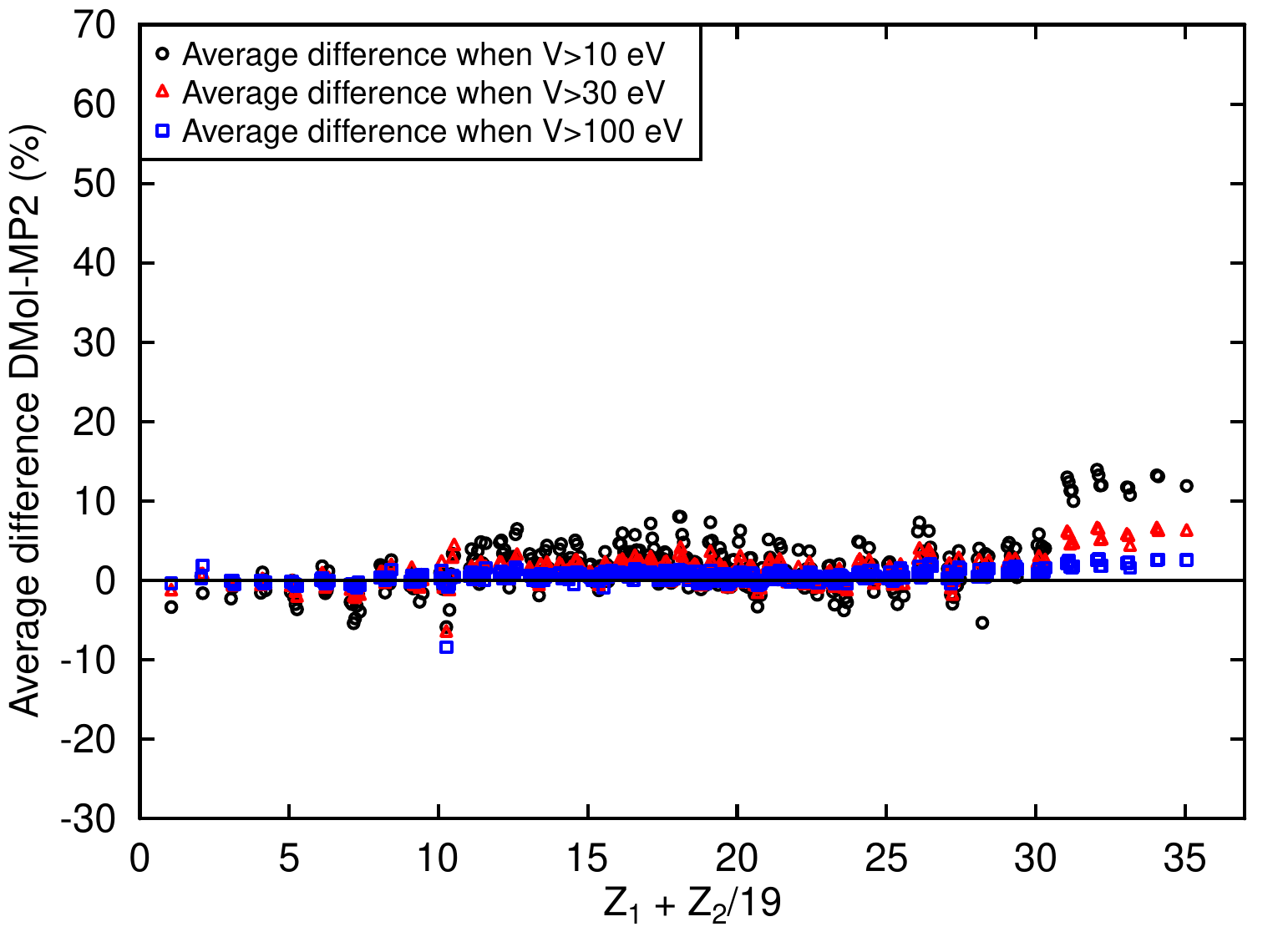} 
 \label{fig:allaverage-dmol-mp2}
    \caption{DMol compared to MP2}
  \end{subfigure}
  \caption{
Average difference  $\Delta_{Z_1,Z_2}$ between (a) ZBL universal, (b) ZBL pair-specific, and (c) DMol potentials with the MP2 potential for all cases within the nonrelativistic limit. The analysis was done at distances where the MP2 potential is either above 10, 30 or 100 eV.
}
\label{fig:allaverage-mp2}
\end{figure}

Results from this comparison are shown for all atom pairs $Z_1 + Z_2 \leq 36$ in \cref{fig:allaverage-mp2}.
To show all data for $\Delta_{Z_1,Z_2}$ in a single plot without overlapping values on the abscissa, the $\Delta$ values are plotted against $Z_1+Z_2/19$; letting $1 \leq Z_1 \leq 36$ and $1 \leq Z_2 \leq 18$ is enough to cover all pairs  $Z_1 + Z_2 \leq 36$.
In agreement with the observations made above for a few specific cases, both the ZBL universal and ZBL pair-specific potentials differ from the MP2 potential by 0--20\% even at energies above 100 eV, while the DMol potential is in much better agreement with the MP2 data.

\begin{table}
    \begin{tabular}{|c|cc|cc|cc|}
\hline
Potential & \multicolumn{2}{c}{$V>10$ eV}  & \multicolumn{2}{|c}{$V>30$ eV}  & \multicolumn{2}{|c|}{$V>100$ eV} \\
          & $\bar \Delta$ & $\sigma_\Delta$ & $\bar \Delta$ & $\sigma_\Delta$ & $\bar \Delta$ & $\sigma_\Delta$ \\
\hline
ZBL universal    & 15 & 23 & 5.0 & 13 & 0.6 & 7.2 \\
ZBL pair-specific & 26 & 32 & 15 & 20 & 8.4  & 11 \\
DMol    & 1.9  & 3.8 & 1.0 & 1.9 & 0.5  & 0.9 \\
NLH  (\cref{sec:analyticalpotential}) & 2.5 & 3.9 & 2.5 & 2.1 & 0.15 & 1.5 \\
\hline
    \end{tabular}
    \caption{Average $\bar \Delta$ and root-mean-square deviation $\sigma_\Delta$ over all element pairs $Z_1+Z_2 \leq 36$ of all potentials compared with the MP2 potential. All numbers are given in \%.}
    \label{tab:allaverage-mp2}
\end{table}

To summarize the comparisons of different systems, we averaged $\Delta_{Z_1,Z_2}$ over all elemental pairs, resulting in the averaged differences $\bar \Delta$ and root-mean-square deviations $\sigma_\Delta$ given in \cref{tab:allaverage-mp2}.
These data confirm that both the universal and pair specific ZBL models differ strongly from the  MP2 references, even at energies above 100 eV which is clearly above chemical interactions. 
However, the DMol potential agrees within $\sim$ 1\% with MP2 for practically all atom pairs above 100 eV.

\subsubsection{DMol reference \label{sec:dmolref}}

\begin{figure}
  \centering
  \begin{subfigure}[b]{0.9\columnwidth}
    \centering
    \includegraphics[width=\textwidth]{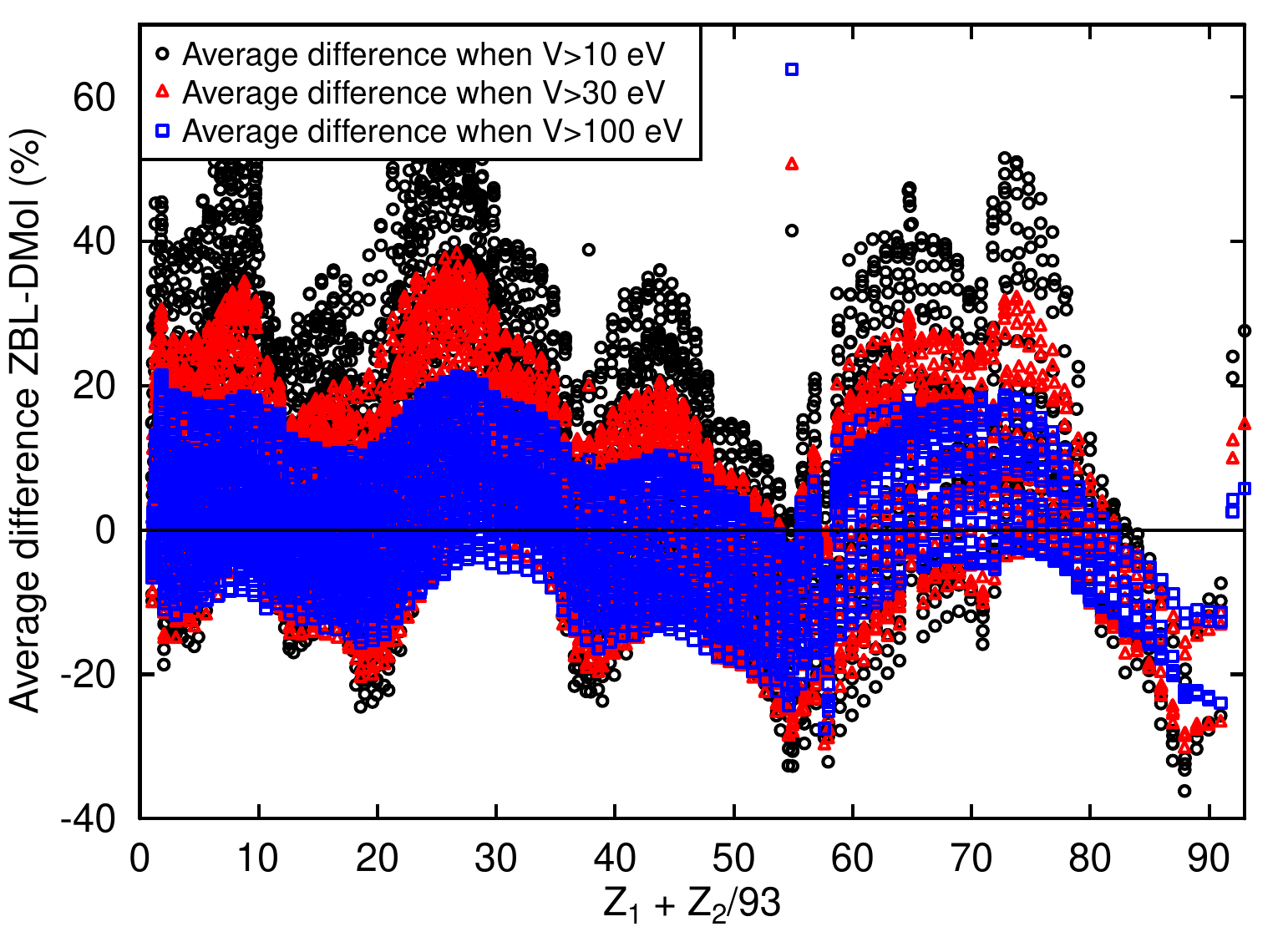}
    \caption{}
    \label{fig:allaverage-zbl-dmol}
  \end{subfigure}
  \hfill
  \begin{subfigure}[b]{0.9\columnwidth}
    \centering
    \includegraphics[width=\textwidth]{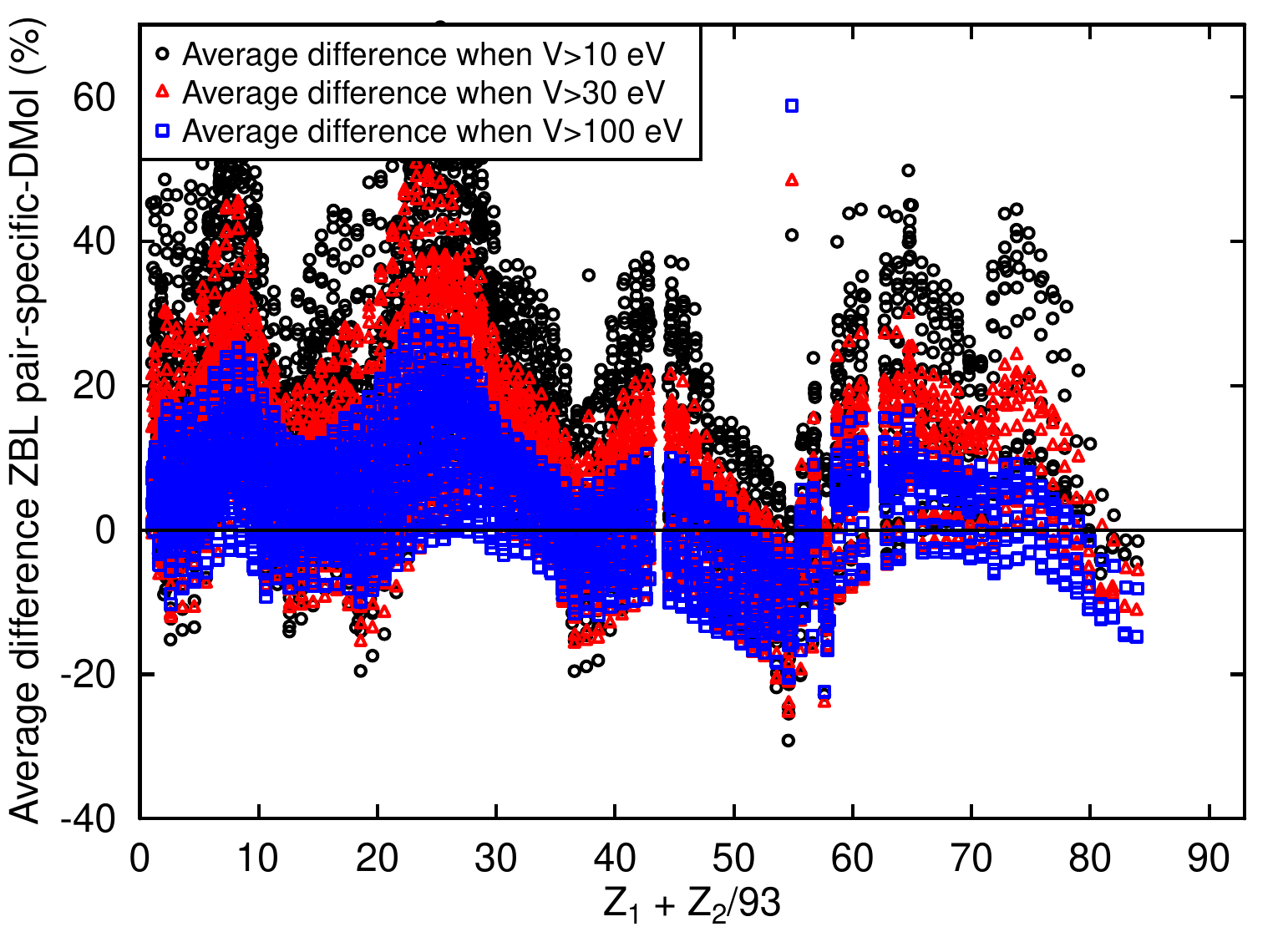}
    \caption{}
    \label{fig:allaverage-zblpair-dmol}
  \end{subfigure}
  \caption{
Average difference  $\Delta_{Z_1,Z_2}$ between (a) ZBL universal and (b) ZBL pair-specific potentials with the DMol potential. 
The analysis was done at distances where the MP2 potential is either above 10, 30 or 100 eV.
(a) ZBL universal compared to DMol, (b) ZBL pair-specific compared to DMol.
}
  \label{fig:allaverage-dmol}
\end{figure}

We also carried out a similar comparison by using the DMol data for all element pairs up to $Z=92$  as reference, as MP2 data is not available for the heavier elements.
The results of this comparison are given in \cref{fig:allaverage-dmol,tab:allaverage-dmol}.
Similarly to the differences to MP2 discussed for lower-$Z$, both the ZBL universal and ZBL pair-specific potentials again differ significantly from the self-consistent calculations.

\begin{table}
    \begin{tabular}{|c|cc|cc|cc|}
\hline
Potential & \multicolumn{2}{c}{$V>10$ eV}  & \multicolumn{2}{|c}{$V>30$ eV}  & \multicolumn{2}{|c|}{$V>100$ eV} \\
          & $\bar \Delta$ & $\sigma_\Delta$ & $\bar \Delta$ & $\sigma_\Delta$ & $\bar \Delta$ & $\sigma_\Delta$ \\
\hline
ZBL universal    & 12 & 22 & 5.6 & 14 & 1.9 & 9.3 \\
ZBL pair-specific & 18 & 25 & 9.7 & 15 & 4.4 & 9.2  \\
NLH (\cref{sec:analyticalpotential})   & 0.7 & 2.0 & $-0.31$ & 1.8 & $-0.13$ & 1.8 \\
\hline
    \end{tabular}
    \caption{Average $\bar \Delta$ and root-mean-square deviation $\sigma_\Delta$ over all element pairs available compared with the DMol potential. All numbers are given in \%.}
    \label{tab:allaverage-dmol}
\end{table}

\section{Analytical potential {\red ("NLH")} fit \label{sec:analyticalpotential}}

The comparison of the potentials showed that the DMol method produces screening functions that are in good agreement with the MP2 reference.
We will now form pair-specific analytical fits of the DMol potentials: we re/express the screening functions in terms of a linear combination of exponential functions.
Such fits have the benefit that the potential can be rapidly evaluated at any internuclear distance, and the fitted potential is guaranteed to be smooth everywhere, as any possible numerical noise at select data points will be eliminated by the fitting procedure.

We have thus produced 3-exponentials fits 
\begin{equation}
    \phi_{\rm NLH}(r) = \sum_{i=1}^{3} a_i \exp(-b_i r) \label{eq:nlh-phi}
\end{equation}
to the screening functions computed with DMol for all combinations of the elements $Z \leq 92$, where the NLH acronym stands for the present authors (Nordlund--Lehtola--Hobler).
Note that the internuclear distance $r$ is not scaled in the exponent, which differs from previous practice \cite{Mol47, Wilson77, ZBL,Bie82,Nak88,Zin11,Zin15}.
The final pair-specific NLH potential is then obtained from \cref{eq:nlh-phi} simply as
\begin{equation}\red
 V_{\rm NLH} (r) = {1 \over 4 \pi \varepsilon_0} {Z_1 e Z_2 e \over r} \phi_{\rm NLH}(r), \label{eq:nlh-v}
\end{equation}
which enables facile evaluation of forces, for instance. {\red The screening function $\phi$ is unitless.}

Any fit to the screening function requires a decision to be made on how errors on various length scales are valued.
Our aim is to produce repulsive force fields that can be combined in the energy range of 10--100~eV with near-equilibrium many-body potentials for MD simulations.
As the low-energy part would be already described by a many-body potential, data points with interaction energies below 10~eV were not considered in the fits. 
A nonlinear least-squares fitting procedure was used to minimize the metric
\begin{align}
 \Vert \phi \Vert = & \sum_{V(r_i) \ge 30 \ {\rm eV}}  \left[{\phi_{\rm NLH}(r_i)-\phi_{\rm DMol}(r_i)} \over {\phi_{\rm DMol}(r_i)} \right]^2   \nonumber \\
 + &  \sum_{10 \ {\rm eV}  \le V(r_i) < 30 \ {\rm eV}} \left[ {\phi_{\rm NLH}(r_i)-\phi_{\rm DMol}(r_i)} \over {\phi_{\rm DMol}(V_{\rm DMol}\!=\!30 {\rm eV})} \right]^2 \label{eq:fittingmetric} .
\end{align}
This metric was chosen to minimize the relative error in the screening function at energies above 30 eV.
At energies below 30 eV, the data points' weight is not allowed to increase further, and an absolute error scaled by the screening function at 30 eV is used instead.

We provide the obtained coefficients $a_i(Z_1,Z_2)$ and exponents $b_i(Z_1,Z_2)$ together with the resulting root mean square errors as open data.
The NLH potential, i.e., the fit coefficients for the element pairs which will be used for simulations below in \cref{sec:range} are given in \cref{tab:coefficients}.
The full set of fit coefficients is available in the supporting information (SI).
The exponents are reported in units of \AA$^{-1}$.
All the fitted potentials are purely repulsive.

\begin{figure}
    \centering
    \includegraphics[width=0.9\columnwidth]{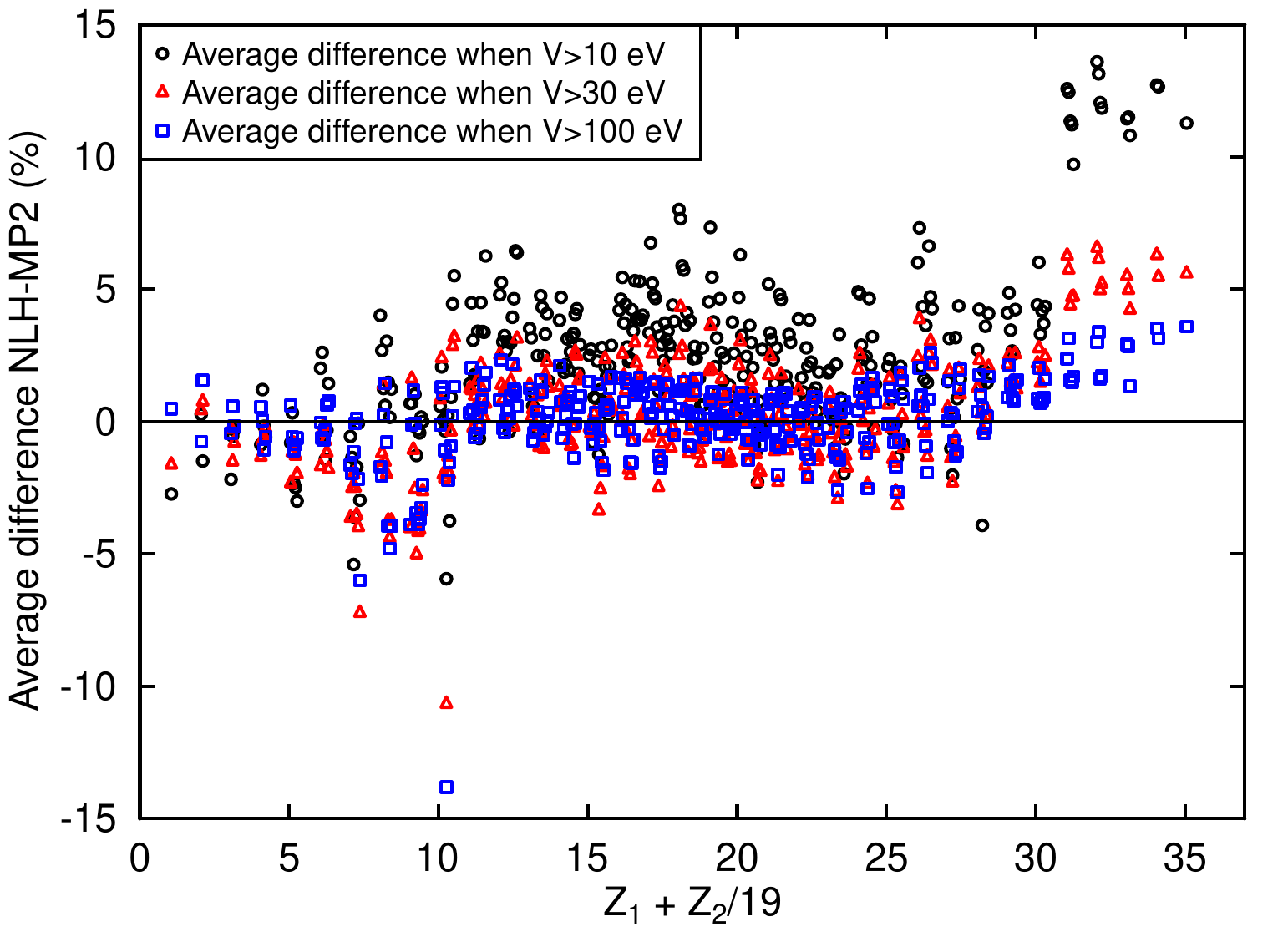} 
   \caption{Average difference $\Delta_{Z_1,Z_2}$ between the NLH and the MP2 potentials for all cases within the nonrelativistic limit. The analysis was done at distances where the MP2 potential is either above 10, 30 or 100 eV. {\red Note that the ordinate scale is 3 times smaller than in \cref{fig:allaverage-mp2}}
}
\label{fig:allaverage-nlh-mp2}
\end{figure}

\begin{figure}
    \centering
    \includegraphics[width=0.9\columnwidth]{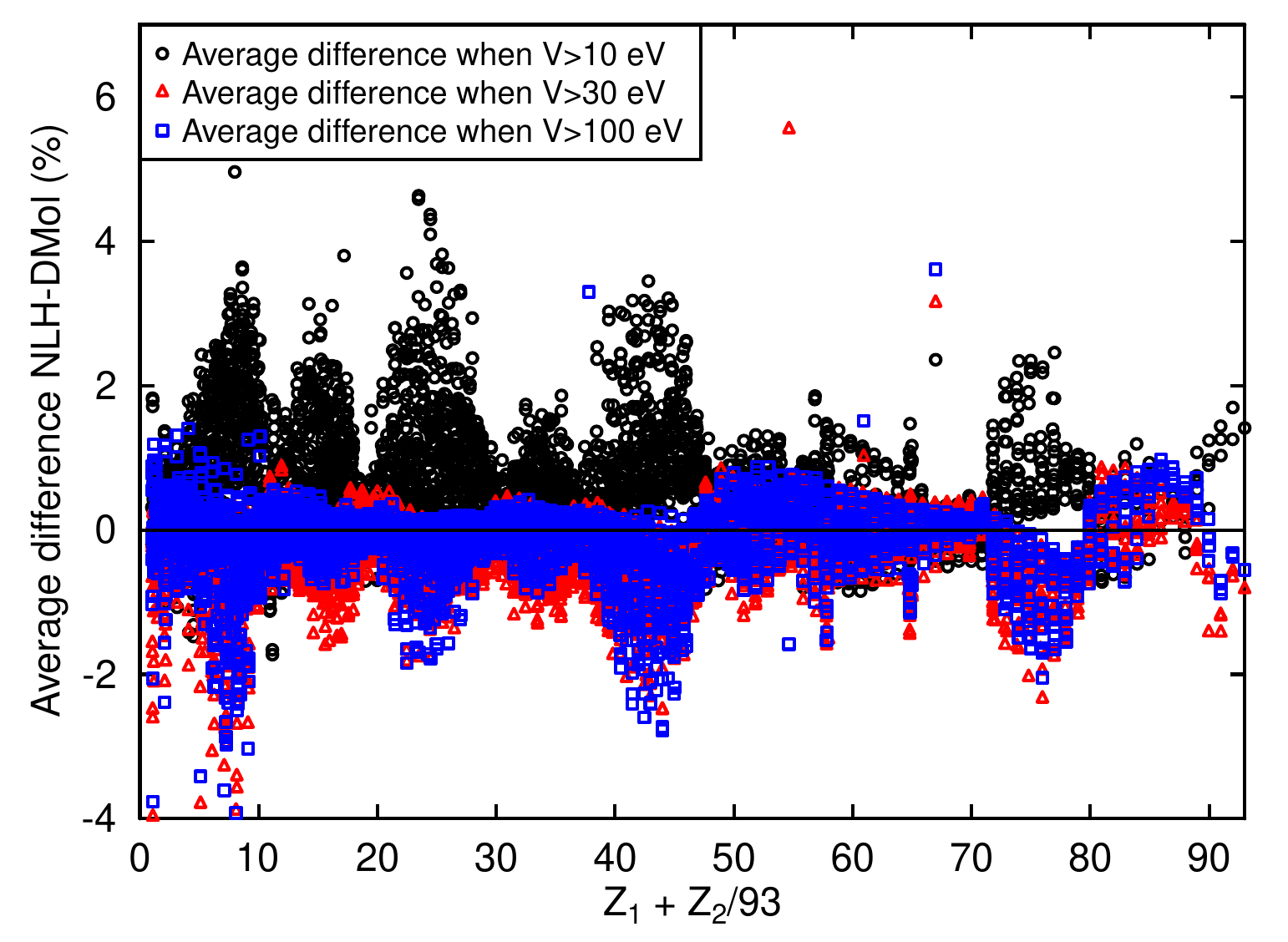} 
    \caption{Average difference $\Delta_{Z_1,Z_2}$ between the NLH and the DMol potentials for all cases within the nonrelativistic limit. The analysis was done at distances where the MP2 potential is either above 10, 30 or 100 eV. Note that the ordinate scale is 10 times smaller than in \cref{fig:allaverage-dmol}.}
    \label{fig:allaverage-nlh-dmol}
\end{figure}

As can be seen from \cref{fig:allaverage-nlh-mp2}, the NLH potentials obtained from fits to the DMol data agree with the MP2 results almost as accurately as the original DMol data sets.
The difference between the NLH and MP2 potentials of all atom combinations has a root-mean-square deviation from zero of 3.95\%,  2.13\% and 1.56\% for energies above 10, 30 and 100~eV, respectively.
The fitting error over all element pairs is shown in \cref{fig:allaverage-nlh-dmol}.
The difference between NLH and DMol potentials of all atom combinations has a root-mean-square deviation from zero of 1.95\%,  1.79\% and 1.86\% for energies above 10, 30 and 100~eV, respectively. Taking these two comparisons together, one can conclude that the NLH fitted potentials agree within $\sim 2$\% with the best available quantum chemical potentials above 30 eV.

\begin{table*}
\begin{tabular}{|rrrrrrrrcc|}
\hline
$Z_1$ & $Z_2$ & $a_1$ & $a_2$ & $a_3$ & $b_1 \times$\AA & $b_2\times$\AA & $b_3\times$\AA & $E_{30}$ (\%) & $E_{10}$ (\%)\\
\hline
   1 & 1 & $-8.99999$ & 9.99999 & 0.00000 & 9.55658 & 8.89086 & 0.00000 & 1.32& 3.14 \\
   1 & 14 & 0.34955 & 0.65045 & 0.00000  & 14.03500 & 3.21949 & 0.00000 & 3.38& 6.01\\
   5 & 14 & 0.21145 & 0.61640 & 0.17215  & 22.46013 & 4.79260 & 2.40710 & 1.65& 3.55\\
   13 & 18 & 0.10006 & 0.59380 & 0.30615 & 38.18078 & 8.03267 & 2.70079 & 3.15& 4.64\\
   14 & 14 & 0.30199 & 0.29621 & 0.40180 & 16.28675 & 6.38346 & 3.20812 & 2.33& 3.45\\
   14 & 33 & 0.16304 & 0.45925 & 0.37771 & 31.18522 & 8.78859 & 3.57348 & 1.74& 4.89\\
   26 & 26 & 0.34794 & 0.65206 & 0.00000 & 19.25771 & 4.81918 & 0.00000 & 2.90& 12.45\\
\hline
\end{tabular}
    \caption{NLH repulsive potential parameters for select element pairs, obtained by fitting the screening functions to DMol data.
    $E$ stands for the root mean square error of the fit above either 10 ($E_{10}$) or 30 ($E_{30}$) eV. Parameters for all the potentials are available in the online open data package.}
    \label{tab:coefficients}
\end{table*}

\section{Effect of potentials on ion penetration depth profiles \label{sec:range}}

The above comparison has shown that there are significant ($\sim 2\%$) differences between the potentials from self-consistent calculations (MP2 and DMol) on the one hand, and even larger ($\sim 10$\%) between the self-consistent calculations and the ZBL potentials on the other hand. 
We next address the question whether these differences affect experimentally measurable quantities. 
As the test case, we consider the depth distribution profiles of ions implanted in solid materials, which can be experimentally measured with several techniques \cite{Eri67, Tor94, Hau95, Cai96, Lar98}.

\subsection{Comparison of potentials with each other}
\label{subsec:interpotentialcomp}

The shape and depth of the range profile is mainly determined by the repulsive interatomic potential when the nuclear stopping power dominates and when the sample temperature is low enough that the implanted ions do not diffuse.
These conditions can be achieved by a suitable choice of the ions and implantation energies.

We compare potentials for four systems, which represent different energy-ion-solid combinations: 10 keV H ions on Si, 10 keV Si ions on Si, 30 keV Ar ions on Al, and 100 keV Fe ions on Fe (for protons the electronic stopping dominates over the nuclear one, but we include it in the comparison because of the wide interest in hydrogen implantations). 
Since for the last case $Z_1+Z_2 > 36$, MP2 potentials are not available for this system.

We simulated range profiles both in the strongest channeling direction \cite{Nor16}---[110] in Si and Al, and [111] in Fe---as well as a good non-channeling direction for all four cases.
The implantation direction with polar angle $\theta=20\degree$ off the [001] surface normal and azimuthal angle $\phi= 20\degree$ measured from the [100] direction was used as the non-channeling direction;
this choice of angles has been shown to lead to range profiles with minimal channeling effects in all common crystal systems \cite{Nor16}.

Since the aim in this subsection is not to compare with experiment, but to assess how sensitive the range profiles are to the repulsive potential, we used in all cases a conventional electronic stopping power model, namely the ZBL parameterization from the 1995 version of the TRIM code \cite{ZBL, TRIM95}. The following subsection presents comparisons with experiments with a more realistic electronic stopping model.

The simulations were carried out in the recoil interaction approximation \cite{Nor04b, Hob00} with the MDRANGE code \cite{Nor94b,MDRANGE}, which has been widely used for studies of ion range profiles \cite{Sil00,Hog02,Pel03e,Nor16}.
The atoms were given random thermal displacements $u_i$ corresponding to 300 K using the quantum level approach that includes contributions from zero-point vibrations 
\cite{Gem74,Nor17d}:
\begin{equation}\label{eq:Debyedispl}                                           
          u_i= {1 \over 2} \sqrt{3 \hbar^2 \over k_B} \sqrt{ \displaystyle {4x^{-1}\Phi(x)} + {1} \over m T_D }                                                                  
\end{equation}
where $i$ denotes the $x$, $y$, or $z$ direction, $x={T_D/T}$, $T$ is the sample temperature, $m$ the mass of the sample atoms, $T_D$ is the Debye temperature of the
material, and the function $\Phi(x)$ is the Debye integral
\begin{equation}\label{eq:Debyeint}                                             
  \Phi(x) = { 1 \over x} \int_0^x {\xi d\xi \over e^\xi -1}.
\end{equation}
Since the integral does not have a closed-form analytical solution, we evaluated it with the Stegun numerical series approximation \cite{Stegun} as described in detail in the appendix of \citeref{Nor17d}.
The Debye temperatures were taken from the tables of \citeref{Ashcroft-Mermin}, except for Si for which the value of 519~K based on more recent experiments was used \cite{Bus97}.

\begin{figure}
\centering
  \begin{subfigure}[b]{0.9\columnwidth}
    \centering
    \includegraphics[width=\textwidth]{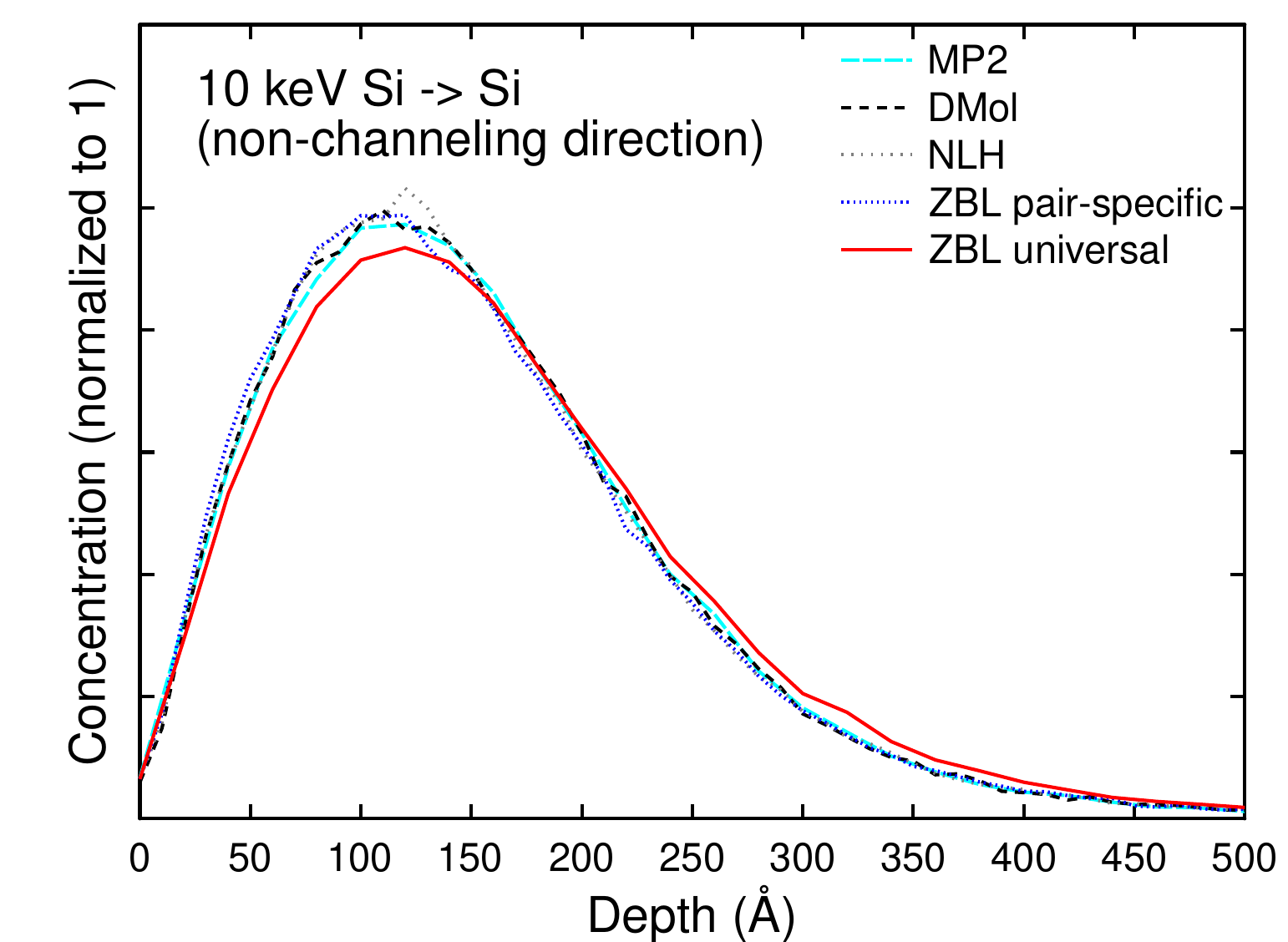} 
    \caption{Non-channeling direction (tilt=twist=20$^\circ$)}
    \label{fig:sisiranges-a}
  \end{subfigure}
  \hfill
  \begin{subfigure}[b]{0.9\columnwidth}
    \centering
    \includegraphics[width=\textwidth]{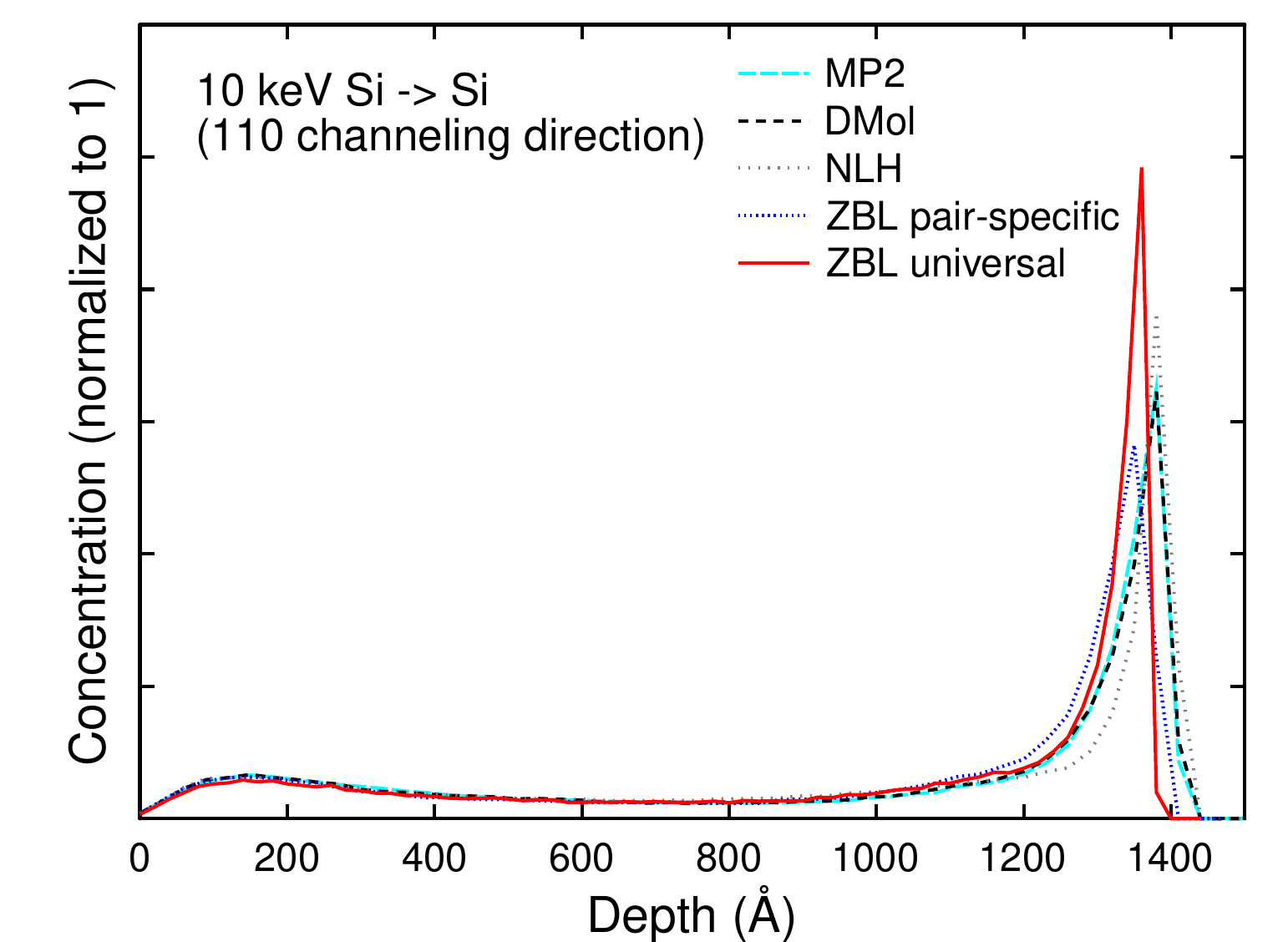}
    \caption{Strongest channeling [110] direction}
    \label{fig:sisiranges-b}
  \end{subfigure}
\caption{
Range profiles for 10 keV Si ions implanted into Si in a non-channeling (\cref{fig:sisiranges-a}) and the strongest channeling [110] direction (\cref{fig:sisiranges-b}).
}
\label{fig:sisiranges}
\end{figure}

\begin{figure}
\centering
  \begin{subfigure}[b]{0.9\columnwidth}
    \centering
    \includegraphics[width=\textwidth]{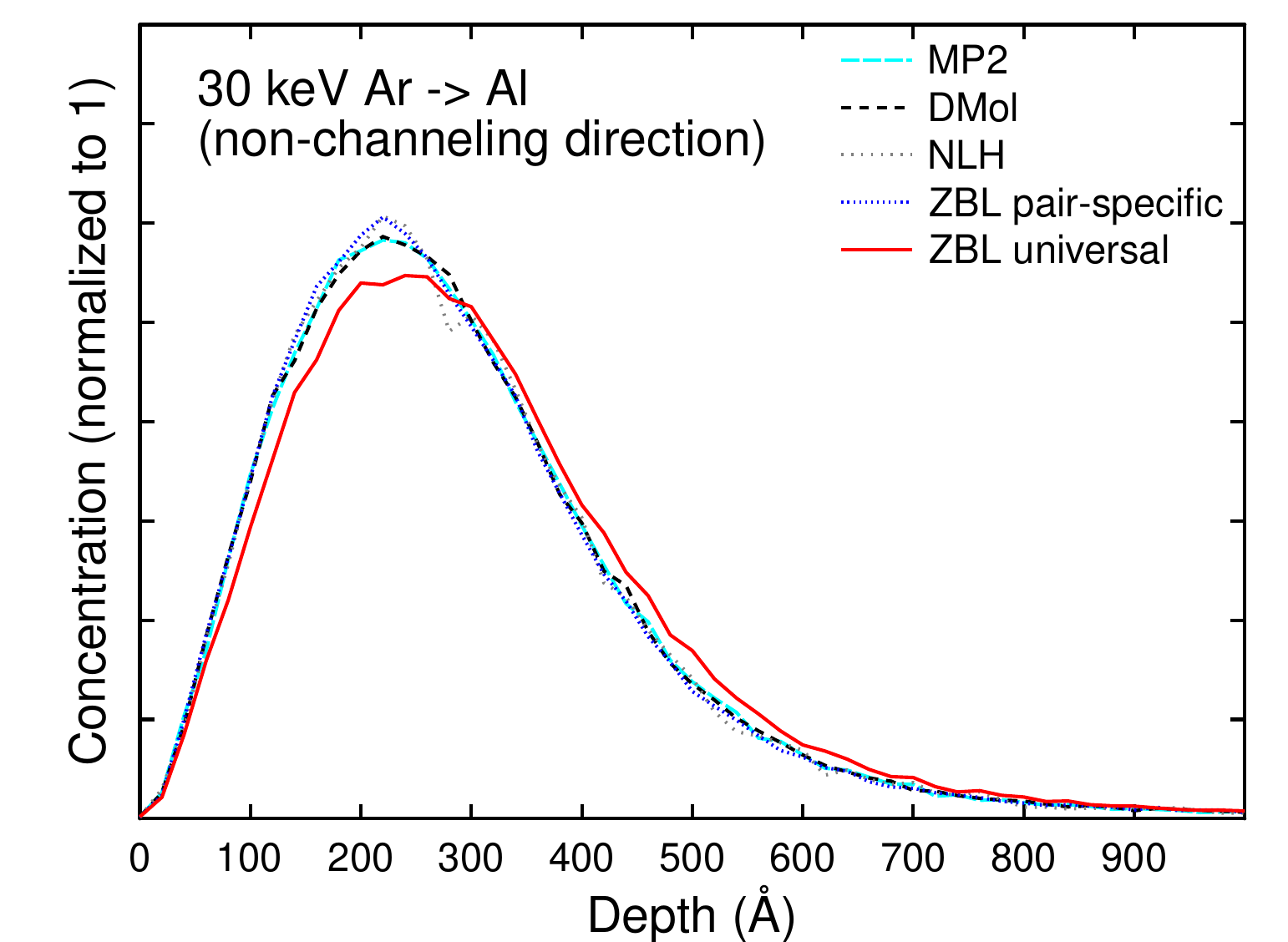}
    \caption{Non-channeling direction (tilt=twist=20$^\circ$)}
    \label{fig:aralranges-a}
  \end{subfigure}
  \hfill
  \begin{subfigure}[b]{0.9\columnwidth}
    \centering
    \includegraphics[width=\textwidth]{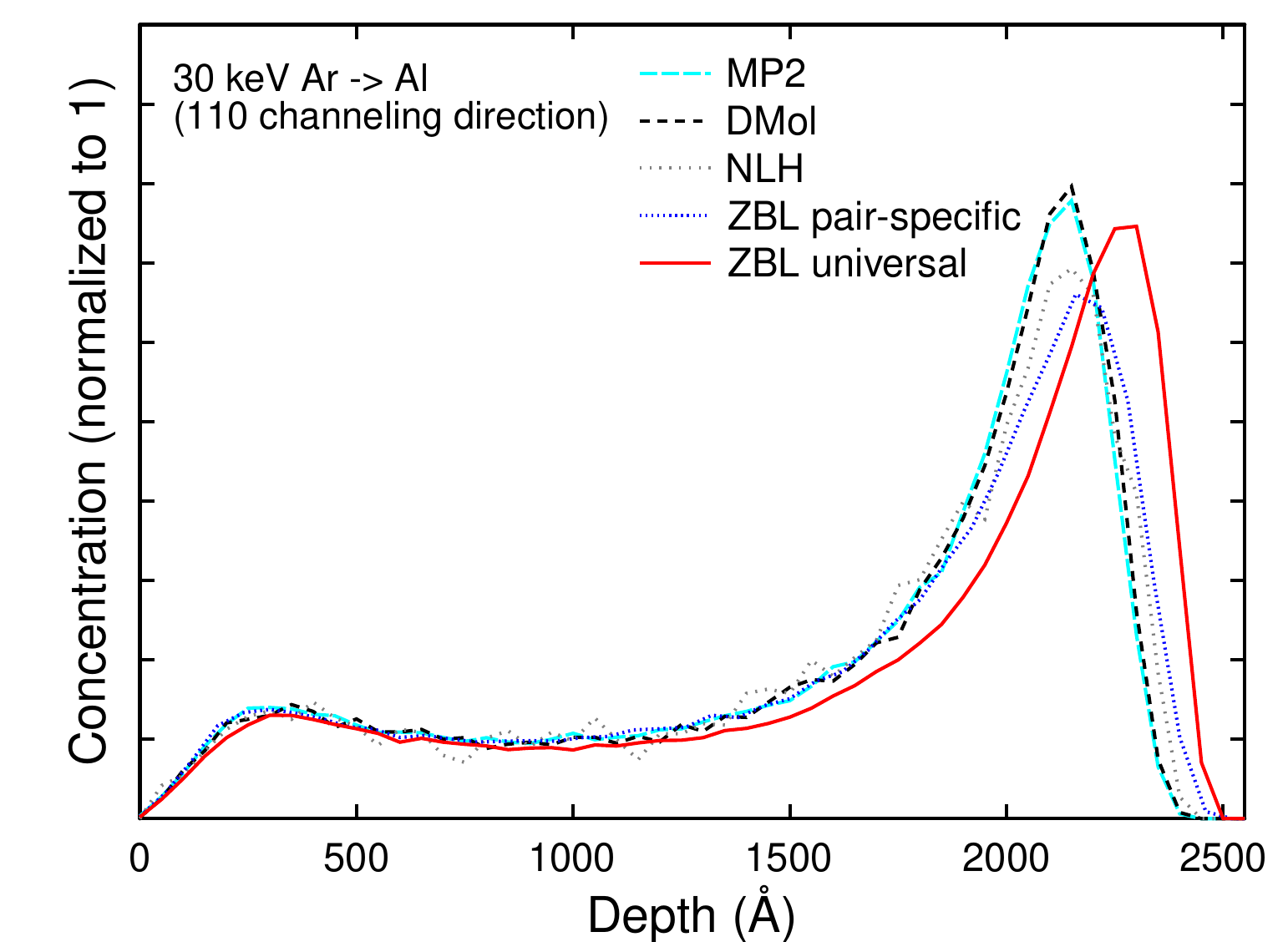}
    \caption{Strongest channeling [110] direction}
    \label{fig:aralranges-b}
  \end{subfigure}
\caption{
Range profiles for 30 keV Ar ions implanted into Al in a non-channeling (\cref{fig:aralranges-a} and the strongest channeling [110] direction (\cref{fig:aralranges-b}).
}
\label{fig:aralranges}
\end{figure}

\begin{figure}
\centering
  \begin{subfigure}[b]{0.9\columnwidth}
    \centering
    \includegraphics[width=\textwidth]{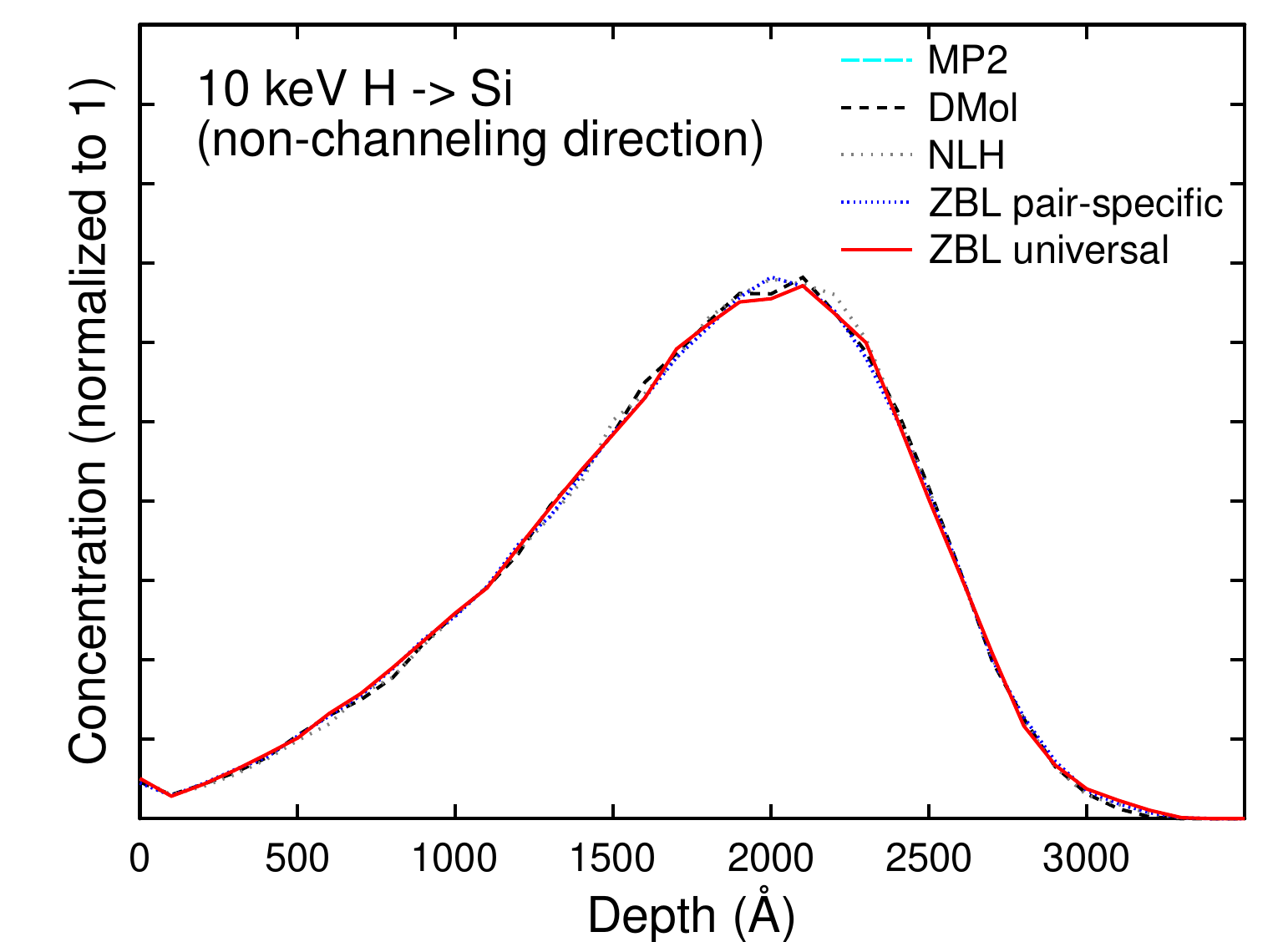}
    \caption{Non-channeling direction (tilt=twist=20$^\circ$)}
    \label{fig:hsiranges-a}
  \end{subfigure}
  \hfill
  \begin{subfigure}[b]{0.9\columnwidth}
    \centering
    \includegraphics[width=\textwidth]{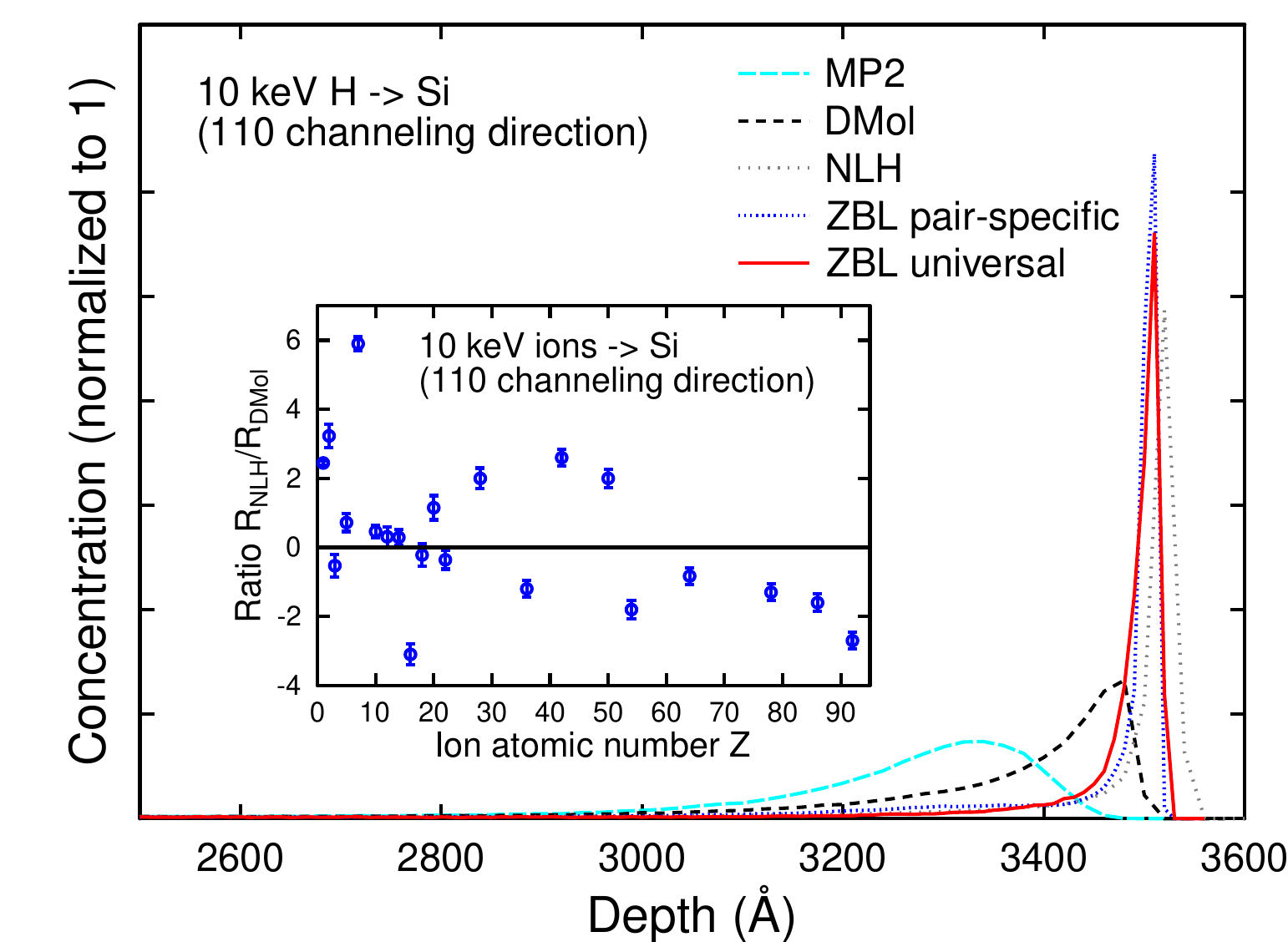}
    \caption{Strongest channeling [110] direction}
    \label{fig:hsiranges-b}
  \end{subfigure}
\caption{
Range profiles for 10 keV H ions implanted into Si in a non-channeling (\cref{fig:hsiranges-a}) and the strongest channeling (\cref{fig:hsiranges-b}) direction. 
Note that in part (b), the abscissa scale is focused at the end of range. 
The inset shows a comparison of the NLH and DMol potential mean ranges of various ions in the range $1 \leq Z \leq 92$ under the same channeling condition.
}
\label{fig:hsiranges}
\end{figure}

\begin{figure}
\centering
  \begin{subfigure}[b]{0.9\columnwidth}
    \centering
    \includegraphics[width=\textwidth]{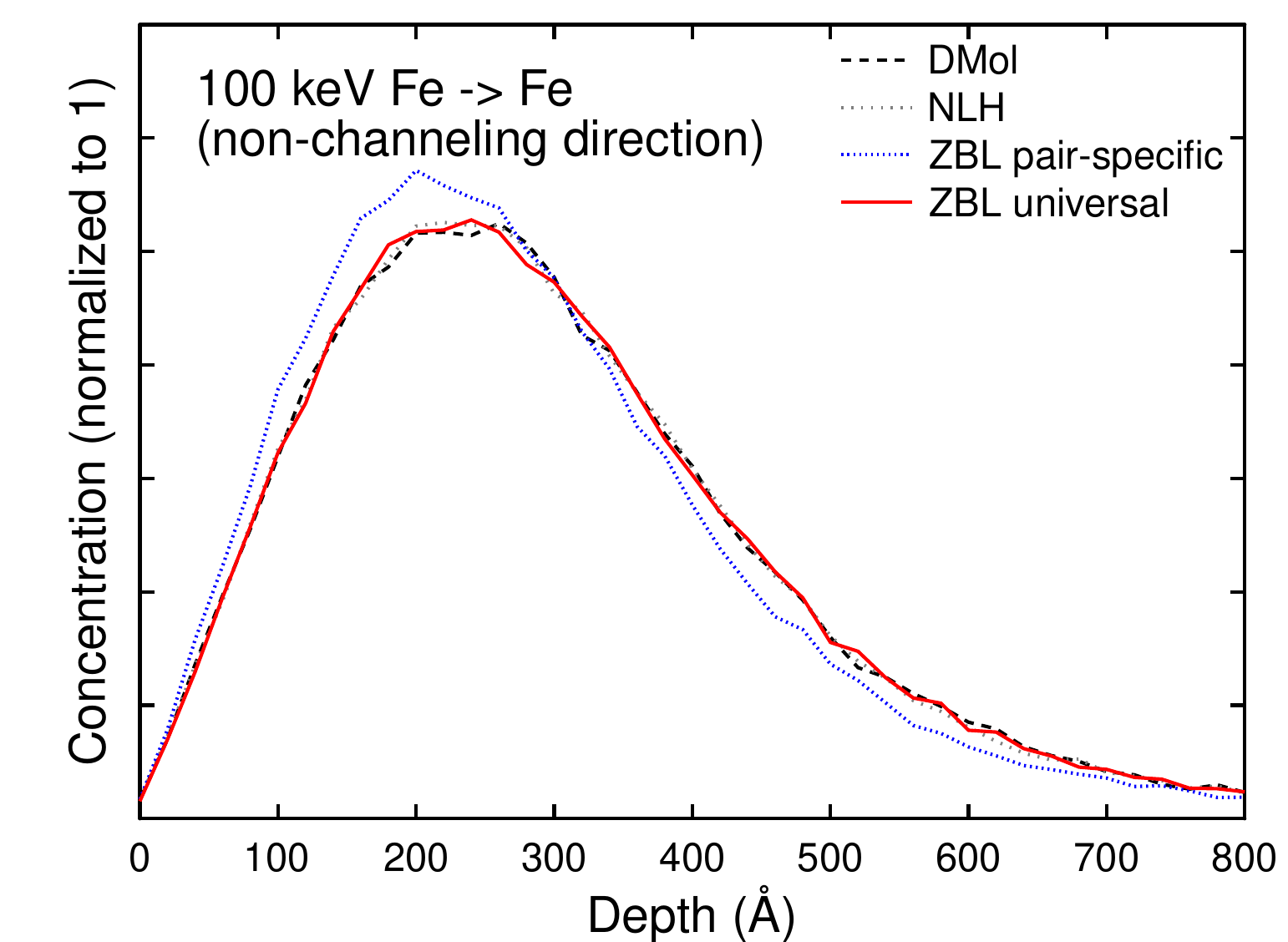}
    \caption{Non-channeling direction (tilt=twist=20$^\circ$)}
    \label{fig:feferanges-a}
  \end{subfigure}
  \hfill
  \begin{subfigure}[b]{0.9\columnwidth}
    \centering
    \includegraphics[width=\textwidth]{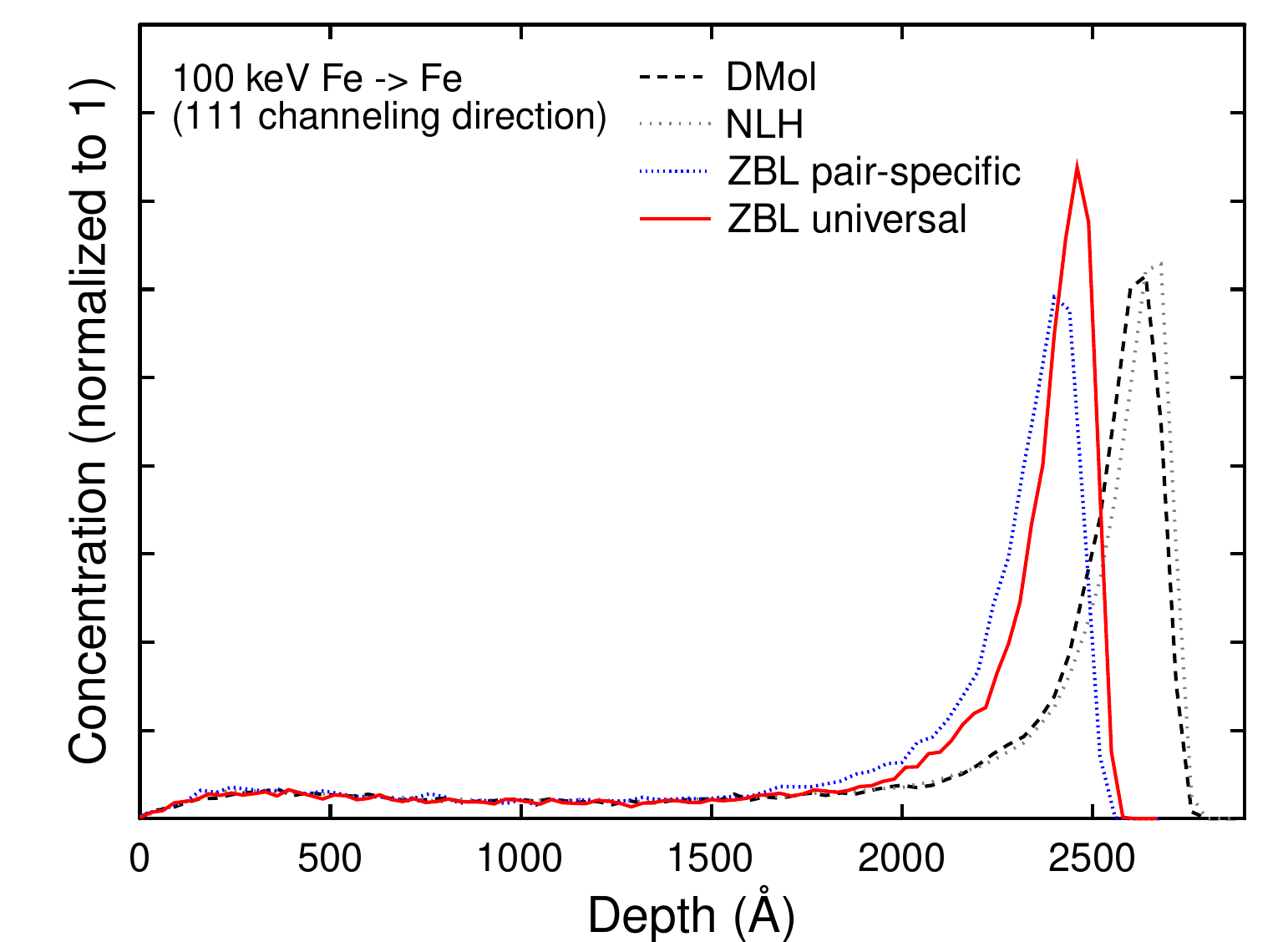}
    \caption{Strongest channeling [111] direction}
    \label{fig:feferanges-b}
  \end{subfigure}
\caption{
Range profiles for 100 keV Fe ions implanted into Fe in a non-channeling (\cref{fig:feferanges-a}) or the strongest channeling (\cref{fig:feferanges-b}) direction.
}
\label{fig:feferanges}
\end{figure}

The results of these calculations are shown in \cref{fig:sisiranges} for the implantation of Si-Si, in \cref{fig:aralranges} for the implantation of Ar-Al, in \cref{fig:hsiranges} for the implantation of H-Si, and in \cref{fig:feferanges} for the implantation of Fe-Fe.
{\red The depth distributions in the plots are area normalized to 1 to be in the form of a probability distribution of the implantation depth}.
A summary of the results is given in \cref{tab:meanranges}.
{\red The error bars were calculated as the standard error of the mean, i.e. the standard deviation of the values of the individual ion range divided by the square root of the number of ions \cite{Bevington}.}
The choice of potential clearly matters: there are {\red statistically significant} differences in the range profiles as well as mean ranges between the ZBL and quantum chemical DMol and MP2 potentials that are comparable to differences in the interatomic potentials (of the order of $5\%$).

\begin{table*}
    \begin{center}
    \begin{tabular}{|ccccccccc|}
    \hline
        Ion & Target & E (keV) & Direction & MP2 & DMol &  NLH & ZBL pair-specific & ZBL universal \\
        \hline
Si & Si & 10 & Non-channeled & 154 $\pm$ 1 & 154 $\pm$ 1 & 153 $\pm$ 1 &  153 $\pm$ 1 & 162 $\pm$ 1 \\
   &    &    & [110]           & 986 $\pm$ 1 & 987 $\pm$ 1 & 986 $\pm$ 1 &  982 $\pm$ 1 & 1001 $\pm$ 2 \\
            \hline 
Ar & Al & 30 & Non-channeled & 295 $\pm$ 1 & 295 $\pm$ 1 & 294 $\pm$ 1 &  291 $\pm$ 1 & 311 $\pm$ 1 \\
   &    &    & [110] & 1609 $\pm$ 2 & 1620 $\pm$ 2 & 1617 $\pm$ 1 &  1630 $\pm$ 2 & 1710 $\pm$ 2 \\
         \hline
H & Si & 10 & Non-channeled & 1753 $\pm$ 2 & 1751 $\pm$ 2 & 1758 $\pm$ 2 &  1750 $\pm$ 2 & 1750 $\pm$ 2 \\
  &    &    & [110]           & 3123 $\pm$ 2 & 3212 $\pm$ 2 & 3310 $\pm$ 2 &  3307 $\pm$ 2 & 3319 $\pm$ 2 \\
         \hline
Fe & Fe & 100 & Non-channeled &  --   & 314 $\pm$ 1  & 312 $\pm$ 1 &  292 $\pm$ 1 & 313 $\pm$ 1 \\
   &    &     & [111]           &  --   & 2180 $\pm$ 4 & 2180 $\pm$ 2 &  1997 $\pm$ 4 & 2072 $\pm$ 4 \\
         \hline
    \end{tabular}
    \end{center}
    \caption{Mean ranges $\bar R$ in \AA{} for the universal ZBL potential, the pair-specific ZBL potential, the DMol potential, and the MP2 potential for the studied test cases.
    Non-channeled means implantation in the non-channeling direction $\theta = \phi = 20^\circ.$ Since the Fe-Fe dimer has $Z_1+Z_2 > 36$, MP2 results are not available in this case (see text).}
    \label{tab:meanranges}
\end{table*}

As expected from the comparison of potentials discussed earlier, the calculations with the DMol and MP2 screening functions lead to results that are practically identical to each other in all cases except for 10 keV H ions implanted in a [110] channel in Si.
Moreover, in many cases one of the ZBL potentials agrees fairly well with the DMol and MP2 potential in the non-channeled case, while the other agrees better in the channeled case.
Also, in all channeling cases there is a noticeable difference between the DMol and NLH potential results, even though they have almost the same high-energy part.

\begin{figure}
\begin{center} 
\includegraphics[width=0.9\columnwidth]{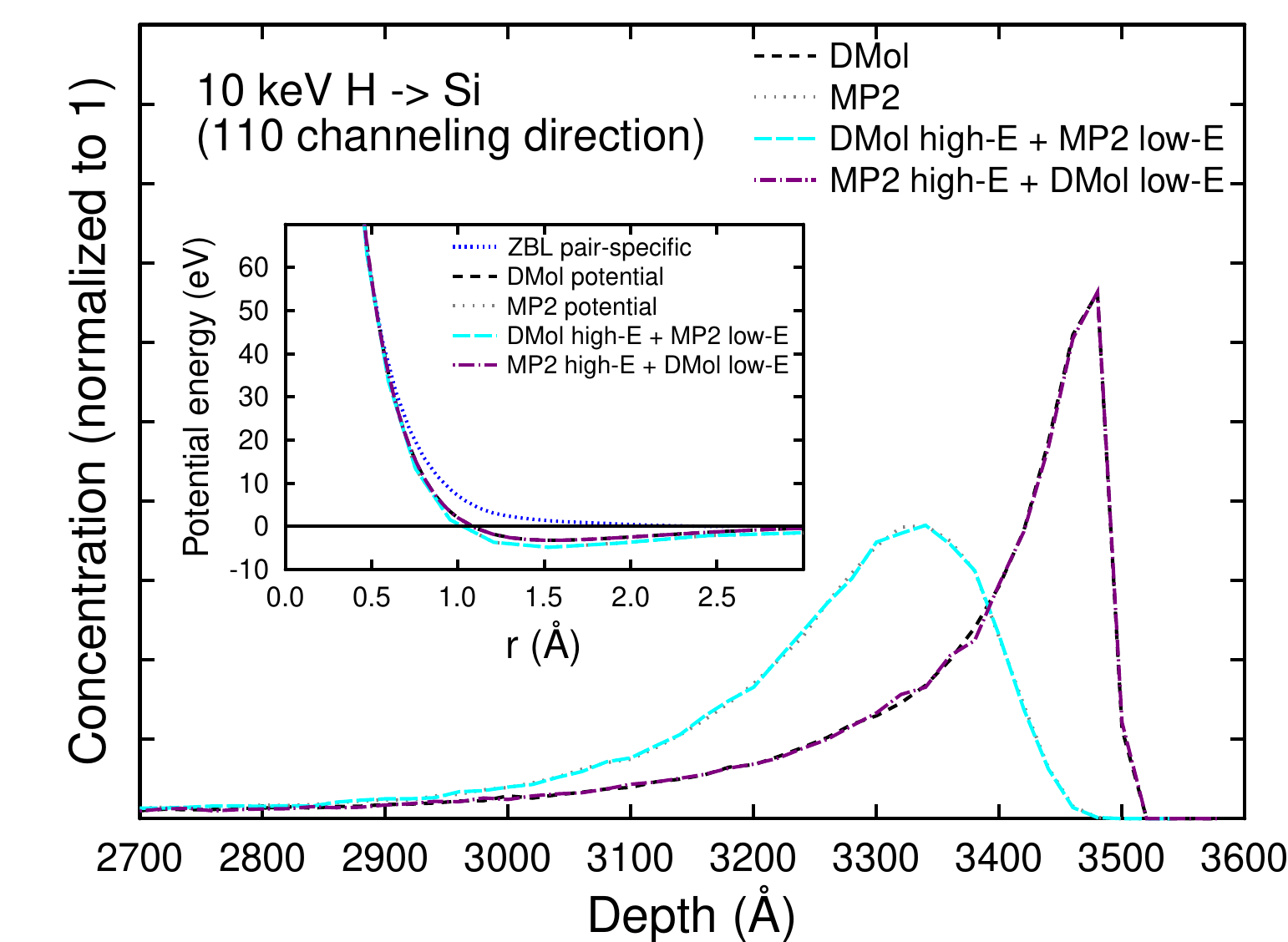} \\
\end{center}
\caption{
  Range profiles for 10 keV H ions implanted on Si in the strongest channeling [110] direction, for the DMol and MP2 potentials as well as artificial test potentials where the DMol and MP2 parts are swapped above and below 100 eV.
  The potentials are shown in the inset.
}
\label{fig:hsiranges_hybridpot}
\end{figure}

These observations indicate that the range distributions under channeling conditions may be particularly sensitive to the low-energy part of the potential. 
To test this hypothesis, we made hybrid potentials for H-Si where we swapped the low-energy and high-energy parts of the DMol and MP2 potentials above and below 100 eV.
Results for this test are illustrated in \cref{fig:hsiranges_hybridpot}, showing that the shape of the range distribution is indeed dominated by the low-energy part of the potential.
This is likely due to the low-energy part affecting the dechanneling of the ion once it has slowed down sufficiently; note that the differences between potentials become significant only near the end of their range.

To get more insight on whether atomic size affects the sensitivity of the range calculation on the interatomic potential, we simulated ions in the whole studied range $1 \leq Z \leq 92$ implanted in the same [110] channeling direction of Si for which a clear sensitivity was observed in the case of H. This comparison was done for mean ranges obtained with the DMol and NLH potentials. 
Since the latter potential is a fit to the DMol data above 30 eV, they agree within $\sim$ 2\% at high energies (see \cref{sec:systematic}).
However, the NLH and DMol potentials are quite different at low energy since
the NLH potential is repulsive at all internuclear distances $r$, while the DMol potential has an attractive part at distances close to the chemical equilibrium.
Data in the inset of \cref{fig:hsiranges} shows that there are sometimes large differences in the mean ranges for small atomic numbers $Z \lesssim$ 10.
For larger atomic numbers, the difference is at most $\sim$ 2\%, which may be explained by the higher interaction energies of heavier ions.
Hence, the results indicate that small atoms of low-atomic number may be sensitive to the choice of attractive potential, when implanted in wide channels such as the [110] channel in Si.
Comparisons of different low-energy interaction parts using the data provided in the Supplementary Information \cite{Supplemental} can be used to test whether a particular channeling condition is indeed sensitive to low-energy interactions.

However, we note that using a pair potential with the attractive part of a diatomic system may also be misleading because at the low, near-equilibrium energies below 100 eV, the interatomic interactions are many-body in character.
In other words, the attractive part of the potential in a solid depends on how many neighbours an atom has in a particular configuration \cite{Abe85,Alb01b}, and hence will be different from that of the diatomic molecule. 
{\red To account for such effects, one can use a many-body analytical or machine-learned potential that also accounts for chemical effects \cite{Ter88,Bre00,Alb01b,Beh16,Byg19b}. We present an approach by which one can join equilibrium potentials to the currently developed repulsive ones without any fitting parameters in Appendix A. }

Taken together, these results show that the common assumption, mentioned in the introduction, that radiation effects can be described as a sequence of binary interatomic collisions governed by a repulsive potential may not be valid under all channeling conditions.

\subsection{Comparison of range profiles with experiments \label{sec:rangeexpt}}

\subsubsection{\red Simulation setup}

Direct comparison of range profiles with experiments is complicated by several factors. Often the thickness of the surface oxide layers is not known, and in many experiments the implantation tilt and twist angles are not accurately controlled or reported. The implantation process may also amorphize semiconductor samples. Moreover, if the sample is polycrystalline, it is difficult to know what crystal directions were facing the beam.

We compare range profiles with the experiments of Cai {\it et al} \cite{Cai96,Cai98}, carried out on single-crystalline Si wafers of (001) surface orientation, at implantation doses low enough such that the samples were not significantly damaged, which in most cases meant a fluence of 10$^{13}$ ions/cm$^2$.
Both the sample tilt and twist angles were reported for non-channeling implantations.
The implantation profiles were measured with the secondary ion mass spectrometry (SIMS) technique, which is well established to measure implantation profiles in Si.

The experimental depth profiles were scanned in and digitized from \citerefs{Cai96,Cai98} with the \textsc{g3data} code \cite{g3data}.
From comparison of repeated digitizations, this process is estimated to introduce errors $\lesssim$ 1\% to the data.
The experimental data may carry larger systematic errors associated with the SIMS method itself.
However, since the original references do not give estimates of experimental errors, error bars are not reported here for the experimental data.

Simulations were carried out with the same MDRANGE approach described earlier, except that a surface oxide layer of 2.2 nm thickness was included in the calculations; the thickness of the surface layer corresponds to 4 Si unit cells.
{\red 
The sample atoms were given thermal displacements corresponding to 300 K using eq. \ref{eq:Debyedispl}, which gave a 1D root-mean-square displacement magnitude of 0.079 Å for the Si atoms.}
The electronic stopping power was described using the density-functional based Puska--Echenique--Nieminen--Ritchie model utilizing a 3D electron density of Si \cite{Sil00}, combined with a Firsov local stopping.
This approach has previously been shown to give excellent agreement between experimental and simulated range profiles in Si \cite{Pel03e}. 

For the case of As implantation into amorphous Si, the nonlocal TRIM 1996 electronic stopping was used \cite{TRIM95}.
This electronic stopping was also used for the amorphous 2.2 nm SiO$_2$ surface layers. 
The amorphous Si was modelled with a 4 nm cube of random atom positions with a minimum separation of 2.1 Å.
This approach has previously been shown to well describe implantation into amorphous material \cite{Nor16}.
The ion-oxygen interactions were modelled with the ZBL universal repulsive potential; since the surface oxide layer is very thin, using any other potential would not make any significant difference.
We verified this by simulating the {\red cases} of 180 keV As implantation into amorphous Si  {\red and 15 keV As into Si in non-channeling direction} also with the NLH potential for the As-O interactions.
As expected, {\red in both cases} the mean ranges agreed within the statistical uncertainty of $\pm$ 1 Å.

The beam in ion implanters has always some angular spread due to the ion optics.
The works presenting the experimental results reported a beam divergence of 1 degree \cite{Cai98}.
We described this in our simulations by a Gaussian distribution in the incoming tilt ($\theta)$ angle with a standard deviation of $1^\circ$ around the nominal angle.

The simulations for which mean ranges are given in \cref{tab:exptcomp} were carried out for 100 000 ions; the statistics is reflected in the {\red standard} error of the mean. 
{\red
At the bottom of the table, 
the average deviation for each potential is the arithmetic average over each individual deviation ($R_{\rm mean,sim,i}/R_{\rm mean,expt,i} -1$), where sim stands for the simulated and expt for the experimental values.
The root-mean-square ($\textrm{rms}$) deviations were calculated as 
\begin{equation}     \red
		     \textrm{rms} = \sqrt{\frac{ \sum_{i=1}^N \left( R_{\rm mean,sim,i}/R_{\rm mean,expt,i} -1  \right)^2 }{ N_{\rm pot} } }
\end{equation}
where $i$ loops over the various cases simulated (ion, energy, direction), and $N_{\rm pot}$ is the total number of cases simulated with this potential (10 for MP2 and 16 for the other potentials).
}

\begin{figure}
\centering
  \begin{subfigure}[b]{0.9\columnwidth}
    \centering
    \includegraphics[width=\textwidth]{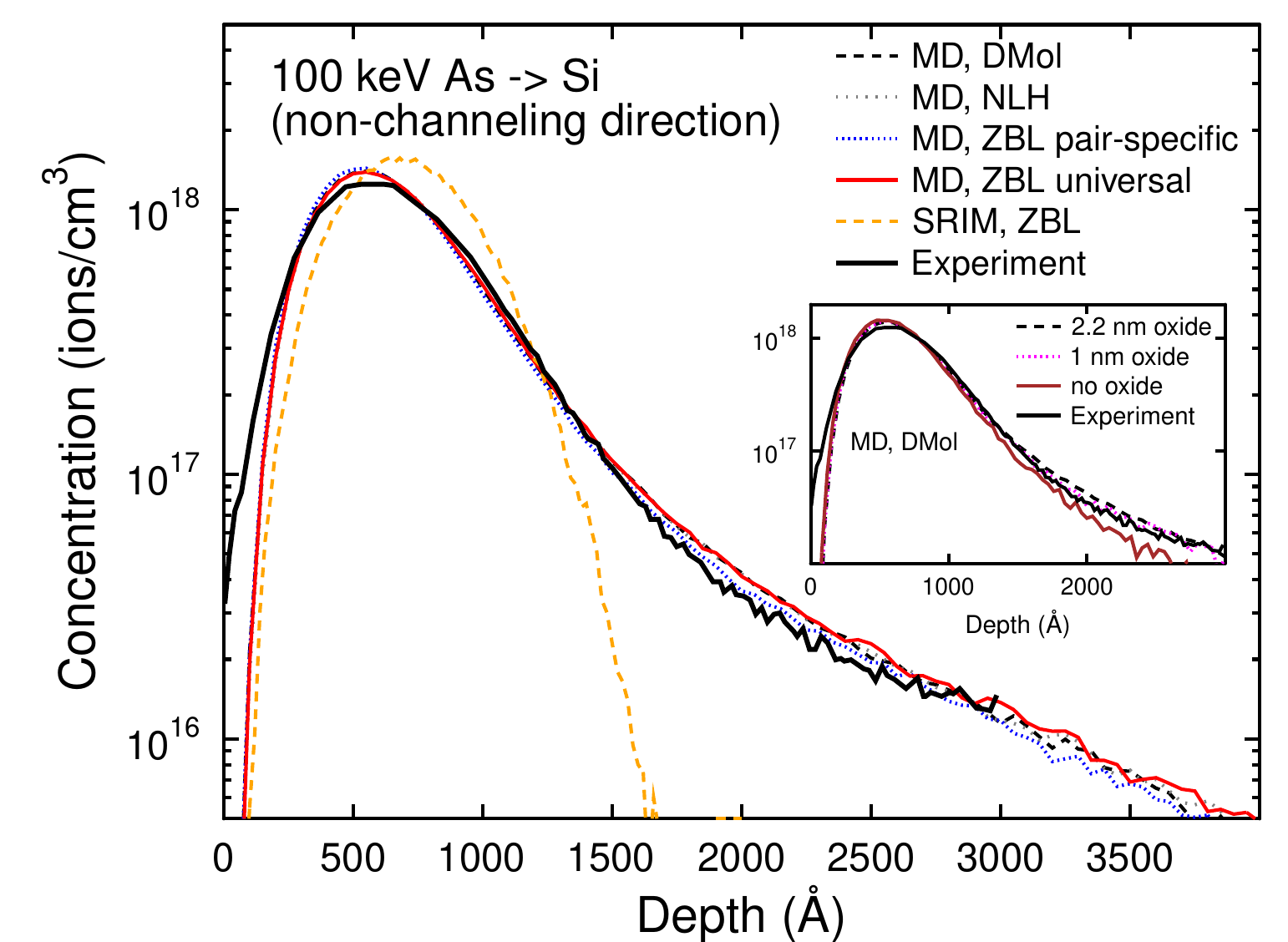} 
    \caption{100 keV As ions into a non-channeling direction (tilt=8$^\circ$, twist=30$^\circ$)}
    \label{fig:assiranges-a}
  \end{subfigure}
  \hfill
  \begin{subfigure}[b]{0.9\columnwidth}
    \centering
    \includegraphics[width=\textwidth]{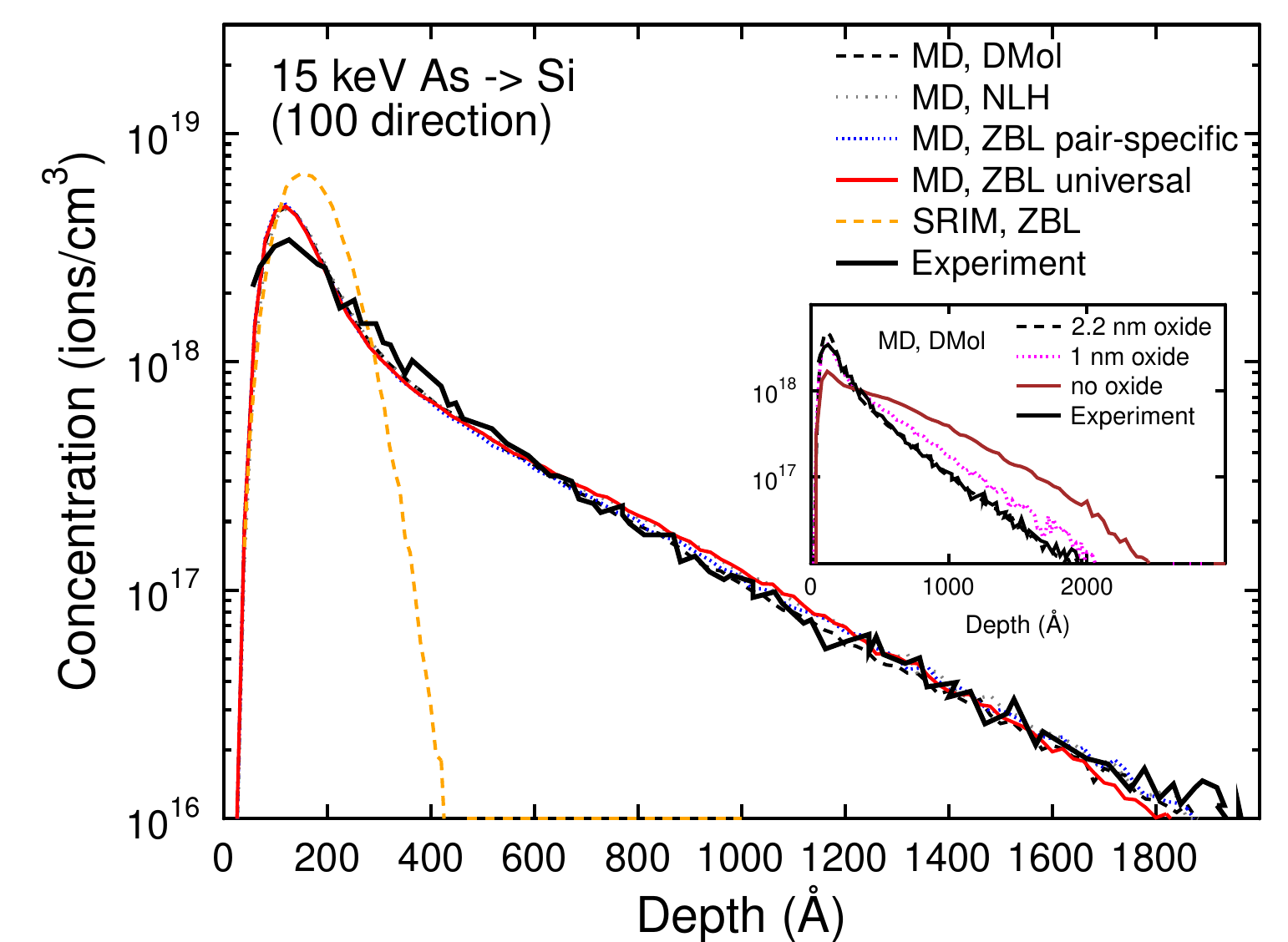}
    \caption{15 keV As ions into the [100] channeling direction}
    \label{fig:assiranges-b}
  \end{subfigure}
\caption{
Range profiles for As ions implanted into Si compared with experimental data \cite{Cai96}. The data labeled "MD" are simulated with the MD-RIA approach with the MDRANGE code, and "SRIM" with the commonly used BCA code SRIM.  {\red The insets show how the range profile depends on the thickness of the amorphous oxide layer.}
}
\label{fig:assiranges}
\end{figure}

\begin{figure}
\centering
  \begin{subfigure}[b]{0.9\columnwidth}
    \centering
    \includegraphics[width=\textwidth]{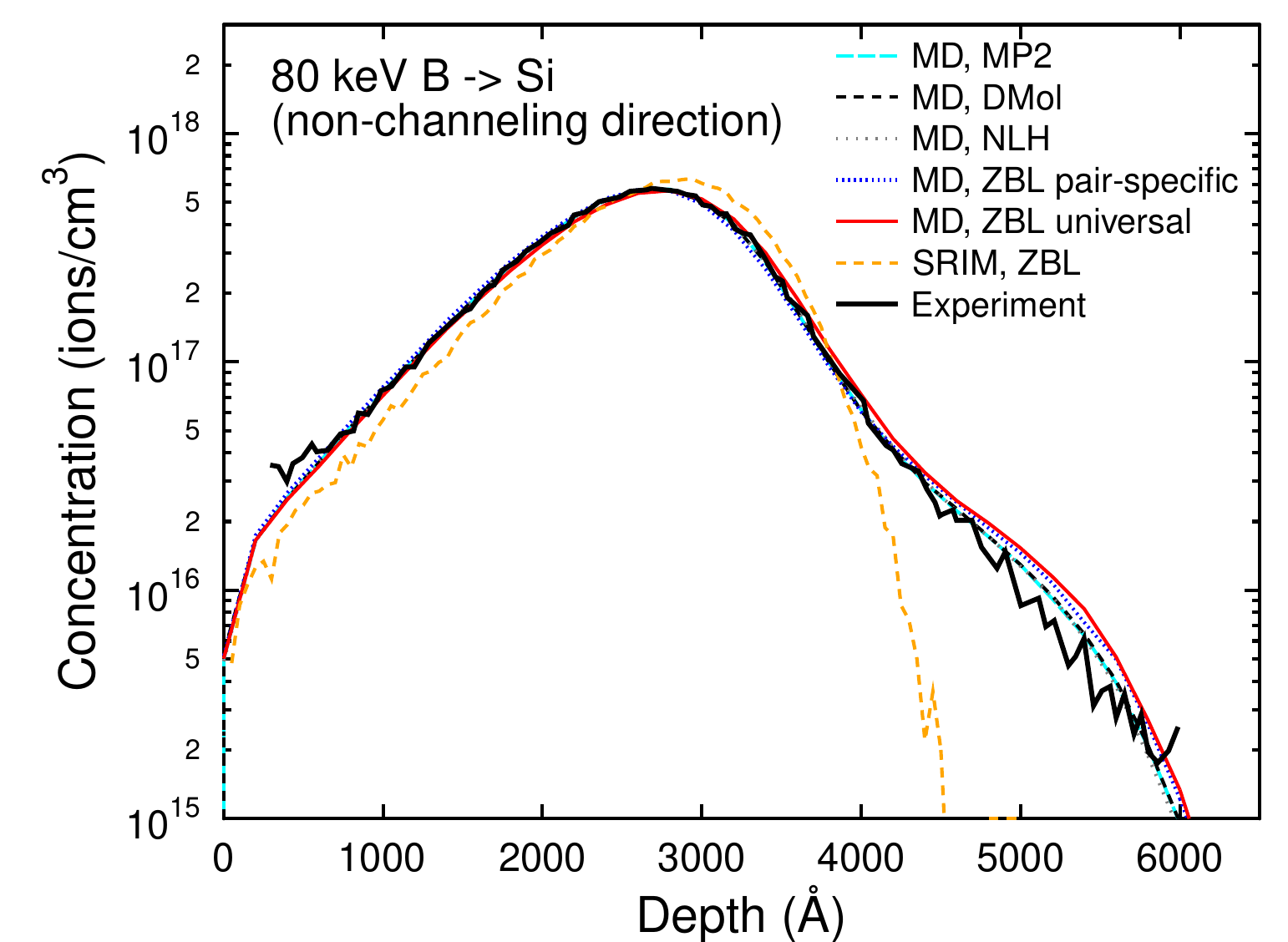} 
    \caption{80 keV B ions into a non-channeling direction (tilt=7$^\circ$, twist=30$^\circ$)}
    \label{fig:bsiranges-a}
  \end{subfigure}
  \hfill
  \begin{subfigure}[b]{0.9\columnwidth}
    \centering
    \includegraphics[width=\textwidth]{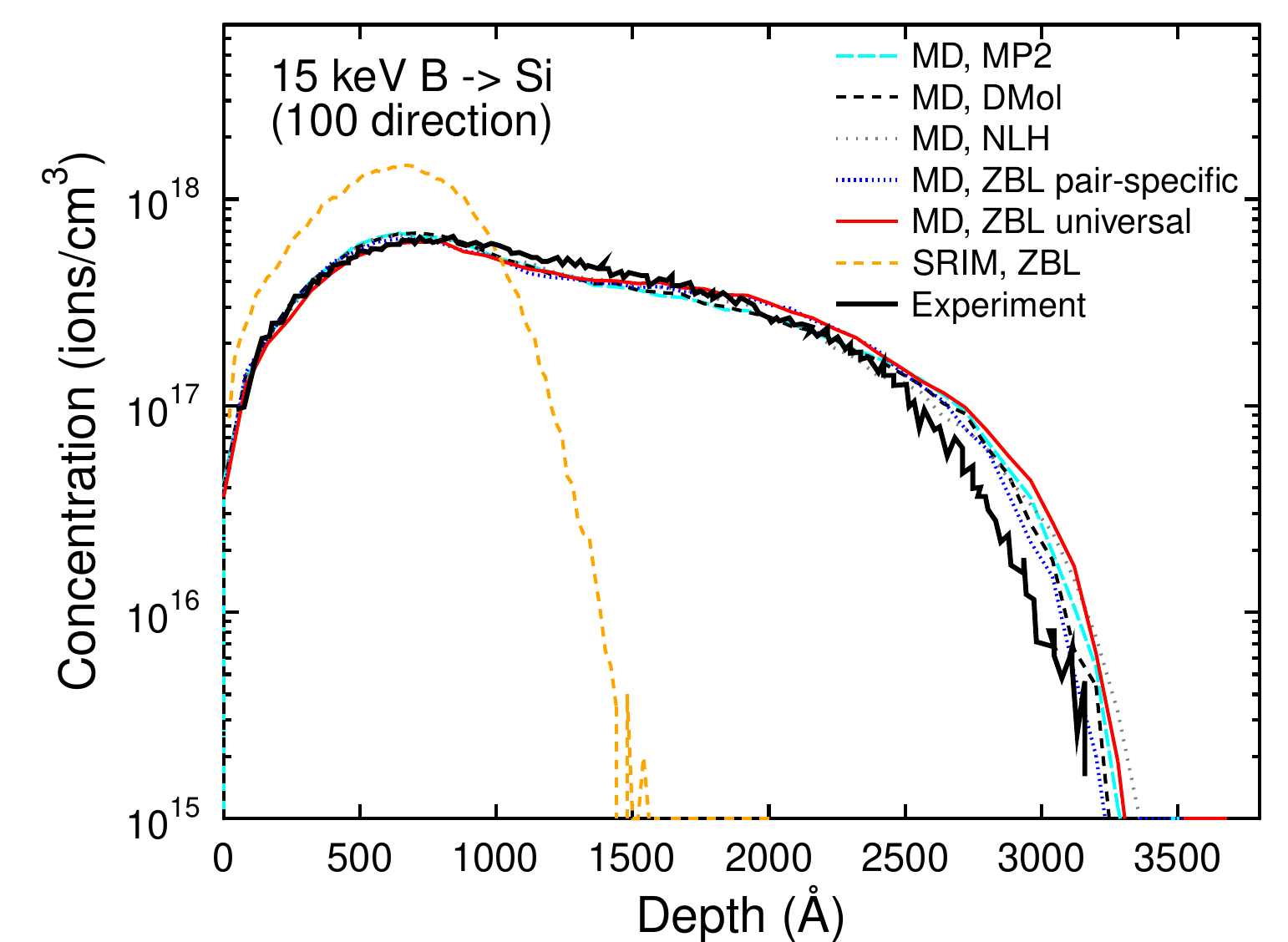}
    \caption{15 keV B ions into the [100] channeling direction}
    \label{fig:bsiranges-b}
  \end{subfigure}
\caption{
Computed range profiles for B ions implanted into Si in a non-channeling direction (\cref{fig:bsiranges-a}) or the channeling 100 direction (\cref{fig:bsiranges-b}).
The data labeled "MD" are simulated with the MD-RIA approach with the MDRANGE code, and "SRIM" with the commonly used BCA code SRIM.
Experimental data \cite{Cai96} shown for comparison.
}
\label{fig:bsiranges}
\end{figure}

\begin{figure}
\centering
  \begin{subfigure}[b]{0.9\columnwidth}
    \centering
    \includegraphics[width=\textwidth]{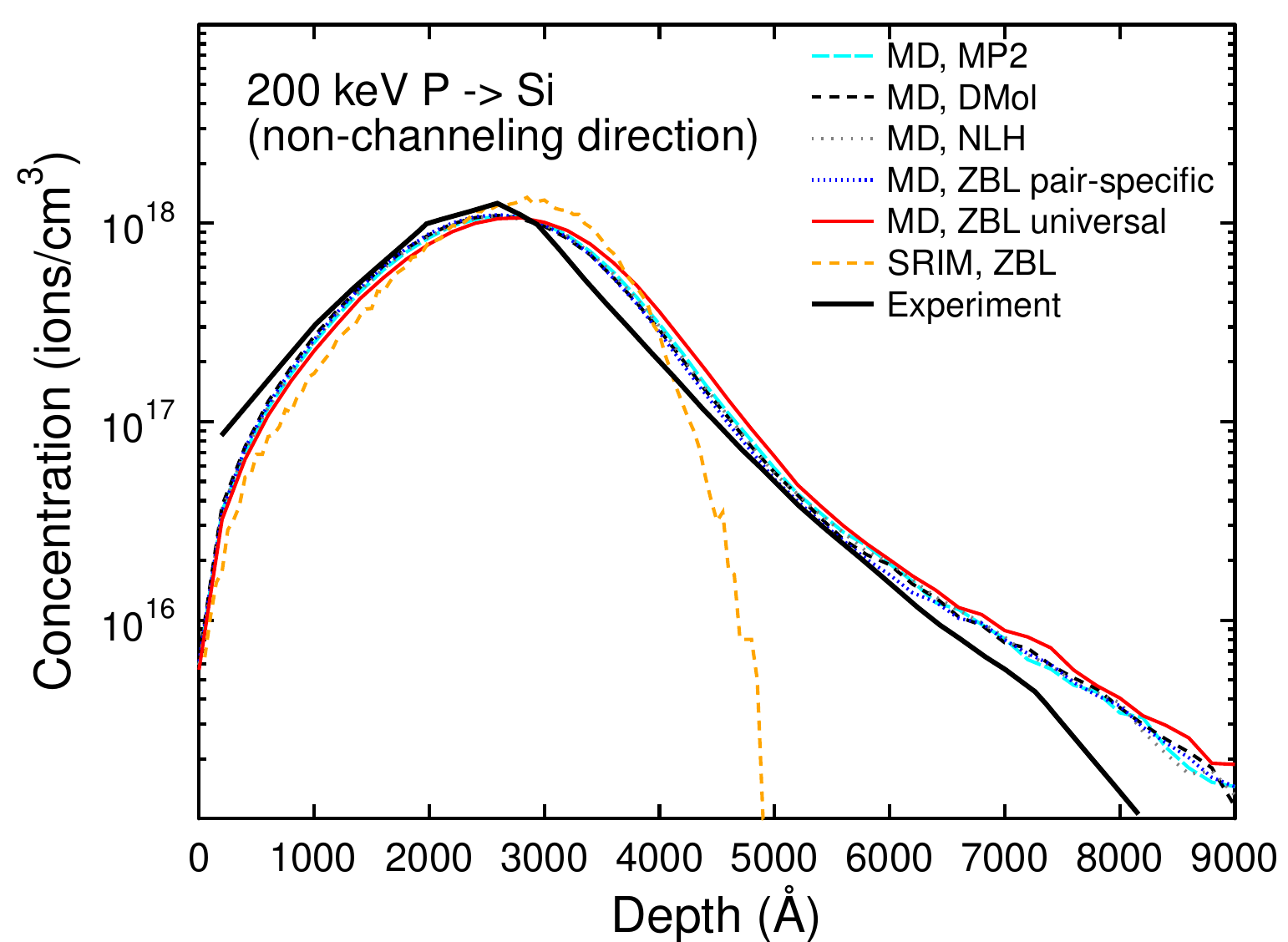} 
    \caption{200 keV P ions into a non-channeling direction (tilt=8$^\circ$, twist=18$^\circ$)}
    \label{fig:psiranges-a}
  \end{subfigure}
  \hfill
  \begin{subfigure}[b]{0.9\columnwidth}
    \centering
    \includegraphics[width=\textwidth]{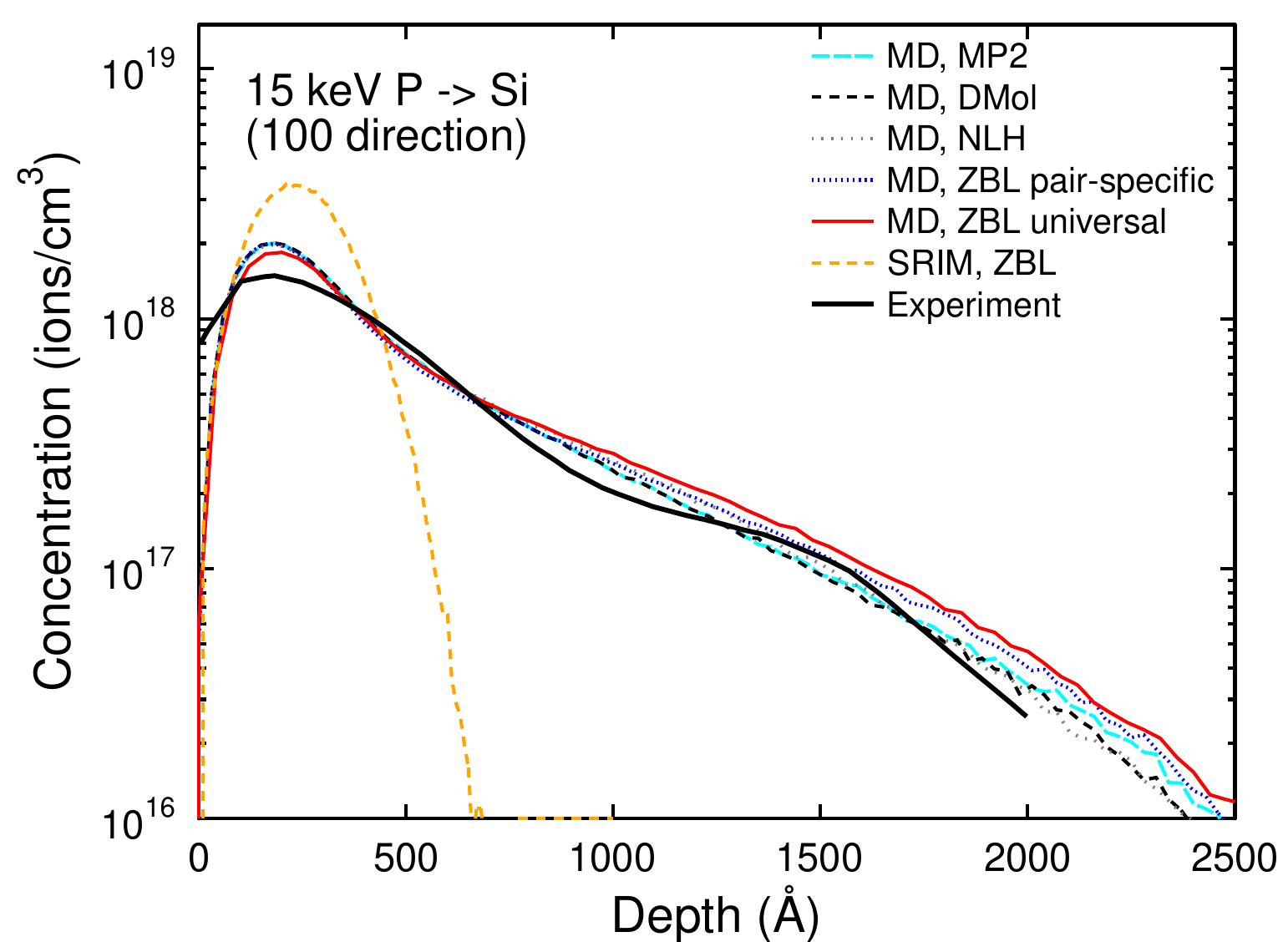}
    \caption{15 keV P ions into the [100] channeling direction}
    \label{fig:psiranges-b}
  \end{subfigure}
\caption{
Computed range profiles for P ions implanted into Si in a non-channeling (\cref{fig:psiranges-a}) or the channeling [100] direction (\cref{fig:psiranges-b}).
The data labeled "MD" are simulated with the MD-RIA approach with the MDRANGE code, and "SRIM" with the commonly used BCA code SRIM.
Experimental data \cite{Cai98} shown for comparison. 
}
\label{fig:psiranges}
\end{figure}

The dose used in the simulations was the same as the experimental fluence of 10$^{13}$ ions/cm$^2$, except for the case of 200 keV P ions, for which it was $2.5 \times$10$^{13}$ ions/cm$^2$. For the cases shown in \crefrange{fig:assiranges}{fig:psiranges}, the simulations were carried out for one million ions for clarity in the plots.
The comparisons with experiments contain no adjustable parameters, and both the depth and concentration scales are comparable in absolute values.

\subsubsection{\red Effect of oxide thickness}

{\red To examine how sensitive the results are to the oxide layer thickness, we ran the case of 15 keV As implantation into the 100 channel and 100 keV As implantation into the non-channeling thickness with only 1 nm oxide thickness as well as without any oxide at all. The results are plotted in the inset of Fig. \ref{fig:assiranges} and the mean ranges given in Table \ref{tab:oxideeffect}. The comparisons show that the ranges are indeed quite sensitive to the oxide. 
In the channeling  case, the ranges are much longer without the oxide. On the other hand, for a 1 nm oxide the results are not too different from the case of 2.2 nm oxide. This shows that even a thin oxide layer is sufficient to scatter many of the ions right at the surface and hence reduce the channeling. In the non-channeling case, the ranges are, on the other hand, slightly shorter without any oxide. This is because the oxide scatters some of the ions into a channel, leading to a longer channeling tail.
In both cases, the results for 2.2 nm oxide thickness clearly agree best with the experiments.

\begin{table}
\caption{\red Comparison of mean ranges $R_{\rm mean}$ for As ranges in Si modelled with the DMol potential for different oxide thicknesses.
The error bar is the standard error of the mean.
Non-channeling means implantation into directions where channeling is avoided, simulated with the same sample tilt and twist angles as in the experiments.
}
\label{tab:oxideeffect}
\begin{tabular}{|lccc|}
\hline
Case    &  \multicolumn{3}{c|}{$R_{\rm mean}$ (Å)} \\
        &  No oxide & 1 nm oxide & 2.2 nm oxide \\
\hline
  100 keV non-channeling & 712$\pm$2 & 763$\pm$2 & 778$\pm$1 \\ 
\hline
  15 keV [100] channel   & 588$\pm$2 & 376$\pm$& 298$\pm$1 \\
\hline
\end{tabular}
\end{table}

This observation also explains why many of the cases that are not compared to experiments (subsection \ref{subsec:interpotentialcomp}) have much stronger channeling effects than the Si cases compared to experiment. In those simulations, the idealized condition of having no surface oxide or angular beam spread maximizes the channeling effects.
}

\subsubsection{\red Comparison of results with experiments}

The results of the above approach are shown in \cref{fig:assiranges} for As-Si implantation, in \cref{fig:bsiranges} for B-Si implantation, and in \cref{fig:psiranges} for P-Si implantation.
To put the results into context, a range profile calculated with the commonly used SRIM code \cite{SRIM-2013,SRIMbook} that does not include channeling are also shown in the figures.
A numerical summary of the results is given in \cref{tab:exptcomp}.
Since the SRIM results obviously do not agree with experiments, SRIM mean ranges are not included in the table.

The results show that the MDRANGE simulation results agree overall well with experiments, which is largely thanks to the experimental electronic stopping power being the physically well-motivated one utilizing a 3D electron density. 
The MP2, DMol, and NLH results are similar, as expected from the agreement of the DMol and MP2 potentials (cf. \cref{fig:allaverage-mp2}) and from the NLH potential being a fit to the DMol data. 
The ZBL pair-specific potential results also tends to agree well with experiment. This is almost certainly due to the ZBL pair-specific potentials relevant for the three cases here happening to be fairly close to the MP2 and DMol data. While many ZBL pair-specific potentials differ from the MP2 and DMol by more than 10\% (\cref{fig:allaverage-zblpair-dmol}), the ones here all differ less than this: for B-Si, P-Si and As-Si the average difference above 100 eV is only 2.2\%, 2.8\% and 3.3\%, respectively.
The ZBL universal potential differs the most from experiments, especially tending to overestimate the range in the channeling tail for B-Si and P-Si cases (\cref{fig:bsiranges,fig:psiranges}).

Inspection of \cref{tab:exptcomp} shows that none of the potentials always gives the best agreement with the experiment.
However, the analysis of the average deviations show that the first principles (MP2 and DMol) approaches on average are closest to the experiments.

A likely reason for the minor deviations from the experiment is inadequacies of the electronic stopping model; when the Puska--Echenique--Nieminen--Ritchie model was developed, different varieties did give somewhat different results especially in the channeling conditions \cite{Sil99,Sil00,Pel03e}.
Also, in the current work we found that the range profiles in channeling conditions are sensitive to the beam divergence, which was not reported in the experimental work.
For instance, for the case of 180 keV As ions implanted in the [100] channel in Si modelled with the DMol potential, the beam divergence of 1$^\circ$ gave a mean range of 3560$\pm$9 \AA, whereas a divergence of 0.8$^\circ$ gave a mean range of 3710$\pm$30 \AA.

While it is clear that SRIM (which does not include crystal structure) strongly disagrees with experiments in the channeling cases, it is noteworthy that it also very clearly disagrees with experiments in the tails of the non-channeling implantation profiles.
This emphasizes the need to include the crystal structure when simulating ion implantation range profiles of crystals regardless of implantation direction.

\begin{table*}
\caption{Comparison of mean ranges $R_{\rm mean}$ of different ions and implantation energies $E$ in Si with experiment.
The error bar is the standard error of the mean.
The experimental values
are determined from profiles digitized from References \cite{Cai96,Cai98}.
To enable a one-to-one comparison, the mean range is determined from data in the depth interval $z_{\rm min}$--$z_{\rm max}$ where experimental data is available. Non-channeling means implantation into directions where channeling is avoided, simulated with the same sample tilt and twist angles as in the experiments. The bottom part of the table shows the average and root-mean-square deviation of the simulated results from the experiments.
}
\label{tab:exptcomp}
\begin{tabular}{|lllrrcccccc|}
\hline
Ion & $E$ (keV) & Direction    & $z_{\rm min}$(Å) & $z_{\rm max}$ (Å) & \multicolumn{6}{c|}{$R_{\rm mean} [\hbox{between } z_{\rm min} \hbox{ and } z_{\rm max}]$ (Å)} \\
    &           &           &    & & Experiment & MP2 & DMol & NLH & ZBLpair & ZBL univ. \\
\hline
As   &  180    & Amorphous Si  & 0  & 2997 & 1212 & -- & 1211$\pm$1 & 1212$\pm$1 & 1187$\pm$1 & 1227$\pm$1 \\ 
As   &  15     & Non-channeling   & 0 & 1563 & 175 & -- & 197$\pm$1 & 204$\pm$1 & 194$\pm$1  & 198$\pm$1  \\ 
As   &  100    & Non-channeling   & 0 & 2986 & 739 & -- & 747$\pm$1 & 750$\pm$1 & 727$\pm$1 & 747$\pm$1 \\ 
As   &  15       & [100] channel   & 55 & 2674 & 325 & -- & 298$\pm$1  & 310$\pm$1 & 298$\pm$1 & 304$\pm$1 \\
As   &  50      & [100] channel   & 41 & 5846 & 1020 & -- & 1100$\pm$3 & 1117$\pm$3 & 1091$\pm$3  & 1127$\pm$3 \\ 
As   &  180      & [100] channel   & 41 & 14936 & 3654 & -- & 3560$\pm$9 & 3591$\pm$9 & 3523$\pm$9 & 3644$\pm$9\\ 
\hline
B   & 15        & Non-channeling & 89 & 2827 &  638 & 668$\pm$1 & 665$\pm$1 & 667$\pm$1 & 673$\pm$1 & 689$\pm$1 \\
B   & 35        & Non-channeling &  284 & 3447 &  1365 & 1335$\pm$2 & 1333$\pm$2 & 1337$\pm$2 & 1332$\pm$2 & 1367$\pm$2 \\
B   & 80        & Non-channeling & 291 & 5989 &  2579 & 2579$\pm$2 & 2579$\pm$3 & 2581$\pm$2 & 2561$\pm$3 & 2621$\pm$3 \\
B   & 15        & [100] channel    & 46 & 3159 & 1185 & 1118$\pm$2 & 1119$\pm$2 & 1115$\pm$2 & 1146$\pm$2 & 1179 $\pm$2 \\
B  & 35 & [100] channel & 271 & 4804 & 2306 & 2289$\pm$3 & 2283$\pm$6 & 2276$\pm$3 & 2343$\pm$6 & 2464$\pm$6  \\
B & 80 & [100] channel & 308 & 7477 & 4346 & 4112$\pm$5 & 4104$\pm$5 & 4091$\pm$5 & 4174$\pm$5 & 4202$\pm$5 \\
\hline 
P   & 100       & Non-channeling  & 133 & 3103 & 1270 & 1299$\pm$2 & 1286$\pm$2 & 1294$\pm$2  & 1278$\pm$2 & 1347$\pm$2 \\
P   & 200       & Non-channeling  & 194 & 8162 & 2438 & 2655$\pm$3 & 2619$\pm$3 & 2637$\pm$3 & 2614$\pm$4 & 2731$\pm$3 \\
P   & 15       & [100] channel  & 0 & 1998 & 479 & 468$\pm$1 & 463$\pm$1 & 476$\pm$1 & 481$\pm$1 & 505$\pm$2 \\
P   & 100       & [100] channel  & 255 & 8393 & 3073 & 3157$\pm$6 & 3125$\pm$6 & 3165$\pm$6 & 3156$\pm$6 & 3243$\pm$6 \\
\hline
\hline
\multicolumn{3}{|l}{\bf Average deviation } & & & & 0.23\% & 0.46\% & 1.4\% & 0.62\% & 3.8\% \\
\multicolumn{3}{|l}{\bf Root-mean-square deviation } & & & & 4.3\% & 5.3\% & 6.0\% & 4.9\% & 6.6\% \\
\hline
\hline
\end{tabular}
\end{table*}

\section{Summary and conclusions \label{sec:summary}}

We have {\red presented and compared results of} repulsive interatomic potential{\red calculations} at three levels of theory.
The one-electron basis set is known to be the most important source of error for repulsive potentials \cite{Nor96c}.
A numerically robust basis set scheme has been recently verified against fully numerical electronic structure calculations at the Hartree--Fock level of theory in \citeref{Lehtola2020_PRA_32504}.
The state of the art was represented in this work by an analogous scheme based on the use of flexible Gaussian-type orbital (GTO) basis sets within the PySCF program \cite{Sun2020_JCP_24109}, in which the total energy was estimated with second-order M\o{}ller--Plesset perturbation theory (MP2) calculations.
The MP2 calculations were carried out for the set of diatomic molecules with $Z_1+Z_2 \leq 36$.

The MP2 calculations served to validate the accuracy of a set of density functional theory (DFT) calculations performed for $Z_1,Z_2\le 92$ with numerical atomic orbitals (NAOs) with the DMol97 program.
These calculations have already been used in several studies in the literature.
We found an excellent level of agreement between the new MP2 and old DFT calculations: the screening functions $\phi(r)$ defined by \cref{eq:screen} agreed to within 1\% for all atom pairs above 100 eV, with the exception of \ce{B\bond{-}Ne} for which the DMol calculation appears incorrect.

The third level of theory considered was orbital-free density functional theory (OF-DFT) evaluated on the superposition of atomic densities, following the approach of Ziegler, Biersack, and Littmark (ZBL) in \citeref{ZBL}.
We found that the ZBL potentials and the reproduced OF-DFT calculations have considerable differences to our more accurate self-consistent quantum chemical calculations: both the ZBL universal potential and the ZBL pair-specific calculations differ from the quantum chemical approaches by {\red $\sim$ 5-10\% above 100 eV.}

Having verified the accuracy of the DMol scheme, we formed pair-specific analytical fits of the DMol screening functions to a sum of three exponentials, yielding the purely repulsive NLH potentials of this work.
By comparison to the MP2 reference data, we showed that the pair-specific NLH potentials agree with the best available quantum chemical data within $\sim 2$\% above 30 eV.
The exponents and fitting coefficients defining the NLH potential are included in the Supplementary Material \cite{Supplemental} and in an Open data set.

Studying models of ion implantation with the various potentials, we found that the range distribution may be quite sensitive to the low-energy (below 100 eV) part of the potentials under some channeling conditions.
In such cases, a description of interaction models based on purely repulsive potentials  may not be sufficient for a highly accurate description of ion implantation.

\section*{Open data and supplementary materials}

The ZBL pair-specific, DMol and MP2 data sets and the NLH potential coefficients are available open
access in the online Supplementary materials of this publication \cite{Supplemental} in the compressed archive package file \verb+nlh_potentials_opendata.tar.gz+ , as well
as at https://zenodo.org/records/14172633 (doi:0.5281/zenodo.14172632).

The online supplementary material also includes the file \verb+nlh_table_all.pdf+ that has a table of all the NLH potential coefficients \cite{Supplemental}.

\section*{Acknowledgements}

We thank Prof. Dage Sundholm for discussions and Clemens Domnanovits for developing the graphical user interface of the software for comparing screening functions, which is distributed in the Supplementary Material.
S.L. thanks the Academy of Finland for financial support under project numbers 350282 and 353749 and K. N. under  project numbers 333225. 
Sone aspects of this work have been carried out within the framework of the EUROfusion Consortium, funded by the European Union via the Euratom Research and Training Programme (Grant Agreement No 101052200 — EUROfusion). Views and opinions expressed are however those of the author(s) only and do not necessarily reflect those of the European Union or the European Commission.
Neither the European Union nor the European Commission can be held responsible for them.
Grants of computer time from the IT Center for Science CSC in Espoo, Finland, are gratefully acknowledged. 

\appendix
\section{Combining high-energy and equilibrium potentials \label{sec:highe_equilibrium}}

Atoms in a solid or molecular system are bound to other atoms by various bonds (covalent, ionic, metallic, van der Waals, etc.). 
This bonding can be modeled computationally within the molecular dynamics method, in which the motion of $N$ atoms is described within the Born--Oppenheimer approximation \cite{Born1927_APB_457} by integrating Newton's equations of motion for all atoms with a small time step \cite{Allen-Tildesley, Leach}. 
According to Newton's II law, the instantaneous acceleration of each atom is given by the net force acting on it; hence conventional MD simulations are enabled by interatomic potentials that describe the interactions in the system.

The slowing down (stopping) of energetic ions and nuclear recoils in materials can be described within the binary collision approximation (BCA) \cite{Nor18b}. 
Energetic atom movement is described in this approach as a sequence of binary collisions for which the scattering angle and energy loss is determined from the two-body scattering integral \cite{BCA}.
As only binary collisions are treated, the pairwise atomic high-energy repulsive potential suffices to describe interatomic interactions. 
Only the repulsive pair potential is needed also in the special case of the recoil interaction approximation variant of MD (MD-RIA) \cite{Nor94b}, which is meant only for high-energy interactions.

We now address the question of how the two types of interactions---near-equilibrium many-body and high-energy repulsive---can be combined.
{\red The importance of doing this in a systematic manner is emphasized by the conclusion of the main text that ranges in channeling conditions are sometimes sensitive to the equilibrium part of the potential (cf. section \ref{subsec:interpotentialcomp}).
To address this issue it is important to be able to join the repulsive and equilibrium potentials without free parameters, as the choice of such parameters could introduce additional uncertainty into the channeling results.}

The key to this combination is that at the highest energies the atom collisions can practically always be handled as two-body interactions, as strong keV energy collisions involve internuclear distances $< 0.5$ \AA, while regular internuclear distances are of the order of 2--3 \AA.
It is thus practically impossible for more than two atoms to simultaneously experience the strongly repulsive part of the potential. 
It is also fully sufficient to use BCA type repulsive pair potentials of the form of \cref{eq:screenedCoulomb} for small internuclear distances in MD simulations.

However, when the energetic particles thermalize, one needs a smooth transition from the highly repulsive potential at small internuclear distances to the equilibrium interatomic potential in full MD.
As has been discussed above, the potentials used in most modern MD simulations contain at least some many-body character. 

Until recently, the short-range repulsive pair potential was joined usually in a rather {\it ad hoc} character to the equilibrium potential.
As the first derivative of the potential gives the force, it is clear from basic physico-chemical considerations that a classical interatomic potential and its first derivatives should be smoothly varying and continuous functions.
A straightforward approach is thus to take the equilibrium potential and join it to the high-energy repulsive expression using some interpolation function that ensures a continuous transition.

Since the strength of a chemical bond is typically in the order of 1--2 eV/bond, the formalism for purely repulsive potentials can be assumed to be valid for distances $r_{\rm rep}$ for which $V_{\rm rep} \gtrsim 10$ eV.
Hence, the range of distances where the transition should take place is $r_{\rm rep} < r < r_{\rm equi}$.

One can easily think of two formalisms for achieving the wanted behavior.
The first is to use an interpolation function $F(r)$ defined at all values of $r$
\begin{equation}\label{eq:Fermi}
	V(r) = F(r) V_{\rm rep}(r) + [1-F(r)] V_{\rm equi}(r) 
\end{equation}
where $0 \le F(r) \le 1$ is some continuous function that has the properties $F(r)\to 1$ when $r \to 0$ and $F(r)\to 0$ when $r\to\infty$.
To achieve a {\red smooth} transition from the many-body to the repulsive potential {\red that does not affect the equilibrium potential}, the function $F(r)$ should transition from 1 to 0 in a {\red fairly} narrow interval below $r_{\rm equi}$; the Fermi function is one possibility \cite{Nor96}. 
This approach has the advantage that it naturally ensures that e.g. angular dependencies associated with chemical bonds are not active at small interatomic separations. 

The second approach is to use a piecewise definition, interpolating only between $r_{\rm rep}$ and $r_{\rm interp}$, which is some distance at which the interpolation is turned off and the potential becomes the many-body potential:
\begin{equation}\label{eq:piecewise}
	V(r) = \left\{ 
	    \begin{array}{ll}
		V_{\rm rep}(r)   & \hbox{when\ } r < r_{\rm rep}  \\
            	V_{\rm interp}(r)  & \hbox{when\ }  r_{\rm rep} \le r \le r_{\rm interp} \\
            	V_{\rm equi}(r)  & \hbox{when\ }  r > r_{\rm interp}  \\
    	    \end{array}\right.
\end{equation}
To ensure correctness, $r_{\rm interp} < r_{\rm equi} - r_{\rm vib}$, where $r_{\rm vib}$ is the magnitude of the equilibrium thermal vibrations around the equilibrium separation in the system.
The interpolation function in \cref{eq:piecewise}, $V_{\rm interp}(r)$, should be at least twice differentiable and thereby yield continuous values for the energy and its two derivatives at the interval boundaries, $r_{\rm rep}$ and $r_{\rm interp}$. 
This can be achieved e.g. with a fifth-order polynomial, whose six coefficients can be solved analytically from the three boundary conditions at either end of the interval.

The first approach, \cref{eq:Fermi}, is typically used with Tersoff-like potentials \cite{Nor96, Alb01b, Jus05}, while the second one, \cref{eq:piecewise}, is often used with embedded-atom-method (EAM) potentials \cite{Gha94, Nor98}. 
In either case, the choice of the joining model parameters and/or interval is somewhat arbitrary, and there is no experimental data to which the joining parameters can be directly fit.
However, the quality of the final potential can be tested e.g. against high-pressure equations of state \cite{Nor98}, threshold displacement energies \cite{Bjo07a, San15a}, or the melting points of materials \cite{Nor98}.

The intermediate energy regime can of course also be tested against DFT data for atoms placed intentionally at shorter-than-normal interatomic separations in crystals or amorphous materials \cite{Bel02, San15a}.
However, this was earlier done more in the way of testing than fitting.

Recently, machine-learned (ML) interatomic potentials have been extended to include also repulsive pair potential fitting in a systematic manner, removing the {\it ad hoc} character of the earlier approaches.
Generally, ML potentials are trained systematically against a large set of DFT data.
However, these data sets do not typically include data for small internuclear distances, and obtaining such data tends to be difficult, as standard solid state codes assume that the inner electronic shells are frozen, via the use of so-called pseudopotentials.
But, we argue that including such data is not necessary, as per the arguments presented above, the high-energy interactions are in any case pair-wise in character and can thereby be obtained from separate calculations on diatomic molecules, as carried out in this work, for example.

Systematic ML potentials can be fit fully consistently with the short-range potentials without introducing additional free parameters in the manner first introduced by \citet{Byg19b}.
In this method, a DFT-derived high-energy interatomic potential $V_{\rm rep}(r)$ is first obtained for a diatomic molecule.
Next, this potential is smoothly set to vanish at some distance $r < r_{\rm equi}$ with a cutoff function $f_c$,
\begin{equation}
	V_{\rm pair}(r) = V_{\rm rep}(r) f_c(r). 
\end{equation} 
The DFT database that would be used to construct an ML equilibrium interatomic potential $V_{\rm ML}$ needs to include some configurations that involve distances $r \lesssim r_{\rm equi}$, but configurations with very small distances $r \ll r_{\rm equi}$ are not needed in the ML databases.
After this, the total repulsive pair energy $V_{\rm pair}$ is calculated for all these configurations, and is subtracted from the DFT energies, yielding the reduced energies:
\begin{equation}
	E_{\rm DFT}^{\rm reduced} = E_{\rm DFT} - \sum_{i<j} V_{\rm pair}
\end{equation}
When the ML potential $V_{\rm ML}^{\rm reduced}$ is then trained against these reduced energies, a smooth fit to the short-range interatomic potential is ensured, and the final energies of atomic
configurations are fully consistent with both the short-range and equilibrium potential DFT data sets. 
Naturally, one needs to ensure that the ML formalism does not diverge to extreme values in the short distance range, i.e. $|V_{\rm ML}^{\rm reduced}|$ should be $\ll V_{\rm rep}$ when $r < r_{\rm equi}$.
The final energies are then reconstructed simply as \cite{Byg19b}:
\begin{equation}
	V_{\rm TOT} = \sum_{i<j} V_{\rm pair} + \sum_{i}^N V_{\rm ML}^{\rm reduced}
\end{equation}

This approach also has the advantage that it tackles a minor deficiency of repulsive potentials calculated for diatomic molecules: diatomic calculations do not fully account for the fact that the outer electronic orbitals are somewhat contracted in the solid state compared to those of a free atom \cite{ZBL}, as discussed in \cref{sec:solidstateresults}.
This effect is included in the formalism for fitting the ML potential described above since the DFT calculations do include the solid state effects.
Hence, if there is a deficiency in the repulsive  potential, the fitting of the reduced DFT data set will compensate for this.

{\red  We also note that the approach to first fix the repulsive potential $V_{\rm rep}$, then construct a many-body potential by fitting the reduced energies $E_{\rm DFT}^{\rm reduced}$ could also be used to remove the ambiguity in potential joining when constructing analytical interatomic potentials.}

This discussion highlighted that energies calculated for a diatomic system offer an excellent starting point for potentials to be used in simulations of radiation effects both within the BCA approximation as well as full MD simulations, the latter class also including the most modern machine-learned approaches.

\section{Computation of the orbital-free density functional potentials \label{sec:orbital-free-appendix}}

In this appendix, we discuss how the orbital-free density functional calculations are carried out using the fixed atomic electron densities.
In order to make the notation unambiguous, we therefore use a different notation in this appendix.
Following the literature on electronic structure calculations on diatomic molecules \cite{Lehtola2019_IJQC_25968}, the internuclear distance is denoted as $R$, points in space by ${\bf r}$, and the distances of the given point ${\bf r}$ to the two nuclei will be denoted as $r_1$ and $r_2$.
The nuclei are placed at coordinates ${\bf R}_1$ and ${\bf R}_2$, and relative coordinates from the two nuclei are analogously denoted as ${\bf r}_1 = {\bf r} - {\bf R_1}$ and ${\bf r}_2 = {\bf r} - {\bf R}_2$, respectively. 
Obviously, $r_1=|{\bf r}_1|$, $r_2=|{\bf r}_2|$, and $R =|{\bf R}_1 - {\bf R}_2|$.

The basic input to the calculations are the electron densities $\rho(r)$, where $\rho$ is given in units of electrons per \AA$^3$ and the distance from the nucleus $r$ is in units of \AA. 
We used the spherically averaged solid-state electron densities as listed in the ZBL book \cite{ZBL} to determine what we call the pair-specific ZBL potentials for all non-radioactive elements, plus Bi and U. 
In addition, we calculated electron densities for many atoms in the gas phase with the Grasp2018 program \cite{Fro19, Sch22} for comparison.

The ZBL electron distribution $\rho(r)$ for each atom is assumed to be constant beyond half the nearest-neighbor distance in the most common crystal structure of the element (``muffin-tin'' radius $r_{\rm MT}$): $\rho(r) = \rho(r_{\rm MT})$ at $r > r_{\rm MT}$. 
To describe the isolated-atom electron densities in the same way as the ZBL electron densities, a pseudo-muffin-tin radius was determined such that the total number of electrons found at $r > r_{\rm MT}$ evaluates to 0.01.
A maximum radius $r_{\rm max}$ results from the requirement of accommodating the total charge of the atom. 

\begin{figure}
    \center
    \includegraphics[width=0.6\columnwidth]{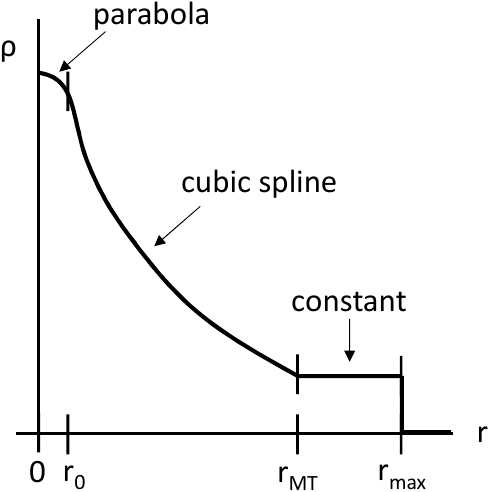}
    \caption{Schematic of the electron density distribution. 
    In $r_0 \le r \le r_{\rm MT}$, $\rho(r)$ is described by a cubic spline. 
    In $0 \le r \le r_0$, $\rho(r)$ is described by a parabola with $\rho'(0) = 0$, and $\rho(r_0)$ and $\rho'(r_0)$ are continuously joined to the spline used for $r \ge r_0$.
    For $r_{\rm MT} \le r \le r_{\rm max}$, $\rho(r) = \rho(r_{\rm MT})$.}
    \label{fig:charge}
\end{figure}

The evaluation of the integrals requires care, because of (i) the discontinuous first derivative of $\rho(r)$ at $r_{\rm MT}$, (ii) the abrupt termination of the electron densities at $r_{\rm max}$ (compare \cref{fig:charge,fig:eldens}), and (iii) the subtraction of the almost equal terms $V_{nn} + V_{ee}$ and $|V_{ne} + V_{en}|$ at large internuclear distances $R$. 
Our implementation differs from the approach of \citet{Wed67} used by ZBL in \citeref{ZBL} and is described in the following.

To calculate the interatomic potential, a piecewise polynomial describing $\rho(r)$ was constructed as follows; see \cref{fig:charge} for an illustration. 
From the $\rho(r)$ values between the smallest listed radius $r=r_0$ and $r = r_{\rm MT}$, a cubic spline with not-a-knot boundary conditions was constructed using the \texttt{CubicSpline} class of the SciPy \cite{SciPy} library.
Since the tabulated values start at $r_0 > 0$, the resulting piecewise polynomial was extended at its lower end towards $r = 0$ by a parabola satisfying $\rho'(0) = 0$ and $C^1$ continuity at $r = r_0$.
At its upper end, the piecewise polynomial was extended with only C$^0$ continuity by the constant $\rho(r) = \rho(r_{\rm MT})$ up to $r = r_{\rm max}$.
This piecewise polynomial was represented by a \texttt{PPoly} object of the SciPy library.

As stated by \cref{eq:ofdft}, the potential energy of the two atoms consists of terms describing the Coulomb interaction and two quantum-mechanical contributions.
The electronic Coulomb term $V_c$ consists of the nucleus-electron, electron-nucleus, and electron-electron contributions,
\begin{equation}
    V_c = V_{ne} + V_{en} + V_{ee},
    \label{eq:Vc}
\end{equation}
which are given by
\begin{align}
    V_{ne} & = - \frac{e^2}{4 \pi \varepsilon_0} 
        \int_{V_2} \frac{Z_1 \rho_2(r_2)}{r_1} \,{\rm d}^3 r
    \label{eq:Vne} \\
    V_{en} & = 
        - \frac{e^2}{4 \pi \varepsilon_0} 
        \int_{V_1} \frac{\rho_1(r_1) Z_2}{r_2} \,{\rm d}^3 r
    \label{eq:Ven} \\
    V_{ee} & = 
        \frac{e^2}{4 \pi \varepsilon_0} 
	    \int_{V_1} \int_{V_2} \frac{\rho_1({\bf x}_1) \rho_2({\bf x}_2)}{|{\bf x}_1 - {\bf x}_2|} \, {\rm d}^3 x_1 \,{\rm d}^3 x_2 .
    \label{eq:Vee}
\end{align}

\Cref{eq:Vne,eq:Ven} describe the interaction of a
charge cloud with a point charge, integrating over all of space, while
\cref{eq:Vee} describes the interaction of two charge clouds with each
other.
In \cref{eq:Vne,eq:Ven}, $r_1$ and $r_2$ denote the distances of the infinitesimal volume element ${\rm d}^3r$ from nuclei 1 and 2, respectively.
\Cref{eq:Vee} is for now written in a general form where ${\bf x}_1$ and ${\bf x}_2$ are coordinates of two volume elements, and $|{\bf x}_{1}-{\bf x}_2|$ is the distance between them.
The numerical value
of $e^2 /(4\pi \epsilon_0)$ was given in the main text after
\cref{eq:Coulomb}.

\begin{figure}
    \center
    \includegraphics[width=0.9\columnwidth]{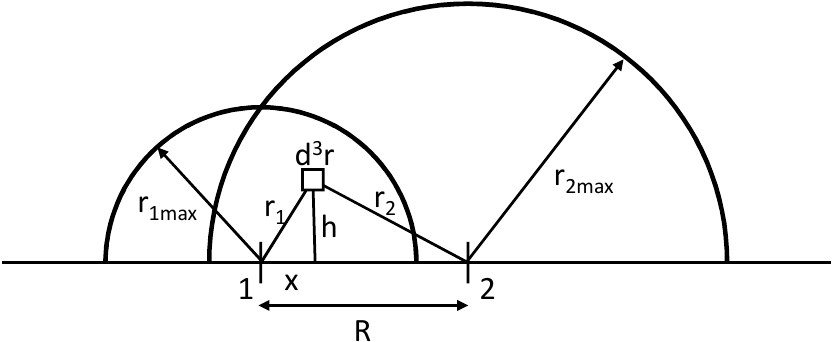}
    \caption{Coordinates describing a position with respect to the two nuclei labeled with ``1'' and ``2''.}
    \label{fig:coordinates}
\end{figure}

As extensively discussed in the main text, the basic assumption for
the orbital-free density functional calculation of the interatomic
potential $V(R)$ of two atoms 1 and 2 with internuclear distance $R$
is that their electron densities $\rho_1(r_1)$ and $\rho_2(r_2)$
remain unchanged relative to their nuclei \cite{Gom49}.  As a
consequence, electrons of atom 1 are always confined to the volume
$V_1$ defined by $r_1 < r_{1{\rm max}}$ and electrons of atom 2 to the
volume $V_2$ defined by $r_2 < r_{2{\rm max}}$, see
\cref{fig:coordinates}.  In the overlap volume, $V_1 \cap V_2$, the
electron densities of the two atoms add up linearly, $\rho = \rho_1 +
\rho_2$.

The Coulomb integrals of \cref{eq:Vne,eq:Ven,eq:Vee} are facile to
evaluate using Green's theorem or the Laplace expansion for the potential \cite{Lehtola2019_IJQC_25968}: a
spherical charge distribution yields a potential which coincides with
that of a point charge
\begin{equation}
    \zeta_i(r) = 
    \int_{0}^{r} 
    4 \pi r'^2 \rho_i(r') \,{\rm d}r' 
    + r \int_{r}^{r_{i{\rm max}}} 
    4 \pi r' \rho_i(r') \,{\rm d}r' 
    \label{eq:zeta}
\end{equation}
where we have assumed that $r$ is within the extent of the charge
cloud, $r < r_{i {\rm max}}$; if $r \ge r_{i {\rm max}}$ the full
charge of the electron cloud is observed, $\zeta_i = Z_i$, as $Z_i$ is
the number of electrons on atom $i$. The Coulomb contributions can
therefore be readily written as
\begin{align}
    V_{ne} & = 
        - \frac{e^2}{4\pi\varepsilon_0}
        \frac{Z_1 \zeta_2(R)}{R}
    \label{eq:Vne1} \\
    V_{en} & = 
        - \frac{e^2}{4\pi\varepsilon_0}
        \frac{\zeta_1(R) Z_2}{R}
    \label{eq:Ven1} 
\end{align}

The electron-electron term requires a bit more thought to
evaluate. Yet, one of the integrals in the electronic Coulomb integral
of \cref{eq:Vee} can be carried out using \cref{eq:zeta} as was done
for the electron-nuclear terms above. Let us consider the interaction
of the electron charge of atom 1 contained in an infinitesimal volume
element ${\rm d}^3 r_1$ with the electrons of atom 2.  The
contribution ${\rm d}V_{ee}$ to the interatomic potential is obtained
by replacing $Z_1$ in \cref{eq:Vne1} with $- \rho_1 {\rm d}^3 r_1$ and
$R$ with $r_2$, and the total electron-electron interaction
energy is obtained by integration over the volume $V_1$,
\begin{equation}
    V_{ee} =
    	\frac{e^2}{4\pi\varepsilon_0} 
        \int_{V_1} \frac{\rho_1(r_1) \zeta_2(r_2)}{r_2}
    \, {\rm d}^3 r
    \label{eq:Vee1}
\end{equation}

The remaining task is thus to carry out the single integrals over all
space in \cref{eq:Vee1}, as well as the analogous integral in the quantum
mechanical terms: the excess kinetic energy
\begin{multline}
    V_{k} = 
        \kappa_{k} \int_{V_1 \cap V_2} \left\{ 
            \left[ \rho_1(r_1) + \rho_2(r_2) \right] ^{5/3} \right. \\ \left. - \rho_1(r_1)^{5/3} - \rho_1(r_2)^{5/3} \right\}
            \,{\rm d}^3 r
    \label{eq:Vk}
\end{multline}
and the excess exchange energy
\begin{multline}
    V_{x} = 
        - \kappa_{x} \int_{V_1 \cap V_2} \left\{
            \left[ \rho_1(r_1) + \rho_2(r_2) \right] ^{4/3} \right. \\ \left. - \rho_1(r_1)^{4/3} - \rho_1(r_2)^{4/3} \right\} 
            \,{\rm d}^3 r ,
    \label{eq:Va}
\end{multline}
where the constants $\kappa_{k}$ and $\kappa_{x}$ were already given
in \cref{eq:kappa-k,eq:kappa-x}, respectively.

Examining \cref{fig:coordinates} we realize that all integrands in
this section are rotationally symmetric about the line connecting the
two nuclei, and that the position of a point in space can be expressed
in cylindrical coordinates by denoting the position on the
internuclear axis by $x$, the distance from this axis $h$, and the
azimuthal angle $\varphi$. Because of the rotational symmetry, the
integrand can be denoted as $g(x,h)$, and the volume element ${\rm d}^3 r
$ can be written as ${\rm d}^3 r = {\rm d}x \, {\rm d}h \, h \, {\rm
  d} \varphi$. In the new coordinates, the distances from the two
nuclei can be expressed as
\begin{align}
    r_1^2 &= x^2 + h^2 \\
    r_2^2 &= (R - x)^2 + h^2,
\end{align}
as is obvious from \cref{fig:coordinates}.

As the various terms only give non-cancelling contributions in the
case the atomic spheres overlap, we only need to evaluate integrals
within atomic spheres. The expression for the integral over the volume
of the first atom can be written as
\begin{align}
  & \int_{V_1} g(x,h) \,{\rm d}^3 r \\
  & =   2 \pi \int_{-r_{1 {\rm max}}}^{r_{1 {\rm max}}} {\rm d}x
        \int_{0}^{\sqrt{r_{1{\rm max}}^2 - x^2}} {\rm d}h \, h \, g(x,h) ,
\end{align}
where the integration over $\varphi$ evaluated to $2\pi$.

Straightforward calculus yields
\begin{equation}
    h \, {\rm d}x \, {\rm d}h = 
        \frac{r_1 r_2}{R} \, {\rm d}r_1  {\rm d}r_2
\end{equation}
and
\begin{multline}
    \int_{V_1} g(x,h) \,{\rm d}^3 r_1 \\ = 
        \frac{2 \pi}{R} 
        \int_{0}^{r_{1 {\rm max}}} {\rm d}r_1 \, r_1 
        \int_{|R-r_1|}^{R+r_1} {\rm d}r_2 \, r_2 \, g(r_1,r_2).
    \label{eq:transformation}
\end{multline}

Using \cref{eq:transformation} to evaluate \cref{eq:Vee1}, we obtain
\begin{multline}    
    V_{ee} =
        \frac{e^2}{4\pi\varepsilon_0}
        \frac{2 \pi}{R}
        \int_{0}^{r_{1 {\rm max}}} r_1 \rho_1(r_1) \\
        \left[ \chi_2(R + r_1) - \chi_2(|R-r_1|) \right] \,{\rm d}r_1
    \label{eq:Vee_final}
\end{multline}
with
\begin{equation}
    \chi_2(r_2) =
        \int_{0}^{r_2} \zeta_2(r) \,{\rm d}r
    \label{eq:chi2}
\end{equation}
Similarly, the excess kinetic and exchange energies evaluate to
\setlength{\multlinegap}{0pt}
\begin{multline}
    V_{k} =
        \kappa_{k} \,\frac{2 \pi}{R} 
        \int_{0}^{r_{1 {\rm max}}} {\rm d}r_1 \, r_1 
        \int_{\min(r_{2 {\rm max}},|R-r_1|)}
        ^{\min(r_{2 {\rm max}},R+r_1)} {\rm d}r_2 \,r_2 \\
    \left\{ \left[ \rho_1(r_1) + \rho_2(r_2) \right] ^{5/3} 
        - \rho_1(r_1)^{5/3} - \rho_1(r_2)^{5/3} \right\}
  \label{eq:Vk_final}
\end{multline}
and
\begin{multline}
    V_{x} =
        - \kappa_{x} \,\frac{2 \pi}{R} 
        \int_{0}^{r_{1 {\rm max}}} {\rm d}r_1 \, r_1 
        \int_{\min(r_{2 {\rm max}},|R-r_1|)}
        ^{\min(r_{2 {\rm max}},R+r_1)} {\rm d}r_2 \,r_2 \\
    \left\{ \left[ \rho_1(r_1) + \rho_2(r_2) \right] 	^{4/3} 
    - \rho_1(r_1)^{4/3} - \rho_1(r_2)^{4/3} \right\} ,
  \label{eq:Vx_final}
\end{multline}
respectively.

Since we approximate the electron densities $\rho_i(r)$ by piecewise polynomials, the integrals in \cref{eq:zeta,eq:chi2} can be carried out exactly, which improves the accuracy of the Coulomb interaction term \cref{eq:Vc}. 
This also speeds up the computation of $V_{c}$, because the integrals can be represented by another piecewise polynomial, which has to be defined only once for each atom species.

The integrals in \cref{eq:Vee_final,eq:Vk_final,eq:Vx_final} were evaluated using the \texttt{quad} function of the SciPy library \cite{SciPy}, using a relative tolerance of $10^{-6}$ for the quantum-mechanical terms and $10^{-7}$ for the Coulomb term $V_{ee}$.
For the quantum-mechanical terms, the tolerance cannot always be reached, but the relative error usually is better than $10^{-4}$.
In addition, the error estimate is not guaranteed to be accurate.
Numerical problems are detected and fixed in the course of choosing the abscissae for the function $V(R)$ as described in the following.

An adaptive algorithm was used with the goal of keeping the interpolation error below a given tolerance.
The interpolation was performed by fitting a cubic spline with not-a-knot boundary conditions to the screening function data $\phi(R) = V(R)/V_{\rm Coul}(R)$.
The interpolation error was specified by a relative and an absolute tolerance for the potential $V$.
Using values of $10^{-3}$ for both tolerances, the interpolation error was intended to be kept below the larger of 0.1\% and 1~meV. The algorithm consists of the following steps:
\begin{enumerate}
    \item Calculate the screening function $\phi$ on an equidistant grid of $R$ values with a spacing of $0.16\mathrm{\,\AA}$, starting from $R = 0$, so that the last value of $R$ does not exceed $R_{\rm max}\!-\!0.04\mathrm{\,\AA}$, where $R_{\rm max} = r_{1 {\rm max}} + r_{2 {\rm max}}$.
    \item Bisect each interval, calculate $\phi$, and evaluate the interpolation error when evaluating the spline through all previously calculated $\phi(R)$ values at the new point. Insert the new data point into the list of $\phi(R)$ values. If the interpolation error is below the tolerance, mark the intervals left and right to the new point as not to be bisected further in subsequent iterations. Iterate until all errors are below the tolerance. At the upper boundary, make sure there is exactly one point between $R_{\rm max}\!-\!0.08\mathrm{\AA}$ and $R_{\rm max}\!-\!0.04\mathrm{\AA}$. 
    \item Starting from the smallest spacing, iterate over all spacings $\Delta R$, and determine points between intervals of equal length $\Delta R$. If there are more than two intervals of equal length in a row, form pairs of adjacent intervals and consider only points in the centers of the pairs. Remove these points, determine the spline of the remaining points and the interpolation error at the removed points. Reinsert the point with the largest interpolation error if above the tolerance, and re-calculate the spline. Repeat until no interpolation error exceeds the tolerance. Continue with the next $\Delta R$ until all $\Delta R$ have been processed in this way.
\end{enumerate}
First including points meeting the tolerance criterion in step 2 and then removing and reinserting points in step 3 leads to more consistent results.
The reason is that spline interpolation is not strictly local, i.e., the approximation in neighboring intervals influences the spline in the interval under consideration.
In the course of the algorithm, initially, there is usually only a poor approximation at small $R$, influencing to some degree the spline at larger $R$.

Step 2 did not converge in four cases (Ca-Kr, In-Dy, Pr-Lu, Dy-Hg).
The reason turned out to be one point in each case that had been calculated inaccurately by the integration function \texttt{quad}.
To handle these cases, the iteration was stopped when the interval length reached 0.00125$\,\rm\AA$, and the inaccurate value of the screening function was replaced by the mean of the neighboring values.
The accumulation of points around the corrected point was then removed automatically in step 3.

Finally, we note that an alternative method could be used to compute the orbital-free density functional potentials, as well: electronic structure calculations on diatomic molecules typically employ the prolate spheroidal coordinate system, as the Coulomb problem can be factorized in this coordinate system \cite{Lehtola2019_IJQC_25944, Lehtola2019_IJQC_25968}.
This coordinate system, defined by $\xi = r_1 + r_2$ and $\eta= r_1 - r_2$, where $r_1$ and $r_2$ are distances to the nuclei 1 and 2 and the angle $\theta$ around the bond axis would also enable facile evaluation of the above integrals.


%

\bibliography{general,susi,gerhard}

\begin{thebibliography}{114}%
\makeatletter
\providecommand \@ifxundefined [1]{%
 \@ifx{#1\undefined}
}%
\providecommand \@ifnum [1]{%
 \ifnum #1\expandafter \@firstoftwo
 \else \expandafter \@secondoftwo
 \fi
}%
\providecommand \@ifx [1]{%
 \ifx #1\expandafter \@firstoftwo
 \else \expandafter \@secondoftwo
 \fi
}%
\providecommand \natexlab [1]{#1}%
\providecommand \enquote  [1]{``#1''}%
\providecommand \bibnamefont  [1]{#1}%
\providecommand \bibfnamefont [1]{#1}%
\providecommand \citenamefont [1]{#1}%
\providecommand \href@noop [0]{\@secondoftwo}%
\providecommand \href [0]{\begingroup \@sanitize@url \@href}%
\providecommand \@href[1]{\@@startlink{#1}\@@href}%
\providecommand \@@href[1]{\endgroup#1\@@endlink}%
\providecommand \@sanitize@url [0]{\catcode `\\12\catcode `\$12\catcode
  `\&12\catcode `\#12\catcode `\^12\catcode `\_12\catcode `\%12\relax}%
\providecommand \@@startlink[1]{}%
\providecommand \@@endlink[0]{}%
\providecommand \url  [0]{\begingroup\@sanitize@url \@url }%
\providecommand \@url [1]{\endgroup\@href {#1}{\urlprefix }}%
\providecommand \urlprefix  [0]{URL }%
\providecommand \Eprint [0]{\href }%
\providecommand \doibase [0]{https://doi.org/}%
\providecommand \selectlanguage [0]{\@gobble}%
\providecommand \bibinfo  [0]{\@secondoftwo}%
\providecommand \bibfield  [0]{\@secondoftwo}%
\providecommand \translation [1]{[#1]}%
\providecommand \BibitemOpen [0]{}%
\providecommand \bibitemStop [0]{}%
\providecommand \bibitemNoStop [0]{.\EOS\space}%
\providecommand \EOS [0]{\spacefactor3000\relax}%
\providecommand \BibitemShut  [1]{\csname bibitem#1\endcsname}%
\let\auto@bib@innerbib\@empty
\bibitem [{\citenamefont {Was}(2012)}]{Was}%
  \BibitemOpen
  \bibfield  {author} {\bibinfo {author} {\bibfnamefont {G.~S.}\ \bibnamefont
  {Was}},\ }\href@noop {} {\emph {\bibinfo {title} {{Fundamentals of Radiation
  Materials Science}}}}\ (\bibinfo  {publisher} {Springer},\ \bibinfo {address}
  {Berlin},\ \bibinfo {year} {2012})\BibitemShut {NoStop}%
\bibitem [{\citenamefont {Bernas}(2010)}]{Bernas}%
  \BibitemOpen
  \bibinfo {editor} {\bibfnamefont {H.}~\bibnamefont {Bernas}},\ ed.,\
  \href@noop {} {\emph {\bibinfo {title} {Materials Science with Ion Beams}}},\
  \bibinfo {series} {Topics in Applied Physics}, Vol.\ \bibinfo {volume} {116}\
  (\bibinfo  {publisher} {Springer},\ \bibinfo {year} {2010})\BibitemShut
  {NoStop}%
\bibitem [{\citenamefont {Williams}(1998)}]{Williams1998_MSEA_8}%
  \BibitemOpen
  \bibfield  {author} {\bibinfo {author} {\bibfnamefont {J.~S.}\ \bibnamefont
  {Williams}},\ }\bibfield  {title} {\bibinfo {title} {Ion implantation of
  semiconductors},\ }\href {https://doi.org/10.1016/s0921-5093(98)00705-9}
  {\bibfield  {journal} {\bibinfo  {journal} {Mater. Sci. Eng. A}\ }\textbf
  {\bibinfo {volume} {253}},\ \bibinfo {pages} {8} (\bibinfo {year}
  {1998})}\BibitemShut {NoStop}%
\bibitem [{\citenamefont {Seeger}(1962)}]{See62}%
  \BibitemOpen
  \bibfield  {author} {\bibinfo {author} {\bibfnamefont {A.}~\bibnamefont
  {Seeger}},\ }\bibinfo {title} {{The Nature of Radiation Damage in Metals}},\
  in\ \href@noop {} {\emph {\bibinfo {booktitle} {Radiation Damage in
  Solids}}},\ Vol.~\bibinfo {volume} {1}\ (\bibinfo  {publisher} {International
  Atomic Energy Agency},\ \bibinfo {address} {Vienna},\ \bibinfo {year}
  {1962})\ pp.\ \bibinfo {pages} {101--127}\BibitemShut {NoStop}%
\bibitem [{\citenamefont {Lindhard}\ \emph {et~al.}(1963)\citenamefont
  {Lindhard}, \citenamefont {Scharff},\ and\ \citenamefont {Schi{\o}tt}}]{LSS}%
  \BibitemOpen
  \bibfield  {author} {\bibinfo {author} {\bibfnamefont {J.}~\bibnamefont
  {Lindhard}}, \bibinfo {author} {\bibfnamefont {M.}~\bibnamefont {Scharff}},\
  and\ \bibinfo {author} {\bibfnamefont {H.~E.}\ \bibnamefont {Schi{\o}tt}},\
  }\bibfield  {title} {\bibinfo {title} {{Range concepts and heavy ion
  ranges}},\ }\href@noop {} {\bibfield  {journal} {\bibinfo  {journal} {Kgl.
  Danske Vid. Selskab, Mat.- Fys. Medd.}\ }\textbf {\bibinfo {volume} {33
  (14)}},\ \bibinfo {pages} {1} (\bibinfo {year} {1963})}\BibitemShut {NoStop}%
\bibitem [{\citenamefont {Nordlund}\ \emph {et~al.}(2018)\citenamefont
  {Nordlund}, \citenamefont {Zinkle}, \citenamefont {Sand}, \citenamefont
  {Granberg}, \citenamefont {Averback}, \citenamefont {Stoller}, \citenamefont
  {Suzudo}, \citenamefont {Malerba}, \citenamefont {Banhart}, \citenamefont
  {Weber}, \citenamefont {Willaime}, \citenamefont {Dudarev},\ and\
  \citenamefont {Simeone}}]{Nor18}%
  \BibitemOpen
  \bibfield  {author} {\bibinfo {author} {\bibfnamefont {K.}~\bibnamefont
  {Nordlund}}, \bibinfo {author} {\bibfnamefont {S.~J.}\ \bibnamefont
  {Zinkle}}, \bibinfo {author} {\bibfnamefont {A.~E.}\ \bibnamefont {Sand}},
  \bibinfo {author} {\bibfnamefont {F.}~\bibnamefont {Granberg}}, \bibinfo
  {author} {\bibfnamefont {R.~S.}\ \bibnamefont {Averback}}, \bibinfo {author}
  {\bibfnamefont {R.}~\bibnamefont {Stoller}}, \bibinfo {author} {\bibfnamefont
  {T.}~\bibnamefont {Suzudo}}, \bibinfo {author} {\bibfnamefont
  {L.}~\bibnamefont {Malerba}}, \bibinfo {author} {\bibfnamefont
  {F.}~\bibnamefont {Banhart}}, \bibinfo {author} {\bibfnamefont {W.~J.}\
  \bibnamefont {Weber}}, \bibinfo {author} {\bibfnamefont {F.}~\bibnamefont
  {Willaime}}, \bibinfo {author} {\bibfnamefont {S.}~\bibnamefont {Dudarev}},\
  and\ \bibinfo {author} {\bibfnamefont {D.}~\bibnamefont {Simeone}},\
  }\bibfield  {title} {\bibinfo {title} {{Primary radiation damage: a review of
  current understanding and models}},\ }\href@noop {} {\bibfield  {journal}
  {\bibinfo  {journal} {J. Nucl. Mater.}\ }\textbf {\bibinfo {volume} {512}},\
  \bibinfo {pages} {450} (\bibinfo {year} {2018})}\BibitemShut {NoStop}%
\bibitem [{\citenamefont {Nordlund}(2019)}]{Nor18b}%
  \BibitemOpen
  \bibfield  {author} {\bibinfo {author} {\bibfnamefont {K.}~\bibnamefont
  {Nordlund}},\ }\bibfield  {title} {\bibinfo {title} {{Historical review of
  computer simulation of radiation effects in materials}},\ }\href
  {https://doi.org/10.1016/j.jnucmat.2019.04.028} {\bibfield  {journal}
  {\bibinfo  {journal} {J. Nucl. Mater.}\ }\textbf {\bibinfo {volume} {520}},\
  \bibinfo {pages} {273} (\bibinfo {year} {2019})},\ \bibinfo {note} {{Invited
  review in Diamond Anniversary issue}}\BibitemShut {NoStop}%
\bibitem [{\citenamefont {Bohr}(1948)}]{Boh48}%
  \BibitemOpen
  \bibfield  {author} {\bibinfo {author} {\bibfnamefont {N.}~\bibnamefont
  {Bohr}},\ }\bibfield  {title} {\bibinfo {title} {The penetration of atomic
  particles through matter},\ }\href@noop {} {\bibfield  {journal} {\bibinfo
  {journal} {Mat. Fys. Medd. Dan. Vid. Selsk.}\ }\textbf {\bibinfo {volume}
  {18}},\ \bibinfo {pages} {1} (\bibinfo {year} {1948})}\BibitemShut {NoStop}%
\bibitem [{\citenamefont {Leach}(2001)}]{Leach}%
  \BibitemOpen
  \bibfield  {author} {\bibinfo {author} {\bibfnamefont {A.~R.}\ \bibnamefont
  {Leach}},\ }\href@noop {} {\emph {\bibinfo {title} {{Molecular modelling:
  principles and applications}}}},\ \bibinfo {edition} {2nd}\ ed.\ (\bibinfo
  {publisher} {Pearson Education},\ \bibinfo {address} {Harlow, England},\
  \bibinfo {year} {2001})\BibitemShut {NoStop}%
\bibitem [{\citenamefont {Robinson}\ and\ \citenamefont {Torrens}(1974)}]{BCA}%
  \BibitemOpen
  \bibfield  {author} {\bibinfo {author} {\bibfnamefont {M.~T.}\ \bibnamefont
  {Robinson}}\ and\ \bibinfo {author} {\bibfnamefont {I.~M.}\ \bibnamefont
  {Torrens}},\ }\bibfield  {title} {\bibinfo {title} {{Computer Simulation of
  atomic-displacement cascades in solids in the binary-collision
  approximation}},\ }\href@noop {} {\bibfield  {journal} {\bibinfo  {journal}
  {Phys. Rev. B}\ }\textbf {\bibinfo {volume} {9}},\ \bibinfo {pages} {5008}
  (\bibinfo {year} {1974})}\BibitemShut {NoStop}%
\bibitem [{\citenamefont {Moli\`ere}(1947)}]{Mol47}%
  \BibitemOpen
  \bibfield  {author} {\bibinfo {author} {\bibfnamefont {G.}~\bibnamefont
  {Moli\`ere}},\ }\bibfield  {title} {\bibinfo {title} {{Theorie der Streuung
  schneller geladener Teilchen .I. Einzelstreuung am abgeschirmten
  Coulomb-Feld}},\ }\href@noop {} {\bibfield  {journal} {\bibinfo  {journal}
  {Z. Naturforsch. A}\ }\textbf {\bibinfo {volume} {2}},\ \bibinfo {pages}
  {133} (\bibinfo {year} {1947})}\BibitemShut {NoStop}%
\bibitem [{\citenamefont {Eckstein}()}]{Eck91a}%
  \BibitemOpen
  \bibfield  {author} {\bibinfo {author} {\bibfnamefont {W.}~\bibnamefont
  {Eckstein}},\ }\href@noop {} {}\bibinfo {note} {In Ref.\ \cite{Eck91}, eq.
  (4.4.3) on p.\ 55}\BibitemShut {NoStop}%
\bibitem [{\citenamefont {Ziegler}\ \emph {et~al.}(1985)\citenamefont
  {Ziegler}, \citenamefont {Biersack},\ and\ \citenamefont {Littmark}}]{ZBL}%
  \BibitemOpen
  \bibfield  {author} {\bibinfo {author} {\bibfnamefont {J.~F.}\ \bibnamefont
  {Ziegler}}, \bibinfo {author} {\bibfnamefont {J.~P.}\ \bibnamefont
  {Biersack}},\ and\ \bibinfo {author} {\bibfnamefont {U.}~\bibnamefont
  {Littmark}},\ }\href@noop {} {\emph {\bibinfo {title} {{The Stopping and
  Range of Ions in Matter}}}}\ (\bibinfo  {publisher} {Pergamon},\ \bibinfo
  {address} {{New York}},\ \bibinfo {year} {1985})\BibitemShut {NoStop}%
\bibitem [{\citenamefont {{A. N. Zinoviev}}(2015)}]{Zin15}%
  \BibitemOpen
  \bibfield  {author} {\bibinfo {author} {\bibnamefont {{A. N. Zinoviev}}},\
  }\bibfield  {title} {\bibinfo {title} {{Electron screening of the Coulomb
  potential at small internuclear distances}},\ }\href@noop {} {\bibfield
  {journal} {\bibinfo  {journal} {{Nuclear Instruments and Methods in Physics
  Research Section B: Beam Interactions with Materials and Atoms}}\ }\textbf
  {\bibinfo {volume} {354}},\ \bibinfo {pages} {308 } (\bibinfo {year}
  {2015})}\BibitemShut {NoStop}%
\bibitem [{\citenamefont {Born}\ and\ \citenamefont
  {Oppenheimer}(1927)}]{Born1927_APB_457}%
  \BibitemOpen
  \bibfield  {author} {\bibinfo {author} {\bibfnamefont {M.}~\bibnamefont
  {Born}}\ and\ \bibinfo {author} {\bibfnamefont {R.}~\bibnamefont
  {Oppenheimer}},\ }\bibfield  {title} {\bibinfo {title} {Zur {Quantentheorie}
  der {Molekeln}},\ }\href {https://doi.org/10.1002/andp.19273892002}
  {\bibfield  {journal} {\bibinfo  {journal} {Ann. Phys. (Berlin)}\ }\textbf
  {\bibinfo {volume} {389}},\ \bibinfo {pages} {457} (\bibinfo {year}
  {1927})}\BibitemShut {NoStop}%
\bibitem [{\citenamefont {Fermi}\ and\ \citenamefont
  {Richtmyer}(1948)}]{Fer48}%
  \BibitemOpen
  \bibfield  {author} {\bibinfo {author} {\bibfnamefont {E.}~\bibnamefont
  {Fermi}}\ and\ \bibinfo {author} {\bibfnamefont {R.~D.}\ \bibnamefont
  {Richtmyer}},\ }\bibfield  {title} {\bibinfo {title} {{Note on census-taking
  in Monte Carlo calculations}},\ }\href@noop {} {\bibfield  {journal}
  {\bibinfo  {journal} {A declassified report by Enrico Fermi. From the Los
  Alamos Archive.}\ } (\bibinfo {year} {1948})}\BibitemShut {NoStop}%
\bibitem [{\citenamefont {Kishinevskii}(1962)}]{Kis62}%
  \BibitemOpen
  \bibfield  {author} {\bibinfo {author} {\bibfnamefont {L.~M.}\ \bibnamefont
  {Kishinevskii}},\ }\bibfield  {title} {\bibinfo {title} {{Cross sections for
  inelastic atomic collisions}},\ }\href@noop {} {\bibfield  {journal}
  {\bibinfo  {journal} {Bull. Acad. Sci. USSR, Phys. Ser.}\ }\textbf {\bibinfo
  {volume} {26}},\ \bibinfo {pages} {1433} (\bibinfo {year}
  {1962})}\BibitemShut {NoStop}%
\bibitem [{\citenamefont {Echenique}\ \emph {et~al.}(1981)\citenamefont
  {Echenique}, \citenamefont {Nieminen},\ and\ \citenamefont
  {Ritchie}}]{Ech81}%
  \BibitemOpen
  \bibfield  {author} {\bibinfo {author} {\bibfnamefont {P.~M.}\ \bibnamefont
  {Echenique}}, \bibinfo {author} {\bibfnamefont {R.~M.}\ \bibnamefont
  {Nieminen}},\ and\ \bibinfo {author} {\bibfnamefont {R.~H.}\ \bibnamefont
  {Ritchie}},\ }\bibfield  {title} {\bibinfo {title} {{Density functional
  calculation of stopping power of an electron gas for slow ions}},\
  }\href@noop {} {\bibfield  {journal} {\bibinfo  {journal} {Solid State
  Comm.}\ }\textbf {\bibinfo {volume} {37}},\ \bibinfo {pages} {779} (\bibinfo
  {year} {1981})}\BibitemShut {NoStop}%
\bibitem [{\citenamefont {Schiwietz}\ and\ \citenamefont
  {Grande}(2011)}]{Sch11}%
  \BibitemOpen
  \bibfield  {author} {\bibinfo {author} {\bibfnamefont {G.}~\bibnamefont
  {Schiwietz}}\ and\ \bibinfo {author} {\bibfnamefont {P.~L.}\ \bibnamefont
  {Grande}},\ }\bibfield  {title} {\bibinfo {title} {Introducing electron
  capture into the unitary-convolution-approximation energy-loss theory at low
  velocities},\ }\href {https://doi.org/10.1103/PhysRevA.84.052703} {\bibfield
  {journal} {\bibinfo  {journal} {Phys. Rev. A}\ }\textbf {\bibinfo {volume}
  {84}},\ \bibinfo {pages} {052703} (\bibinfo {year} {2011})}\BibitemShut
  {NoStop}%
\bibitem [{\citenamefont {Kittel}(1968)}]{Kittel}%
  \BibitemOpen
  \bibfield  {author} {\bibinfo {author} {\bibfnamefont {C.}~\bibnamefont
  {Kittel}},\ }\href@noop {} {\emph {\bibinfo {title} {{Introduction to Solid
  State Physics}}}},\ \bibinfo {edition} {3rd}\ ed.\ (\bibinfo  {publisher}
  {John Wiley \& Sons},\ \bibinfo {address} {New York},\ \bibinfo {year}
  {1968})\BibitemShut {NoStop}%
\bibitem [{\citenamefont {Nordlund}\ \emph {et~al.}(2017)\citenamefont
  {Nordlund}, \citenamefont {Sundholm}, \citenamefont {Pyykk{\"o}},
  \citenamefont {{Martinez Zambrano}},\ and\ \citenamefont
  {Djurabekova}}]{Nor17}%
  \BibitemOpen
  \bibfield  {author} {\bibinfo {author} {\bibfnamefont {K.}~\bibnamefont
  {Nordlund}}, \bibinfo {author} {\bibfnamefont {D.}~\bibnamefont {Sundholm}},
  \bibinfo {author} {\bibfnamefont {P.}~\bibnamefont {Pyykk{\"o}}}, \bibinfo
  {author} {\bibfnamefont {D.}~\bibnamefont {{Martinez Zambrano}}},\ and\
  \bibinfo {author} {\bibfnamefont {F.}~\bibnamefont {Djurabekova}},\
  }\bibfield  {title} {\bibinfo {title} {{Nuclear stopping power of
  antiprotons}},\ }\href@noop {} {\bibfield  {journal} {\bibinfo  {journal}
  {Phys. Rev. A}\ }\textbf {\bibinfo {volume} {96}},\ \bibinfo {pages} {042717}
  (\bibinfo {year} {2017})}\BibitemShut {NoStop}%
\bibitem [{\citenamefont {Wilson}\ \emph {et~al.}(1977)\citenamefont {Wilson},
  \citenamefont {Haggmark},\ and\ \citenamefont {Biersack}}]{Wilson77}%
  \BibitemOpen
  \bibfield  {author} {\bibinfo {author} {\bibfnamefont {W.~D.}\ \bibnamefont
  {Wilson}}, \bibinfo {author} {\bibfnamefont {L.~G.}\ \bibnamefont
  {Haggmark}},\ and\ \bibinfo {author} {\bibfnamefont {J.~P.}\ \bibnamefont
  {Biersack}},\ }\bibfield  {title} {\bibinfo {title} {Calculations of nuclear
  stopping, ranges, and straggling in the low-energy region},\ }\href
  {https://doi.org/10.1103/PhysRevB.15.2458} {\bibfield  {journal} {\bibinfo
  {journal} {Phys. Rev. B}\ }\textbf {\bibinfo {volume} {15}},\ \bibinfo
  {pages} {2458} (\bibinfo {year} {1977})},\ \bibinfo {note} {publisher:
  American Physical Society}\BibitemShut {NoStop}%
\bibitem [{\citenamefont {Jensen}(1932)}]{Jen32}%
  \BibitemOpen
  \bibfield  {author} {\bibinfo {author} {\bibfnamefont {H.}~\bibnamefont
  {Jensen}},\ }\bibfield  {title} {\bibinfo {title} {Die {Ladungsverteilung} in
  {Ionen} und die {Gitterkonstante} des {Rubidiumbromids} nach der
  statistischen {Methode}},\ }\href {https://doi.org/10.1007/BF01342151}
  {\bibfield  {journal} {\bibinfo  {journal} {Z. Physik}\ }\textbf {\bibinfo
  {volume} {77}},\ \bibinfo {pages} {722} (\bibinfo {year} {1932})}\BibitemShut
  {NoStop}%
\bibitem [{\citenamefont {Sommerfeld}(1932)}]{Som32}%
  \BibitemOpen
  \bibfield  {author} {\bibinfo {author} {\bibfnamefont {A.}~\bibnamefont
  {Sommerfeld}},\ }\bibfield  {title} {\bibinfo {title} {Asymptotische
  {Integration} der {Differentialgleichung} des {Thomas}-{Fermischen}
  {Atoms}},\ }\href {https://doi.org/10.1007/BF01342197} {\bibfield  {journal}
  {\bibinfo  {journal} {Z. Physik}\ }\textbf {\bibinfo {volume} {78}},\
  \bibinfo {pages} {283} (\bibinfo {year} {1932})}\BibitemShut {NoStop}%
\bibitem [{\citenamefont {Biersack}\ and\ \citenamefont
  {Ziegler}(1982)}]{Bie82}%
  \BibitemOpen
  \bibfield  {author} {\bibinfo {author} {\bibfnamefont {J.~P.}\ \bibnamefont
  {Biersack}}\ and\ \bibinfo {author} {\bibfnamefont {J.~F.}\ \bibnamefont
  {Ziegler}},\ }\bibfield  {title} {\bibinfo {title} {{Refined universal
  potentials in atomic collisions}},\ }\href@noop {} {\bibfield  {journal}
  {\bibinfo  {journal} {Nucl. Instr. Meth.}\ }\textbf {\bibinfo {volume}
  {194}},\ \bibinfo {pages} {93} (\bibinfo {year} {1982})}\BibitemShut
  {NoStop}%
\bibitem [{\citenamefont {Nakagawa}\ and\ \citenamefont
  {Yamamura}(1988)}]{Nak88}%
  \BibitemOpen
  \bibfield  {author} {\bibinfo {author} {\bibfnamefont {S.~T.}\ \bibnamefont
  {Nakagawa}}\ and\ \bibinfo {author} {\bibfnamefont {Y.}~\bibnamefont
  {Yamamura}},\ }\bibfield  {title} {\bibinfo {title} {Interatomic potential in
  solids and its application to range calculations},\ }\href@noop {} {\bibfield
   {journal} {\bibinfo  {journal} {Radiat. Eff.}\ }\textbf {\bibinfo {volume}
  {105}},\ \bibinfo {pages} {239} (\bibinfo {year} {1988})}\BibitemShut
  {NoStop}%
\bibitem [{\citenamefont {Zinoviev}(2011)}]{Zin11}%
  \BibitemOpen
  \bibfield  {author} {\bibinfo {author} {\bibfnamefont {A.~N.}\ \bibnamefont
  {Zinoviev}},\ }\bibfield  {title} {\bibinfo {title} {Interaction potentials
  for modeling of ion-surface scattering},\ }\href
  {https://doi.org/10.1016/j.nimb.2010.11.074} {\bibfield  {journal} {\bibinfo
  {journal} {Nucl. Instrum. Meth. Phys. Res. B}\ }\textbf {\bibinfo {volume}
  {269}},\ \bibinfo {pages} {829} (\bibinfo {year} {2011})}\BibitemShut
  {NoStop}%
\bibitem [{\citenamefont {Thomas}(1927)}]{Thomas1927_MPCPS_542}%
  \BibitemOpen
  \bibfield  {author} {\bibinfo {author} {\bibfnamefont {L.~H.}\ \bibnamefont
  {Thomas}},\ }\bibfield  {title} {\bibinfo {title} {{The calculation of atomic
  fields}},\ }\href {https://doi.org/10.1017/S0305004100011683} {\bibfield
  {journal} {\bibinfo  {journal} {Math. Proc. Cambridge Philos. Soc.}\ }\textbf
  {\bibinfo {volume} {23}},\ \bibinfo {pages} {542} (\bibinfo {year}
  {1927})}\BibitemShut {NoStop}%
\bibitem [{\citenamefont {Fermi}(1927)}]{Fermi1927_RdNdL_602}%
  \BibitemOpen
  \bibfield  {author} {\bibinfo {author} {\bibfnamefont {E.}~\bibnamefont
  {Fermi}},\ }\bibfield  {title} {\bibinfo {title} {Un metodo statistico per la
  determinazione di alcune propriet\'{a} dell'atomo.},\ }\href@noop {}
  {\bibfield  {journal} {\bibinfo  {journal} {Rendiconti dell'Accademia
  Nazionale dei Lincei}\ }\textbf {\bibinfo {volume} {6}},\ \bibinfo {pages}
  {602} (\bibinfo {year} {1927})}\BibitemShut {NoStop}%
\bibitem [{\citenamefont {Bloch}(1929)}]{Bloch1929_ZfuP_545}%
  \BibitemOpen
  \bibfield  {author} {\bibinfo {author} {\bibfnamefont {F.}~\bibnamefont
  {Bloch}},\ }\bibfield  {title} {\bibinfo {title} {{Bemerkung zur
  Elektronentheorie des Ferromagnetismus und der elektrischen
  Leitf{\"{a}}higkeit}},\ }\href {https://doi.org/10.1007/BF01340281}
  {\bibfield  {journal} {\bibinfo  {journal} {Z. Phys.}\ }\textbf {\bibinfo
  {volume} {57}},\ \bibinfo {pages} {545} (\bibinfo {year} {1929})}\BibitemShut
  {NoStop}%
\bibitem [{\citenamefont {Dirac}(1930)}]{Dirac1930_MPCPS_376}%
  \BibitemOpen
  \bibfield  {author} {\bibinfo {author} {\bibfnamefont {P.~A.~M.}\
  \bibnamefont {Dirac}},\ }\bibfield  {title} {\bibinfo {title} {Note on
  exchange phenomena in the {Thomas} atom},\ }\href
  {https://doi.org/10.1017/S0305004100016108} {\bibfield  {journal} {\bibinfo
  {journal} {Math. Proc. Cambridge Philos. Soc.}\ }\textbf {\bibinfo {volume}
  {26}},\ \bibinfo {pages} {376} (\bibinfo {year} {1930})}\BibitemShut
  {NoStop}%
\bibitem [{\citenamefont {Mi}\ \emph {et~al.}(2023)\citenamefont {Mi},
  \citenamefont {Luo}, \citenamefont {Trickey},\ and\ \citenamefont
  {Pavanello}}]{Mi2023_CR_12039}%
  \BibitemOpen
  \bibfield  {author} {\bibinfo {author} {\bibfnamefont {W.}~\bibnamefont
  {Mi}}, \bibinfo {author} {\bibfnamefont {K.}~\bibnamefont {Luo}}, \bibinfo
  {author} {\bibfnamefont {S.~B.}\ \bibnamefont {Trickey}},\ and\ \bibinfo
  {author} {\bibfnamefont {M.}~\bibnamefont {Pavanello}},\ }\bibfield  {title}
  {\bibinfo {title} {Orbital-free density functional theory: An attractive
  electronic structure method for large-scale first-principles simulations},\
  }\href {https://doi.org/10.1021/acs.chemrev.2c00758} {\bibfield  {journal}
  {\bibinfo  {journal} {Chem. Rev.}\ }\textbf {\bibinfo {volume} {123}},\
  \bibinfo {pages} {12039} (\bibinfo {year} {2023})}\BibitemShut {NoStop}%
\bibitem [{\citenamefont {Alml{\"{o}}f}\ \emph {et~al.}(1982)\citenamefont
  {Alml{\"{o}}f}, \citenamefont {Faegri},\ and\ \citenamefont
  {Korsell}}]{Almloef1982_JCC_385}%
  \BibitemOpen
  \bibfield  {author} {\bibinfo {author} {\bibfnamefont {J.}~\bibnamefont
  {Alml{\"{o}}f}}, \bibinfo {author} {\bibfnamefont {K.}~\bibnamefont
  {Faegri}},\ and\ \bibinfo {author} {\bibfnamefont {K.}~\bibnamefont
  {Korsell}},\ }\bibfield  {title} {\bibinfo {title} {Principles for a direct
  {SCF} approach to {LCAO}-{MO} ab-initio calculations},\ }\href
  {https://doi.org/10.1002/jcc.540030314} {\bibfield  {journal} {\bibinfo
  {journal} {J. Comput. Chem.}\ }\textbf {\bibinfo {volume} {3}},\ \bibinfo
  {pages} {385} (\bibinfo {year} {1982})}\BibitemShut {NoStop}%
\bibitem [{\citenamefont {{Van Lenthe}}\ \emph {et~al.}(2006)\citenamefont
  {{Van Lenthe}}, \citenamefont {Zwaans}, \citenamefont {{Van Dam}},\ and\
  \citenamefont {Guest}}]{VanLenthe2006_JCC_32}%
  \BibitemOpen
  \bibfield  {author} {\bibinfo {author} {\bibfnamefont {J.~H.}\ \bibnamefont
  {{Van Lenthe}}}, \bibinfo {author} {\bibfnamefont {R.}~\bibnamefont
  {Zwaans}}, \bibinfo {author} {\bibfnamefont {H.~J.~J.}\ \bibnamefont {{Van
  Dam}}},\ and\ \bibinfo {author} {\bibfnamefont {M.~F.}\ \bibnamefont
  {Guest}},\ }\bibfield  {title} {\bibinfo {title} {{Starting SCF calculations
  by superposition of atomic densities.}},\ }\href
  {https://doi.org/10.1002/jcc.20393} {\bibfield  {journal} {\bibinfo
  {journal} {J. Comput. Chem.}\ }\textbf {\bibinfo {volume} {27}},\ \bibinfo
  {pages} {926} (\bibinfo {year} {2006})}\BibitemShut {NoStop}%
\bibitem [{\citenamefont
  {Lehtola}(2019{\natexlab{a}})}]{Lehtola2019_JCTC_1593}%
  \BibitemOpen
  \bibfield  {author} {\bibinfo {author} {\bibfnamefont {S.}~\bibnamefont
  {Lehtola}},\ }\bibfield  {title} {\bibinfo {title} {Assessment of initial
  guesses for self-consistent field calculations. superposition of atomic
  potentials: Simple yet efficient},\ }\href
  {https://doi.org/10.1021/acs.jctc.8b01089} {\bibfield  {journal} {\bibinfo
  {journal} {J. Chem. Theory Comput.}\ }\textbf {\bibinfo {volume} {15}},\
  \bibinfo {pages} {1593} (\bibinfo {year} {2019}{\natexlab{a}})},\ \Eprint
  {https://arxiv.org/abs/1810.11659} {arXiv:1810.11659} \BibitemShut {NoStop}%
\bibitem [{\citenamefont {Sabelli}\ \emph {et~al.}(1979)\citenamefont
  {Sabelli}, \citenamefont {Benedek},\ and\ \citenamefont
  {Gilbert}}]{Sabelli1979_PRA_677}%
  \BibitemOpen
  \bibfield  {author} {\bibinfo {author} {\bibfnamefont {N.~H.}\ \bibnamefont
  {Sabelli}}, \bibinfo {author} {\bibfnamefont {R.}~\bibnamefont {Benedek}},\
  and\ \bibinfo {author} {\bibfnamefont {T.~L.}\ \bibnamefont {Gilbert}},\
  }\bibfield  {title} {\bibinfo {title} {Ground-state potential curves for
  \ce{Al2} and \ce{Al2^{6+}} in the repulsive region},\ }\href
  {https://doi.org/10.1103/PhysRevA.20.677} {\bibfield  {journal} {\bibinfo
  {journal} {Phys. Rev. A}\ }\textbf {\bibinfo {volume} {20}},\ \bibinfo
  {pages} {677} (\bibinfo {year} {1979})}\BibitemShut {NoStop}%
\bibitem [{\citenamefont {Lehtola}(2020)}]{Lehtola2020_PRA_32504}%
  \BibitemOpen
  \bibfield  {author} {\bibinfo {author} {\bibfnamefont {S.}~\bibnamefont
  {Lehtola}},\ }\bibfield  {title} {\bibinfo {title} {Accurate reproduction of
  strongly repulsive interatomic potentials},\ }\href
  {https://doi.org/10.1103/PhysRevA.101.032504} {\bibfield  {journal} {\bibinfo
   {journal} {Phys. Rev. A}\ }\textbf {\bibinfo {volume} {101}},\ \bibinfo
  {pages} {032504} (\bibinfo {year} {2020})},\ \Eprint
  {https://arxiv.org/abs/1912.12624} {arXiv:1912.12624} \BibitemShut {NoStop}%
\bibitem [{\citenamefont {Pathak}\ and\ \citenamefont
  {Thakkar}(1987)}]{Pathak1987_JCP_2186}%
  \BibitemOpen
  \bibfield  {author} {\bibinfo {author} {\bibfnamefont {R.~K.}\ \bibnamefont
  {Pathak}}\ and\ \bibinfo {author} {\bibfnamefont {A.~J.}\ \bibnamefont
  {Thakkar}},\ }\bibfield  {title} {\bibinfo {title} {{Very short‐range
  interatomic potentials}},\ }\href {https://doi.org/10.1063/1.453144}
  {\bibfield  {journal} {\bibinfo  {journal} {J. Chem. Phys.}\ }\textbf
  {\bibinfo {volume} {87}},\ \bibinfo {pages} {2186} (\bibinfo {year}
  {1987})}\BibitemShut {NoStop}%
\bibitem [{\citenamefont {Sabelli}\ \emph {et~al.}(1978)\citenamefont
  {Sabelli}, \citenamefont {Kantor}, \citenamefont {Benedek},\ and\
  \citenamefont {Gilbert}}]{Sabelli1978_JCP_2767}%
  \BibitemOpen
  \bibfield  {author} {\bibinfo {author} {\bibfnamefont {N.~H.}\ \bibnamefont
  {Sabelli}}, \bibinfo {author} {\bibfnamefont {M.}~\bibnamefont {Kantor}},
  \bibinfo {author} {\bibfnamefont {R.}~\bibnamefont {Benedek}},\ and\ \bibinfo
  {author} {\bibfnamefont {T.~L.}\ \bibnamefont {Gilbert}},\ }\bibfield
  {title} {\bibinfo {title} {{SCF} potential curves for {AlH} and \ce{AlH+} in
  the attractive and repulsive regions},\ }\href
  {https://doi.org/10.1063/1.436068} {\bibfield  {journal} {\bibinfo  {journal}
  {J. Chem. Phys.}\ }\textbf {\bibinfo {volume} {68}},\ \bibinfo {pages} {2767}
  (\bibinfo {year} {1978})}\BibitemShut {NoStop}%
\bibitem [{\citenamefont {Mulliken}(1932)}]{Mulliken1932_RMP_1}%
  \BibitemOpen
  \bibfield  {author} {\bibinfo {author} {\bibfnamefont {R.~S.}\ \bibnamefont
  {Mulliken}},\ }\bibfield  {title} {\bibinfo {title} {The interpretation of
  band spectra part {III}. electron quantum numbers and states of molecules and
  their atoms},\ }\href {https://doi.org/10.1103/RevModPhys.4.1} {\bibfield
  {journal} {\bibinfo  {journal} {Rev. Mod. Phys.}\ }\textbf {\bibinfo {volume}
  {4}},\ \bibinfo {pages} {1} (\bibinfo {year} {1932})}\BibitemShut {NoStop}%
\bibitem [{\citenamefont {Hirao}(2011)}]{Hirao2011_WCMS_337}%
  \BibitemOpen
  \bibfield  {author} {\bibinfo {author} {\bibfnamefont {H.}~\bibnamefont
  {Hirao}},\ }\bibfield  {title} {\bibinfo {title} {Correlation diagram
  approach as a tool for interpreting chemistry: an introductory overview},\
  }\href {https://doi.org/10.1002/wcms.20} {\bibfield  {journal} {\bibinfo
  {journal} {WIREs Comput. Mol. Sci.}\ }\textbf {\bibinfo {volume} {1}},\
  \bibinfo {pages} {337} (\bibinfo {year} {2011})}\BibitemShut {NoStop}%
\bibitem [{\citenamefont {Hait}\ \emph {et~al.}(2019)\citenamefont {Hait},
  \citenamefont {Tubman}, \citenamefont {Levine}, \citenamefont {Whaley},\ and\
  \citenamefont {Head-Gordon}}]{Hait2019_JCTC_5370}%
  \BibitemOpen
  \bibfield  {author} {\bibinfo {author} {\bibfnamefont {D.}~\bibnamefont
  {Hait}}, \bibinfo {author} {\bibfnamefont {N.~M.}\ \bibnamefont {Tubman}},
  \bibinfo {author} {\bibfnamefont {D.~S.}\ \bibnamefont {Levine}}, \bibinfo
  {author} {\bibfnamefont {K.~B.}\ \bibnamefont {Whaley}},\ and\ \bibinfo
  {author} {\bibfnamefont {M.}~\bibnamefont {Head-Gordon}},\ }\bibfield
  {title} {\bibinfo {title} {{What Levels of Coupled Cluster Theory Are
  Appropriate for Transition Metal Systems? A Study Using Near-Exact Quantum
  Chemical Values for 3d Transition Metal Binary Compounds}},\ }\href
  {https://doi.org/10.1021/acs.jctc.9b00674} {\bibfield  {journal} {\bibinfo
  {journal} {J. Chem. Theory Comput.}\ }\textbf {\bibinfo {volume} {15}},\
  \bibinfo {pages} {5370} (\bibinfo {year} {2019})}\BibitemShut {NoStop}%
\bibitem [{\citenamefont {Nordlund}\ \emph {et~al.}(1997)\citenamefont
  {Nordlund}, \citenamefont {Runeberg},\ and\ \citenamefont
  {Sundholm}}]{Nor96c}%
  \BibitemOpen
  \bibfield  {author} {\bibinfo {author} {\bibfnamefont {K.}~\bibnamefont
  {Nordlund}}, \bibinfo {author} {\bibfnamefont {N.}~\bibnamefont {Runeberg}},\
  and\ \bibinfo {author} {\bibfnamefont {D.}~\bibnamefont {Sundholm}},\
  }\bibfield  {title} {\bibinfo {title} {{Repulsive interatomic potentials
  calculated using Hartree-Fock and density-functional theory methods}},\
  }\href@noop {} {\bibfield  {journal} {\bibinfo  {journal} {Nucl. Instr. Meth.
  Phys. Res. B}\ }\textbf {\bibinfo {volume} {132}},\ \bibinfo {pages} {45}
  (\bibinfo {year} {1997})}\BibitemShut {NoStop}%
\bibitem [{\citenamefont {Froese~Fischer}\ \emph {et~al.}(2019)\citenamefont
  {Froese~Fischer}, \citenamefont {Gaigalas}, \citenamefont {Jönsson},\ and\
  \citenamefont {Bieroń}}]{Fro19}%
  \BibitemOpen
  \bibfield  {author} {\bibinfo {author} {\bibfnamefont {C.}~\bibnamefont
  {Froese~Fischer}}, \bibinfo {author} {\bibfnamefont {G.}~\bibnamefont
  {Gaigalas}}, \bibinfo {author} {\bibfnamefont {P.}~\bibnamefont {Jönsson}},\
  and\ \bibinfo {author} {\bibfnamefont {J.}~\bibnamefont {Bieroń}},\
  }\bibfield  {title} {\bibinfo {title} {{GRASP2018}—{A} {Fortran} 95 version
  of the {General} {Relativistic} {Atomic} {Structure} {Package}},\ }\href
  {https://doi.org/10.1016/j.cpc.2018.10.032} {\bibfield  {journal} {\bibinfo
  {journal} {Computer Physics Communications}\ }\textbf {\bibinfo {volume}
  {237}},\ \bibinfo {pages} {184} (\bibinfo {year} {2019})}\BibitemShut
  {NoStop}%
\bibitem [{\citenamefont {Schiffmann}\ \emph {et~al.}(2022)\citenamefont
  {Schiffmann}, \citenamefont {Li}, \citenamefont {Ekman}, \citenamefont
  {Gaigalas}, \citenamefont {Godefroid}, \citenamefont {Jönsson},\ and\
  \citenamefont {Bieroń}}]{Sch22}%
  \BibitemOpen
  \bibfield  {author} {\bibinfo {author} {\bibfnamefont {S.}~\bibnamefont
  {Schiffmann}}, \bibinfo {author} {\bibfnamefont {J.~G.}\ \bibnamefont {Li}},
  \bibinfo {author} {\bibfnamefont {J.}~\bibnamefont {Ekman}}, \bibinfo
  {author} {\bibfnamefont {G.}~\bibnamefont {Gaigalas}}, \bibinfo {author}
  {\bibfnamefont {M.}~\bibnamefont {Godefroid}}, \bibinfo {author}
  {\bibfnamefont {P.}~\bibnamefont {Jönsson}},\ and\ \bibinfo {author}
  {\bibfnamefont {J.}~\bibnamefont {Bieroń}},\ }\bibfield  {title} {\bibinfo
  {title} {Relativistic radial electron density functions and natural orbitals
  from {GRASP2018}},\ }\href {https://doi.org/10.1016/j.cpc.2022.108403}
  {\bibfield  {journal} {\bibinfo  {journal} {Computer Physics Communications}\
  }\textbf {\bibinfo {volume} {278}},\ \bibinfo {pages} {108403} (\bibinfo
  {year} {2022})}\BibitemShut {NoStop}%
\bibitem [{\citenamefont {Boys}\ and\ \citenamefont
  {Bernardi}(1970)}]{Boys1970_MP_553}%
  \BibitemOpen
  \bibfield  {author} {\bibinfo {author} {\bibfnamefont {S.~F.}\ \bibnamefont
  {Boys}}\ and\ \bibinfo {author} {\bibfnamefont {F.}~\bibnamefont
  {Bernardi}},\ }\bibfield  {title} {\bibinfo {title} {{The calculation of
  small molecular interactions by the differences of separate total energies.
  Some procedures with reduced errors}},\ }\href
  {https://doi.org/10.1080/00268977000101561} {\bibfield  {journal} {\bibinfo
  {journal} {Mol. Phys.}\ }\textbf {\bibinfo {volume} {19}},\ \bibinfo {pages}
  {553} (\bibinfo {year} {1970})}\BibitemShut {NoStop}%
\bibitem [{\citenamefont {Jensen}(2001)}]{Jensen2001_JCP_9113}%
  \BibitemOpen
  \bibfield  {author} {\bibinfo {author} {\bibfnamefont {F.}~\bibnamefont
  {Jensen}},\ }\bibfield  {title} {\bibinfo {title} {Polarization consistent
  basis sets: Principles},\ }\href {https://doi.org/10.1063/1.1413524}
  {\bibfield  {journal} {\bibinfo  {journal} {J. Chem. Phys.}\ }\textbf
  {\bibinfo {volume} {115}},\ \bibinfo {pages} {9113} (\bibinfo {year}
  {2001})}\BibitemShut {NoStop}%
\bibitem [{\citenamefont
  {Lehtola}(2019{\natexlab{b}})}]{Lehtola2019_JCP_241102}%
  \BibitemOpen
  \bibfield  {author} {\bibinfo {author} {\bibfnamefont {S.}~\bibnamefont
  {Lehtola}},\ }\bibfield  {title} {\bibinfo {title} {Curing basis set
  overcompleteness with pivoted {Cholesky} decompositions},\ }\href
  {https://doi.org/10.1063/1.5139948} {\bibfield  {journal} {\bibinfo
  {journal} {J. Chem. Phys.}\ }\textbf {\bibinfo {volume} {151}},\ \bibinfo
  {pages} {241102} (\bibinfo {year} {2019}{\natexlab{b}})},\ \Eprint
  {https://arxiv.org/abs/1911.10372} {arXiv:1911.10372} \BibitemShut {NoStop}%
\bibitem [{\citenamefont
  {Lehtola}(2019{\natexlab{c}})}]{Lehtola2019_IJQC_25968}%
  \BibitemOpen
  \bibfield  {author} {\bibinfo {author} {\bibfnamefont {S.}~\bibnamefont
  {Lehtola}},\ }\bibfield  {title} {\bibinfo {title} {A review on
  non-relativistic, fully numerical electronic structure calculations on atoms
  and diatomic molecules},\ }\href {https://doi.org/10.1002/qua.25968}
  {\bibfield  {journal} {\bibinfo  {journal} {Int. J. Quantum Chem.}\ }\textbf
  {\bibinfo {volume} {119}},\ \bibinfo {pages} {e25968} (\bibinfo {year}
  {2019}{\natexlab{c}})},\ \Eprint {https://arxiv.org/abs/1902.01431}
  {arXiv:1902.01431} \BibitemShut {NoStop}%
\bibitem [{\citenamefont
  {Lehtola}(2019{\natexlab{d}})}]{Lehtola2019_IJQC_25944}%
  \BibitemOpen
  \bibfield  {author} {\bibinfo {author} {\bibfnamefont {S.}~\bibnamefont
  {Lehtola}},\ }\bibfield  {title} {\bibinfo {title} {Fully numerical
  {Hartree}--{Fock} and density functional calculations. {II}. {Diatomic}
  molecules},\ }\href {https://doi.org/10.1002/qua.25944} {\bibfield  {journal}
  {\bibinfo  {journal} {Int. J. Quantum Chem.}\ }\textbf {\bibinfo {volume}
  {119}},\ \bibinfo {pages} {e25944} (\bibinfo {year} {2019}{\natexlab{d}})},\
  \Eprint {https://arxiv.org/abs/1810.11653} {arXiv:1810.11653} \BibitemShut
  {NoStop}%
\bibitem [{\citenamefont {Sun}\ \emph {et~al.}(2020)\citenamefont {Sun},
  \citenamefont {Zhang}, \citenamefont {Banerjee}, \citenamefont {Bao},
  \citenamefont {Barbry}, \citenamefont {Blunt}, \citenamefont {Bogdanov},
  \citenamefont {Booth}, \citenamefont {Chen}, \citenamefont {Cui},
  \citenamefont {Eriksen}, \citenamefont {Gao}, \citenamefont {Guo},
  \citenamefont {Hermann}, \citenamefont {Hermes}, \citenamefont {Koh},
  \citenamefont {Koval}, \citenamefont {Lehtola}, \citenamefont {Li},
  \citenamefont {Liu}, \citenamefont {Mardirossian}, \citenamefont {McClain},
  \citenamefont {Motta}, \citenamefont {Mussard}, \citenamefont {Pham},
  \citenamefont {Pulkin}, \citenamefont {Purwanto}, \citenamefont {Robinson},
  \citenamefont {Ronca}, \citenamefont {Sayfutyarova}, \citenamefont
  {Scheurer}, \citenamefont {Schurkus}, \citenamefont {Smith}, \citenamefont
  {Sun}, \citenamefont {Sun}, \citenamefont {Upadhyay}, \citenamefont {Wagner},
  \citenamefont {Wang}, \citenamefont {White}, \citenamefont {Whitfield},
  \citenamefont {Williamson}, \citenamefont {Wouters}, \citenamefont {Yang},
  \citenamefont {Yu}, \citenamefont {Zhu}, \citenamefont {Berkelbach},
  \citenamefont {Sharma}, \citenamefont {Sokolov},\ and\ \citenamefont
  {Chan}}]{Sun2020_JCP_24109}%
  \BibitemOpen
  \bibfield  {author} {\bibinfo {author} {\bibfnamefont {Q.}~\bibnamefont
  {Sun}}, \bibinfo {author} {\bibfnamefont {X.}~\bibnamefont {Zhang}}, \bibinfo
  {author} {\bibfnamefont {S.}~\bibnamefont {Banerjee}}, \bibinfo {author}
  {\bibfnamefont {P.}~\bibnamefont {Bao}}, \bibinfo {author} {\bibfnamefont
  {M.}~\bibnamefont {Barbry}}, \bibinfo {author} {\bibfnamefont {N.~S.}\
  \bibnamefont {Blunt}}, \bibinfo {author} {\bibfnamefont {N.~A.}\ \bibnamefont
  {Bogdanov}}, \bibinfo {author} {\bibfnamefont {G.~H.}\ \bibnamefont {Booth}},
  \bibinfo {author} {\bibfnamefont {J.}~\bibnamefont {Chen}}, \bibinfo {author}
  {\bibfnamefont {Z.-H.}\ \bibnamefont {Cui}}, \bibinfo {author} {\bibfnamefont
  {J.~J.}\ \bibnamefont {Eriksen}}, \bibinfo {author} {\bibfnamefont
  {Y.}~\bibnamefont {Gao}}, \bibinfo {author} {\bibfnamefont {S.}~\bibnamefont
  {Guo}}, \bibinfo {author} {\bibfnamefont {J.}~\bibnamefont {Hermann}},
  \bibinfo {author} {\bibfnamefont {M.~R.}\ \bibnamefont {Hermes}}, \bibinfo
  {author} {\bibfnamefont {K.}~\bibnamefont {Koh}}, \bibinfo {author}
  {\bibfnamefont {P.}~\bibnamefont {Koval}}, \bibinfo {author} {\bibfnamefont
  {S.}~\bibnamefont {Lehtola}}, \bibinfo {author} {\bibfnamefont
  {Z.}~\bibnamefont {Li}}, \bibinfo {author} {\bibfnamefont {J.}~\bibnamefont
  {Liu}}, \bibinfo {author} {\bibfnamefont {N.}~\bibnamefont {Mardirossian}},
  \bibinfo {author} {\bibfnamefont {J.~D.}\ \bibnamefont {McClain}}, \bibinfo
  {author} {\bibfnamefont {M.}~\bibnamefont {Motta}}, \bibinfo {author}
  {\bibfnamefont {B.}~\bibnamefont {Mussard}}, \bibinfo {author} {\bibfnamefont
  {H.~Q.}\ \bibnamefont {Pham}}, \bibinfo {author} {\bibfnamefont
  {A.}~\bibnamefont {Pulkin}}, \bibinfo {author} {\bibfnamefont
  {W.}~\bibnamefont {Purwanto}}, \bibinfo {author} {\bibfnamefont {P.~J.}\
  \bibnamefont {Robinson}}, \bibinfo {author} {\bibfnamefont {E.}~\bibnamefont
  {Ronca}}, \bibinfo {author} {\bibfnamefont {E.~R.}\ \bibnamefont
  {Sayfutyarova}}, \bibinfo {author} {\bibfnamefont {M.}~\bibnamefont
  {Scheurer}}, \bibinfo {author} {\bibfnamefont {H.~F.}\ \bibnamefont
  {Schurkus}}, \bibinfo {author} {\bibfnamefont {J.~E.~T.}\ \bibnamefont
  {Smith}}, \bibinfo {author} {\bibfnamefont {C.}~\bibnamefont {Sun}}, \bibinfo
  {author} {\bibfnamefont {S.-N.}\ \bibnamefont {Sun}}, \bibinfo {author}
  {\bibfnamefont {S.}~\bibnamefont {Upadhyay}}, \bibinfo {author}
  {\bibfnamefont {L.~K.}\ \bibnamefont {Wagner}}, \bibinfo {author}
  {\bibfnamefont {X.}~\bibnamefont {Wang}}, \bibinfo {author} {\bibfnamefont
  {A.}~\bibnamefont {White}}, \bibinfo {author} {\bibfnamefont {J.~D.}\
  \bibnamefont {Whitfield}}, \bibinfo {author} {\bibfnamefont {M.~J.}\
  \bibnamefont {Williamson}}, \bibinfo {author} {\bibfnamefont
  {S.}~\bibnamefont {Wouters}}, \bibinfo {author} {\bibfnamefont
  {J.}~\bibnamefont {Yang}}, \bibinfo {author} {\bibfnamefont {J.~M.}\
  \bibnamefont {Yu}}, \bibinfo {author} {\bibfnamefont {T.}~\bibnamefont
  {Zhu}}, \bibinfo {author} {\bibfnamefont {T.~C.}\ \bibnamefont {Berkelbach}},
  \bibinfo {author} {\bibfnamefont {S.}~\bibnamefont {Sharma}}, \bibinfo
  {author} {\bibfnamefont {A.~Y.}\ \bibnamefont {Sokolov}},\ and\ \bibinfo
  {author} {\bibfnamefont {G.~K.-L.}\ \bibnamefont {Chan}},\ }\bibfield
  {title} {\bibinfo {title} {Recent developments in the {\sc pyscf} program
  package},\ }\href {https://doi.org/10.1063/5.0006074} {\bibfield  {journal}
  {\bibinfo  {journal} {J. Chem. Phys.}\ }\textbf {\bibinfo {volume} {153}},\
  \bibinfo {pages} {024109} (\bibinfo {year} {2020})},\ \Eprint
  {https://arxiv.org/abs/2002.12531} {arXiv:2002.12531} \BibitemShut {NoStop}%
\bibitem [{\citenamefont {Seeger}\ and\ \citenamefont
  {Pople}(1977)}]{Seeger1977_JCP_3045}%
  \BibitemOpen
  \bibfield  {author} {\bibinfo {author} {\bibfnamefont {R.}~\bibnamefont
  {Seeger}}\ and\ \bibinfo {author} {\bibfnamefont {J.~A.}\ \bibnamefont
  {Pople}},\ }\bibfield  {title} {\bibinfo {title} {{Self-consistent molecular
  orbital methods. XVIII. Constraints and stability in Hartree--Fock theory}},\
  }\href {https://doi.org/10.1063/1.434318} {\bibfield  {journal} {\bibinfo
  {journal} {J. Chem. Phys.}\ }\textbf {\bibinfo {volume} {66}},\ \bibinfo
  {pages} {3045} (\bibinfo {year} {1977})}\BibitemShut {NoStop}%
\bibitem [{\citenamefont {M{\o}ller}\ and\ \citenamefont
  {Plesset}(1934)}]{Moller1934_PR_618}%
  \BibitemOpen
  \bibfield  {author} {\bibinfo {author} {\bibfnamefont {C.}~\bibnamefont
  {M{\o}ller}}\ and\ \bibinfo {author} {\bibfnamefont {M.~S.~M.}\ \bibnamefont
  {Plesset}},\ }\bibfield  {title} {\bibinfo {title} {Note on an approximation
  treatment for many-electron systems},\ }\href
  {https://doi.org/10.1103/PhysRev.46.618} {\bibfield  {journal} {\bibinfo
  {journal} {Phys. Rev.}\ }\textbf {\bibinfo {volume} {46}},\ \bibinfo {pages}
  {618} (\bibinfo {year} {1934})}\BibitemShut {NoStop}%
\bibitem [{\citenamefont {Tarus}\ \emph {et~al.}(1999)\citenamefont {Tarus},
  \citenamefont {Nordlund}, \citenamefont {Sillanp{\"a}{\"a}},\ and\
  \citenamefont {Keinonen}}]{Tar98b}%
  \BibitemOpen
  \bibfield  {author} {\bibinfo {author} {\bibfnamefont {J.}~\bibnamefont
  {Tarus}}, \bibinfo {author} {\bibfnamefont {K.}~\bibnamefont {Nordlund}},
  \bibinfo {author} {\bibfnamefont {J.}~\bibnamefont {Sillanp{\"a}{\"a}}},\
  and\ \bibinfo {author} {\bibfnamefont {J.}~\bibnamefont {Keinonen}},\
  }\bibfield  {title} {\bibinfo {title} {{Heat spike and ballistic
  contributions to mixing in Si}},\ }\href@noop {} {\bibfield  {journal}
  {\bibinfo  {journal} {Nucl. Instr. Meth. Phys. Res. B}\ }\textbf {\bibinfo
  {volume} {153}},\ \bibinfo {pages} {378} (\bibinfo {year}
  {1999})}\BibitemShut {NoStop}%
\bibitem [{\citenamefont {Peltola}\ \emph {et~al.}(2003)\citenamefont
  {Peltola}, \citenamefont {Nordlund},\ and\ \citenamefont
  {Keinonen}}]{Pel03d}%
  \BibitemOpen
  \bibfield  {author} {\bibinfo {author} {\bibfnamefont {J.}~\bibnamefont
  {Peltola}}, \bibinfo {author} {\bibfnamefont {K.}~\bibnamefont {Nordlund}},\
  and\ \bibinfo {author} {\bibfnamefont {J.}~\bibnamefont {Keinonen}},\
  }\bibfield  {title} {\bibinfo {title} {{Molecular dynamics study on stopping
  powers of channeled He and Li ions in Si}},\ }\href@noop {} {\bibfield
  {journal} {\bibinfo  {journal} {Nucl. Instr. Meth. Phys. Res. B}\ }\textbf
  {\bibinfo {volume} {217}},\ \bibinfo {pages} {25} (\bibinfo {year}
  {2003})}\BibitemShut {NoStop}%
\bibitem [{\citenamefont {Henriksson}\ \emph {et~al.}(2005)\citenamefont
  {Henriksson}, \citenamefont {Nordlund}, \citenamefont {Krasheninnikov},\ and\
  \citenamefont {Keinonen}}]{Hen05a}%
  \BibitemOpen
  \bibfield  {author} {\bibinfo {author} {\bibfnamefont {K.~O.~E.}\
  \bibnamefont {Henriksson}}, \bibinfo {author} {\bibfnamefont
  {K.}~\bibnamefont {Nordlund}}, \bibinfo {author} {\bibfnamefont
  {A.}~\bibnamefont {Krasheninnikov}},\ and\ \bibinfo {author} {\bibfnamefont
  {J.}~\bibnamefont {Keinonen}},\ }\bibfield  {title} {\bibinfo {title}
  {{Differences in hydrogen and helium cluster formation}},\ }\href@noop {}
  {\bibfield  {journal} {\bibinfo  {journal} {Appl. Phys. Lett.}\ }\textbf
  {\bibinfo {volume} {87}},\ \bibinfo {pages} {163113} (\bibinfo {year}
  {2005})}\BibitemShut {NoStop}%
\bibitem [{\citenamefont {Kotakoski}\ \emph {et~al.}(2005)\citenamefont
  {Kotakoski}, \citenamefont {Krasheninnikov},\ and\ \citenamefont
  {Nordlund}}]{Kot05b}%
  \BibitemOpen
  \bibfield  {author} {\bibinfo {author} {\bibfnamefont {J.}~\bibnamefont
  {Kotakoski}}, \bibinfo {author} {\bibfnamefont {A.~V.}\ \bibnamefont
  {Krasheninnikov}},\ and\ \bibinfo {author} {\bibfnamefont {K.}~\bibnamefont
  {Nordlund}},\ }\bibfield  {title} {\bibinfo {title} {{A molecular dynamics
  study of the clustering of ion irradiation induced potassium in multiwalled
  carbon nanotubes}},\ }\href@noop {} {\bibfield  {journal} {\bibinfo
  {journal} {Nucl. Instr. Meth. Phys. Res. B}\ }\textbf {\bibinfo {volume}
  {240}},\ \bibinfo {pages} {810} (\bibinfo {year} {2005})}\BibitemShut
  {NoStop}%
\bibitem [{\citenamefont {Juslin}\ and\ \citenamefont
  {Nordlund}(2008)}]{Jus07}%
  \BibitemOpen
  \bibfield  {author} {\bibinfo {author} {\bibfnamefont {N.}~\bibnamefont
  {Juslin}}\ and\ \bibinfo {author} {\bibfnamefont {K.}~\bibnamefont
  {Nordlund}},\ }\bibfield  {title} {\bibinfo {title} {{Pair potential for
  Fe-He}},\ }\href@noop {} {\bibfield  {journal} {\bibinfo  {journal} {J. Nucl.
  Mater.}\ }\textbf {\bibinfo {volume} {382}},\ \bibinfo {pages} {143}
  (\bibinfo {year} {2008})}\BibitemShut {NoStop}%
\bibitem [{\citenamefont {Nordlund}\ \emph {et~al.}(2016)\citenamefont
  {Nordlund}, \citenamefont {Djurabekova},\ and\ \citenamefont
  {Hobler}}]{Nor16}%
  \BibitemOpen
  \bibfield  {author} {\bibinfo {author} {\bibfnamefont {K.}~\bibnamefont
  {Nordlund}}, \bibinfo {author} {\bibfnamefont {F.}~\bibnamefont
  {Djurabekova}},\ and\ \bibinfo {author} {\bibfnamefont {G.}~\bibnamefont
  {Hobler}},\ }\bibfield  {title} {\bibinfo {title} {{Large fraction of crystal
  directions leads to ion channeling}},\ }\href@noop {} {\bibfield  {journal}
  {\bibinfo  {journal} {Phys. Rev. B}\ }\textbf {\bibinfo {volume} {94}},\
  \bibinfo {pages} {214109} (\bibinfo {year} {2016})}\BibitemShut {NoStop}%
\bibitem [{\citenamefont {Delley}(1990)}]{Del90}%
  \BibitemOpen
  \bibfield  {author} {\bibinfo {author} {\bibfnamefont {J.}~\bibnamefont
  {Delley}},\ }\href@noop {} {\bibfield  {journal} {\bibinfo  {journal} {J.
  Chem. Phys.}\ }\textbf {\bibinfo {volume} {92}},\ \bibinfo {pages} {508}
  (\bibinfo {year} {1990})}\BibitemShut {NoStop}%
\bibitem [{DMo()}]{DMol}%
  \BibitemOpen
  \href@noop {} {}\bibinfo {note} {DMol is a trademark of AccelRys.
  Inc.}\BibitemShut {Stop}%
\bibitem [{DMo(1993)}]{DMolManual}%
  \BibitemOpen
  \href@noop {} {\emph {\bibinfo {title} {{DMol User Guide}}}},\ \bibinfo
  {organization} {Biosym Technologies Inc.},\ \bibinfo {address} {San Diego,
  California},\ \bibinfo {edition} {v. 2.3.5}\ ed. (\bibinfo {year}
  {1993})\BibitemShut {NoStop}%
\bibitem [{\citenamefont {Hohenberg}\ and\ \citenamefont
  {Kohn}(1964)}]{Hohenberg1964_PR_864}%
  \BibitemOpen
  \bibfield  {author} {\bibinfo {author} {\bibfnamefont {P.}~\bibnamefont
  {Hohenberg}}\ and\ \bibinfo {author} {\bibfnamefont {W.}~\bibnamefont
  {Kohn}},\ }\bibfield  {title} {\bibinfo {title} {Inhomogeneous electron
  gas},\ }\href {https://doi.org/10.1103/PhysRev.136.B864} {\bibfield
  {journal} {\bibinfo  {journal} {Phys. Rev.}\ }\textbf {\bibinfo {volume}
  {136}},\ \bibinfo {pages} {B864} (\bibinfo {year} {1964})}\BibitemShut
  {NoStop}%
\bibitem [{\citenamefont {Kohn}\ and\ \citenamefont
  {Sham}(1965)}]{Kohn1965_PR_1133}%
  \BibitemOpen
  \bibfield  {author} {\bibinfo {author} {\bibfnamefont {W.}~\bibnamefont
  {Kohn}}\ and\ \bibinfo {author} {\bibfnamefont {L.~J.}\ \bibnamefont
  {Sham}},\ }\bibfield  {title} {\bibinfo {title} {Self-consistent equations
  including exchange and correlation effects},\ }\href
  {https://doi.org/10.1103/PhysRev.140.A1133} {\bibfield  {journal} {\bibinfo
  {journal} {Phys. Rev.}\ }\textbf {\bibinfo {volume} {140}},\ \bibinfo {pages}
  {A1133} (\bibinfo {year} {1965})}\BibitemShut {NoStop}%
\bibitem [{\citenamefont {Keinonen}\ \emph {et~al.}(1994)\citenamefont
  {Keinonen}, \citenamefont {Kuronen}, \citenamefont {Nordlund}, \citenamefont
  {Nieminen},\ and\ \citenamefont {Seitsonen}}]{Kei94}%
  \BibitemOpen
  \bibfield  {author} {\bibinfo {author} {\bibfnamefont {J.}~\bibnamefont
  {Keinonen}}, \bibinfo {author} {\bibfnamefont {A.}~\bibnamefont {Kuronen}},
  \bibinfo {author} {\bibfnamefont {K.}~\bibnamefont {Nordlund}}, \bibinfo
  {author} {\bibfnamefont {R.~M.}\ \bibnamefont {Nieminen}},\ and\ \bibinfo
  {author} {\bibfnamefont {A.~P.}\ \bibnamefont {Seitsonen}},\ }\bibfield
  {title} {\bibinfo {title} {{First-Principles Simulation of Collision Cascades
  in Si to Test Pair-Potential for Si-Si Interaction at 10 eV -- 5 keV}},\
  }\href@noop {} {\bibfield  {journal} {\bibinfo  {journal} {Nucl. Instr. Meth.
  Phys. Res. B}\ }\textbf {\bibinfo {volume} {88}},\ \bibinfo {pages} {382}
  (\bibinfo {year} {1994})}\BibitemShut {NoStop}%
\bibitem [{\citenamefont {Vosko}\ \emph {et~al.}(1980)\citenamefont {Vosko},
  \citenamefont {Wilk},\ and\ \citenamefont {Nusair}}]{Vosko1980_CJP_1200}%
  \BibitemOpen
  \bibfield  {author} {\bibinfo {author} {\bibfnamefont {S.~H.}\ \bibnamefont
  {Vosko}}, \bibinfo {author} {\bibfnamefont {L.}~\bibnamefont {Wilk}},\ and\
  \bibinfo {author} {\bibfnamefont {M.}~\bibnamefont {Nusair}},\ }\bibfield
  {title} {\bibinfo {title} {{Accurate spin-dependent electron liquid
  correlation energies for local spin density calculations: a critical
  analysis}},\ }\href {https://doi.org/10.1139/p80-159} {\bibfield  {journal}
  {\bibinfo  {journal} {Can. J. Phys.}\ }\textbf {\bibinfo {volume} {58}},\
  \bibinfo {pages} {1200} (\bibinfo {year} {1980})}\BibitemShut {NoStop}%
\bibitem [{\citenamefont {Park}\ \emph {et~al.}(1991)\citenamefont {Park},
  \citenamefont {Klein}, \citenamefont {Tasch},\ and\ \citenamefont
  {Ziegler}}]{Park91}%
  \BibitemOpen
  \bibfield  {author} {\bibinfo {author} {\bibfnamefont {C.}~\bibnamefont
  {Park}}, \bibinfo {author} {\bibfnamefont {K.~M.}\ \bibnamefont {Klein}},
  \bibinfo {author} {\bibfnamefont {A.~F.}\ \bibnamefont {Tasch}},\ and\
  \bibinfo {author} {\bibfnamefont {J.~F.}\ \bibnamefont {Ziegler}},\
  }\bibfield  {title} {\bibinfo {title} {Critical angles for channeling of
  boron ions implanted into single-crystal silicon},\ }\href@noop {} {\bibfield
   {journal} {\bibinfo  {journal} {J. Electrochem. Soc.}\ }\textbf {\bibinfo
  {volume} {138}},\ \bibinfo {pages} {2107} (\bibinfo {year}
  {1991})}\BibitemShut {NoStop}%
\bibitem [{\citenamefont {Hobler}(1993)}]{Hob93}%
  \BibitemOpen
  \bibfield  {author} {\bibinfo {author} {\bibfnamefont {G.}~\bibnamefont
  {Hobler}},\ }\bibfield  {title} {\bibinfo {title} {The role of lattice
  vibrations and interatomic potentials in the simulation of ion
  implantation},\ }in\ \href@noop {} {\emph {\bibinfo {booktitle} {{NASECODE}
  {IX}}}},\ \bibinfo {editor} {edited by\ \bibinfo {editor} {\bibfnamefont
  {J.~J.~H.}\ \bibnamefont {Miller}}}\ (\bibinfo  {publisher} {Front Range
  Press, Boulder, Colorado},\ \bibinfo {year} {1993})\ pp.\ \bibinfo {pages}
  {55--56}\BibitemShut {NoStop}%
\bibitem [{\citenamefont {Hobler}(1995)}]{Hob95a}%
  \BibitemOpen
  \bibfield  {author} {\bibinfo {author} {\bibfnamefont {G.}~\bibnamefont
  {Hobler}},\ }\bibfield  {title} {\bibinfo {title} {Monte {Carlo} simulation
  of two-dimensional implanted dopant distributions at mask edges},\ }\href
  {https://doi.org/10.1016/0168-583X(94)00476-5} {\bibfield  {journal}
  {\bibinfo  {journal} {Nucl. Instrum. Meth. Phys. Res. B}\ }\textbf {\bibinfo
  {volume} {96}},\ \bibinfo {pages} {155} (\bibinfo {year} {1995})}\BibitemShut
  {NoStop}%
\bibitem [{\citenamefont {Wedepohl}(1967)}]{Wed67}%
  \BibitemOpen
  \bibfield  {author} {\bibinfo {author} {\bibfnamefont {P.~T.}\ \bibnamefont
  {Wedepohl}},\ }\bibfield  {title} {\bibinfo {title} {Influence of electron
  distribution on atomic interaction potentials},\ }\bibfield  {journal}
  {\bibinfo  {journal} {Proc. Phys. Soc.}\ }\textbf {\bibinfo {volume} {92}},\
  \href {https://doi.org/10.1088/0370-1328/92/1/313}
  {10.1088/0370-1328/92/1/313} (\bibinfo {year} {1967})\BibitemShut {NoStop}%
\bibitem [{\citenamefont {Teller}(1962)}]{Teller1962_RMP_627}%
  \BibitemOpen
  \bibfield  {author} {\bibinfo {author} {\bibfnamefont {E.}~\bibnamefont
  {Teller}},\ }\bibfield  {title} {\bibinfo {title} {On the stability of
  molecules in the {Thomas}--{Fermi} theory},\ }\href
  {https://doi.org/10.1103/RevModPhys.34.627} {\bibfield  {journal} {\bibinfo
  {journal} {Rev. Mod. Phys.}\ }\textbf {\bibinfo {volume} {34}},\ \bibinfo
  {pages} {627} (\bibinfo {year} {1962})}\BibitemShut {NoStop}%
\bibitem [{\citenamefont {Press}\ \emph {et~al.}(1995)\citenamefont {Press},
  \citenamefont {Teukolsky}, \citenamefont {Vetterling},\ and\ \citenamefont
  {Flannery}}]{NumericalRecipes}%
  \BibitemOpen
  \bibfield  {author} {\bibinfo {author} {\bibfnamefont {W.~H.}\ \bibnamefont
  {Press}}, \bibinfo {author} {\bibfnamefont {S.~A.}\ \bibnamefont
  {Teukolsky}}, \bibinfo {author} {\bibfnamefont {W.~T.}\ \bibnamefont
  {Vetterling}},\ and\ \bibinfo {author} {\bibfnamefont {B.~P.}\ \bibnamefont
  {Flannery}},\ }\href@noop {} {\emph {\bibinfo {title} {{Numerical Recipes in
  C; The Art of Scientific Computing}}}},\ \bibinfo {edition} {2nd}\ ed.\
  (\bibinfo  {publisher} {Cambridge University Press},\ \bibinfo {address} {New
  York},\ \bibinfo {year} {1995})\BibitemShut {NoStop}%
\bibitem [{\citenamefont {{L. Eriksson and J. A. Davies and P.
  Jespersgaard}}(1967)}]{Eri67}%
  \BibitemOpen
  \bibfield  {author} {\bibinfo {author} {\bibnamefont {{L. Eriksson and J. A.
  Davies and P. Jespersgaard}}},\ }\bibfield  {title} {\bibinfo {title} {{Range
  Measurements in Oriented Tungsten Single Crystals (0.1-1.0 MeV). I.
  Electronic and Nuclear Stopping Powers}},\ }\href@noop {} {\bibfield
  {journal} {\bibinfo  {journal} {Phys. Rev.}\ }\textbf {\bibinfo {volume}
  {161}},\ \bibinfo {pages} {219} (\bibinfo {year} {1967})}\BibitemShut
  {NoStop}%
\bibitem [{\citenamefont {Torri}\ \emph {et~al.}(1994)\citenamefont {Torri},
  \citenamefont {Keinonen},\ and\ \citenamefont {Nordlund}}]{Tor94}%
  \BibitemOpen
  \bibfield  {author} {\bibinfo {author} {\bibfnamefont {P.}~\bibnamefont
  {Torri}}, \bibinfo {author} {\bibfnamefont {J.}~\bibnamefont {Keinonen}},\
  and\ \bibinfo {author} {\bibfnamefont {K.}~\bibnamefont {Nordlund}},\
  }\bibfield  {title} {\bibinfo {title} {{A low-level detection system for
  hydrogen analysis with the reaction $\rm ^1H(^{15}N,\alpha\gamma)^{12}C$}},\
  }\href@noop {} {\bibfield  {journal} {\bibinfo  {journal} {Nucl. Instr. Meth.
  Phys. Res. B}\ }\textbf {\bibinfo {volume} {84}},\ \bibinfo {pages} {105}
  (\bibinfo {year} {1994})}\BibitemShut {NoStop}%
\bibitem [{\citenamefont {Haussalo}\ \emph {et~al.}(1996)\citenamefont
  {Haussalo}, \citenamefont {Nordlund},\ and\ \citenamefont
  {Keinonen}}]{Hau95}%
  \BibitemOpen
  \bibfield  {author} {\bibinfo {author} {\bibfnamefont {P.}~\bibnamefont
  {Haussalo}}, \bibinfo {author} {\bibfnamefont {K.}~\bibnamefont {Nordlund}},\
  and\ \bibinfo {author} {\bibfnamefont {J.}~\bibnamefont {Keinonen}},\
  }\bibfield  {title} {\bibinfo {title} {{Stopping of 5 -- 100 keV helium in
  tantalum, niobium, tungsten, and AISI 316L steel}},\ }\href@noop {}
  {\bibfield  {journal} {\bibinfo  {journal} {Nucl. Instr. Meth. Phys. Res. B}\
  }\textbf {\bibinfo {volume} {111}},\ \bibinfo {pages} {1} (\bibinfo {year}
  {1996})}\BibitemShut {NoStop}%
\bibitem [{\citenamefont {Cai}\ \emph {et~al.}(1996)\citenamefont {Cai},
  \citenamefont {Gr{\o}nbech-Jensen}, \citenamefont {Snell},\ and\
  \citenamefont {Beardmore}}]{Cai96}%
  \BibitemOpen
  \bibfield  {author} {\bibinfo {author} {\bibfnamefont {D.}~\bibnamefont
  {Cai}}, \bibinfo {author} {\bibfnamefont {N.}~\bibnamefont
  {Gr{\o}nbech-Jensen}}, \bibinfo {author} {\bibfnamefont {C.~M.}\ \bibnamefont
  {Snell}},\ and\ \bibinfo {author} {\bibfnamefont {K.~M.}\ \bibnamefont
  {Beardmore}},\ }\bibfield  {title} {\bibinfo {title} {{Phenomenological
  electronic stopping-power model for molecular dynamics and Monte Carlo
  simulation of ion implantation into silicon}},\ }\href@noop {} {\bibfield
  {journal} {\bibinfo  {journal} {Phys. Rev. B}\ }\textbf {\bibinfo {volume}
  {54}},\ \bibinfo {pages} {17147} (\bibinfo {year} {1996})}\BibitemShut
  {NoStop}%
\bibitem [{\citenamefont {Larson}\ \emph {et~al.}(1998)\citenamefont {Larson},
  \citenamefont {Petford-Long}, \citenamefont {Cerezo}, \citenamefont {Smith},
  \citenamefont {Foord},\ and\ \citenamefont {Anthony}}]{Lar98}%
  \BibitemOpen
  \bibfield  {author} {\bibinfo {author} {\bibfnamefont {D.~J.}\ \bibnamefont
  {Larson}}, \bibinfo {author} {\bibfnamefont {A.~K.}\ \bibnamefont
  {Petford-Long}}, \bibinfo {author} {\bibfnamefont {A.}~\bibnamefont
  {Cerezo}}, \bibinfo {author} {\bibfnamefont {G.~D.~W.}\ \bibnamefont
  {Smith}}, \bibinfo {author} {\bibfnamefont {D.~T.}\ \bibnamefont {Foord}},\
  and\ \bibinfo {author} {\bibfnamefont {T.~C.}\ \bibnamefont {Anthony}},\
  }\bibfield  {title} {\bibinfo {title} {{Three-dimensional atom probe
  field-ion microscopy observation of Cu/Co multilayer film structures}},\
  }\href@noop {} {\bibfield  {journal} {\bibinfo  {journal} {Appl. Phys.
  Lett.}\ }\textbf {\bibinfo {volume} {73}},\ \bibinfo {pages} {1125} (\bibinfo
  {year} {1998})}\BibitemShut {NoStop}%
\bibitem [{\citenamefont {Ziegler}(1995)}]{TRIM95}%
  \BibitemOpen
  \bibfield  {author} {\bibinfo {author} {\bibfnamefont {J.~F.}\ \bibnamefont
  {Ziegler}},\ }\href@noop {} {} (\bibinfo {year} {1995}),\ \bibinfo {note}
  {{{TRIM-95 computer code, private communication}}}\BibitemShut {NoStop}%
\bibitem [{\citenamefont {Nordlund}\ and\ \citenamefont
  {Averback}(2005)}]{Nor04b}%
  \BibitemOpen
  \bibfield  {author} {\bibinfo {author} {\bibfnamefont {K.}~\bibnamefont
  {Nordlund}}\ and\ \bibinfo {author} {\bibfnamefont {R.~S.}\ \bibnamefont
  {Averback}},\ }\bibinfo {title} {{Handbook of Materials Modeling}}\ (\bibinfo
   {publisher} {Kluwer Academic},\ \bibinfo {address} {Dordrecht, The
  Netherlands},\ \bibinfo {year} {2005})\ Chap.\ \bibinfo {chapter} {6.2. Point
  defects in metals}\BibitemShut {NoStop}%
\bibitem [{\citenamefont {Hobler}\ and\ \citenamefont {Betz}(2001)}]{Hob00}%
  \BibitemOpen
  \bibfield  {author} {\bibinfo {author} {\bibfnamefont {G.}~\bibnamefont
  {Hobler}}\ and\ \bibinfo {author} {\bibfnamefont {G.}~\bibnamefont {Betz}},\
  }\bibfield  {title} {\bibinfo {title} {{On the useful range of application of
  molecular dynamics simulations in the recoil interaction approximation}},\
  }\href@noop {} {\bibfield  {journal} {\bibinfo  {journal} {Nucl. Instr. Meth.
  Phys. Res. B}\ }\textbf {\bibinfo {volume} {180}},\ \bibinfo {pages} {203}
  (\bibinfo {year} {2001})}\BibitemShut {NoStop}%
\bibitem [{\citenamefont {Nordlund}(1995)}]{Nor94b}%
  \BibitemOpen
  \bibfield  {author} {\bibinfo {author} {\bibfnamefont {K.}~\bibnamefont
  {Nordlund}},\ }\bibfield  {title} {\bibinfo {title} {{Molecular dynamics
  simulation of ion ranges in the 1 -- 100 keV energy range}},\ }\href@noop {}
  {\bibfield  {journal} {\bibinfo  {journal} {Comput. Mater. Sci.}\ }\textbf
  {\bibinfo {volume} {3}},\ \bibinfo {pages} {448} (\bibinfo {year}
  {1995})}\BibitemShut {NoStop}%
\bibitem [{MDR()}]{MDRANGE}%
  \BibitemOpen
  \href@noop {} {}\bibinfo {note} {Open source code available at {\sc
  https://gitlab.com/acclab/MDRANGE}. A presentation of the {\sc MDRANGE}
  computer code is available at {{\footnotesize
  http://beam.mv.helsinki.fi/$\sim$knordlun/mdh/mdh\_program.html}
  \relax}}\BibitemShut {NoStop}%
\bibitem [{\citenamefont {Sillanp{\"a}{\"a}}\ \emph {et~al.}(2001)\citenamefont
  {Sillanp{\"a}{\"a}}, \citenamefont {Peltola}, \citenamefont {Nordlund},
  \citenamefont {Keinonen},\ and\ \citenamefont {Puska}}]{Sil00}%
  \BibitemOpen
  \bibfield  {author} {\bibinfo {author} {\bibfnamefont {J.}~\bibnamefont
  {Sillanp{\"a}{\"a}}}, \bibinfo {author} {\bibfnamefont {J.}~\bibnamefont
  {Peltola}}, \bibinfo {author} {\bibfnamefont {K.}~\bibnamefont {Nordlund}},
  \bibinfo {author} {\bibfnamefont {J.}~\bibnamefont {Keinonen}},\ and\
  \bibinfo {author} {\bibfnamefont {M.~J.}\ \bibnamefont {Puska}},\ }\bibfield
  {title} {\bibinfo {title} {{Electronic stopping calculated using explicit
  phase shift factors}},\ }\href@noop {} {\bibfield  {journal} {\bibinfo
  {journal} {Phys. Rev. B}\ }\textbf {\bibinfo {volume} {63}},\ \bibinfo
  {pages} {134113} (\bibinfo {year} {2001})}\BibitemShut {NoStop}%
\bibitem [{\citenamefont {Hogg}\ \emph {et~al.}(2016)\citenamefont {Hogg},
  \citenamefont {Pipeleers}, \citenamefont {Vantomme},\ and\ \citenamefont
  {Swartz}}]{Hog02}%
  \BibitemOpen
  \bibfield  {author} {\bibinfo {author} {\bibfnamefont {S.~M.}\ \bibnamefont
  {Hogg}}, \bibinfo {author} {\bibfnamefont {B.}~\bibnamefont {Pipeleers}},
  \bibinfo {author} {\bibfnamefont {A.}~\bibnamefont {Vantomme}},\ and\
  \bibinfo {author} {\bibfnamefont {M.}~\bibnamefont {Swartz}},\ }\bibfield
  {title} {\bibinfo {title} {{Channeling of low energy heavy ions: Er in Si
  <111>}},\ }\href@noop {} {\bibfield  {journal} {\bibinfo  {journal} {Appl.
  Phys. Lett.}\ }\textbf {\bibinfo {volume} {80}},\ \bibinfo {pages} {4363}
  (\bibinfo {year} {2016})}\BibitemShut {NoStop}%
\bibitem [{\citenamefont {Peltola}\ \emph {et~al.}(2006)\citenamefont
  {Peltola}, \citenamefont {Nordlund},\ and\ \citenamefont
  {Keinonen}}]{Pel03e}%
  \BibitemOpen
  \bibfield  {author} {\bibinfo {author} {\bibfnamefont {J.}~\bibnamefont
  {Peltola}}, \bibinfo {author} {\bibfnamefont {K.}~\bibnamefont {Nordlund}},\
  and\ \bibinfo {author} {\bibfnamefont {J.}~\bibnamefont {Keinonen}},\
  }\bibfield  {title} {\bibinfo {title} {{Electronic stopping power calculation
  method for molecular dynamics simulations using local Firsov and free
  electron-gas models}},\ }\href@noop {} {\bibfield  {journal} {\bibinfo
  {journal} {Rad. Eff. \& Def. in Sol.}\ }\textbf {\bibinfo {volume} {161}},\
  \bibinfo {pages} {511} (\bibinfo {year} {2006})}\BibitemShut {NoStop}%
\bibitem [{\citenamefont {{D. S. Gemmell}}(1974)}]{Gem74}%
  \BibitemOpen
  \bibfield  {author} {\bibinfo {author} {\bibnamefont {{D. S. Gemmell}}},\
  }\bibfield  {title} {\bibinfo {title} {{Channeling and related effects in the
  motion of charged particles through crystals}},\ }\href@noop {} {\bibfield
  {journal} {\bibinfo  {journal} {Rev. Mod. Phys.}\ }\textbf {\bibinfo {volume}
  {46}},\ \bibinfo {pages} {129} (\bibinfo {year} {1974})}\BibitemShut
  {NoStop}%
\bibitem [{\citenamefont {Nordlund}\ and\ \citenamefont
  {Hobler}(2018)}]{Nor17d}%
  \BibitemOpen
  \bibfield  {author} {\bibinfo {author} {\bibfnamefont {K.}~\bibnamefont
  {Nordlund}}\ and\ \bibinfo {author} {\bibfnamefont {G.}~\bibnamefont
  {Hobler}},\ }\bibfield  {title} {\bibinfo {title} {{Dependence of ion
  channeling on relative atomic number in compounds}},\ }\href@noop {}
  {\bibfield  {journal} {\bibinfo  {journal} {Nucl. Instr. Meth. Phys. Res. B}\
  }\textbf {\bibinfo {volume} {435}},\ \bibinfo {pages} {61} (\bibinfo {year}
  {2018})}\BibitemShut {NoStop}%
\bibitem [{\citenamefont {Stegun}(1964)}]{Stegun}%
  \BibitemOpen
  \bibfield  {author} {\bibinfo {author} {\bibfnamefont {I.~A.}\ \bibnamefont
  {Stegun}},\ }\href@noop {} {\emph {\bibinfo {title} {Handbook of Mathematical
  Functions}}}\ (\bibinfo  {publisher} {National Bureau of Standards},\
  \bibinfo {address} {Washington, DC, USA},\ \bibinfo {year} {1964})\ \bibinfo
  {note} {p. 998}\BibitemShut {NoStop}%
\bibitem [{\citenamefont {Ashcroft}\ and\ \citenamefont
  {Mermin}(1976)}]{Ashcroft-Mermin}%
  \BibitemOpen
  \bibfield  {author} {\bibinfo {author} {\bibfnamefont {N.~W.}\ \bibnamefont
  {Ashcroft}}\ and\ \bibinfo {author} {\bibfnamefont {N.~D.}\ \bibnamefont
  {Mermin}},\ }\href@noop {} {\emph {\bibinfo {title} {{Solid State
  Physics}}}}\ (\bibinfo  {publisher} {Saunders College},\ \bibinfo {address}
  {Philadelphia},\ \bibinfo {year} {1976})\BibitemShut {NoStop}%
\bibitem [{\citenamefont {Buschorn}\ \emph {et~al.}(1997)\citenamefont
  {Buschorn}, \citenamefont {Diedrich}, \citenamefont {Kufner}, \citenamefont
  {M.Rzepka}, \citenamefont {Hoffmann-Stascheck},\ and\ \citenamefont
  {Richter}}]{Bus97}%
  \BibitemOpen
  \bibfield  {author} {\bibinfo {author} {\bibfnamefont {G.}~\bibnamefont
  {Buschorn}}, \bibinfo {author} {\bibfnamefont {E.}~\bibnamefont {Diedrich}},
  \bibinfo {author} {\bibfnamefont {W.}~\bibnamefont {Kufner}}, \bibinfo
  {author} {\bibfnamefont {H.~G.}\ \bibnamefont {M.Rzepka}}, \bibinfo {author}
  {\bibfnamefont {P.}~\bibnamefont {Hoffmann-Stascheck}},\ and\ \bibinfo
  {author} {\bibfnamefont {A.}~\bibnamefont {Richter}},\ }\bibfield  {title}
  {\bibinfo {title} {{}},\ }\href@noop {} {\bibfield  {journal} {\bibinfo
  {journal} {Phys. Rev. B}\ }\textbf {\bibinfo {volume} {55}},\ \bibinfo
  {pages} {6196} (\bibinfo {year} {1997})}\BibitemShut {NoStop}%
\bibitem [{\citenamefont {Bevington}(1992)}]{Bevington}%
  \BibitemOpen
  \bibfield  {author} {\bibinfo {author} {\bibfnamefont {P.~R.}\ \bibnamefont
  {Bevington}},\ }\href@noop {} {\emph {\bibinfo {title} {{Data reduction and
  error analysis for the physical sciences}}}}\ (\bibinfo  {publisher}
  {McGraw-Hill},\ \bibinfo {address} {New York},\ \bibinfo {year}
  {1992})\BibitemShut {NoStop}%
\bibitem [{Sup()}]{Supplemental}%
  \BibitemOpen
  \href@noop {} {}\bibinfo {note} {See Supplemental Material at [URL will be
  inserted by publisher] for the ZBL pair-specific, DMol and MP2 data sets and
  the NLH potential coefficients, which are available in the compressed archive
  package file {\tt nlh\_potentials\_opendata.tar.gz }}\BibitemShut {NoStop}%
\bibitem [{\citenamefont {Abell}(1985)}]{Abe85}%
  \BibitemOpen
  \bibfield  {author} {\bibinfo {author} {\bibfnamefont {G.~C.}\ \bibnamefont
  {Abell}},\ }\bibfield  {title} {\bibinfo {title} {Empirical chemical
  pseudopotential theory of molecular and metallic bonding},\ }\href@noop {}
  {\bibfield  {journal} {\bibinfo  {journal} {Phys. Rev. B}\ }\textbf {\bibinfo
  {volume} {31}},\ \bibinfo {pages} {6184} (\bibinfo {year}
  {1985})}\BibitemShut {NoStop}%
\bibitem [{\citenamefont {Albe}\ \emph {et~al.}(2002)\citenamefont {Albe},
  \citenamefont {Nordlund},\ and\ \citenamefont {Averback}}]{Alb01b}%
  \BibitemOpen
  \bibfield  {author} {\bibinfo {author} {\bibfnamefont {K.}~\bibnamefont
  {Albe}}, \bibinfo {author} {\bibfnamefont {K.}~\bibnamefont {Nordlund}},\
  and\ \bibinfo {author} {\bibfnamefont {R.~S.}\ \bibnamefont {Averback}},\
  }\bibfield  {title} {\bibinfo {title} {{Modeling metal-semiconductor
  interaction: Analytical bond-order potential for platinum-carbon}},\
  }\href@noop {} {\bibfield  {journal} {\bibinfo  {journal} {Phys. Rev. B}\
  }\textbf {\bibinfo {volume} {65}},\ \bibinfo {pages} {195124} (\bibinfo
  {year} {2002})}\BibitemShut {NoStop}%
\bibitem [{\citenamefont {Tersoff}(1988)}]{Ter88}%
  \BibitemOpen
  \bibfield  {author} {\bibinfo {author} {\bibfnamefont {J.}~\bibnamefont
  {Tersoff}},\ }\bibfield  {title} {\bibinfo {title} {{New Empirical approach
  for the structure and energy of covalent systems}},\ }\href@noop {}
  {\bibfield  {journal} {\bibinfo  {journal} {Phys. Rev. B}\ }\textbf {\bibinfo
  {volume} {37}},\ \bibinfo {pages} {6991} (\bibinfo {year}
  {1988})}\BibitemShut {NoStop}%
\bibitem [{\citenamefont {Brenner}(2000)}]{Bre00}%
  \BibitemOpen
  \bibfield  {author} {\bibinfo {author} {\bibfnamefont {D.~W.}\ \bibnamefont
  {Brenner}},\ }\bibfield  {title} {\bibinfo {title} {{The art and science of
  an analytical potential}},\ }\href@noop {} {\bibfield  {journal} {\bibinfo
  {journal} {physica status solidi (b)}\ }\textbf {\bibinfo {volume} {217}},\
  \bibinfo {pages} {23} (\bibinfo {year} {2000})}\BibitemShut {NoStop}%
\bibitem [{\citenamefont {{J. Behler}}(2016)}]{Beh16}%
  \BibitemOpen
  \bibfield  {author} {\bibinfo {author} {\bibnamefont {{J. Behler}}},\
  }\bibfield  {title} {\bibinfo {title} {{Perspective: Machine learning
  potentials for atomistic simulations}},\ }\href@noop {} {\bibfield  {journal}
  {\bibinfo  {journal} {J. Chem. Phys.}\ }\textbf {\bibinfo {volume} {145}},\
  \bibinfo {pages} {170901} (\bibinfo {year} {2016})}\BibitemShut {NoStop}%
\bibitem [{\citenamefont {Byggm{\"a}star}\ \emph {et~al.}(2019)\citenamefont
  {Byggm{\"a}star}, \citenamefont {Hamedani}, \citenamefont {Nordlund},\ and\
  \citenamefont {Djurabekova}}]{Byg19b}%
  \BibitemOpen
  \bibfield  {author} {\bibinfo {author} {\bibfnamefont {J.}~\bibnamefont
  {Byggm{\"a}star}}, \bibinfo {author} {\bibfnamefont {A.}~\bibnamefont
  {Hamedani}}, \bibinfo {author} {\bibfnamefont {K.}~\bibnamefont {Nordlund}},\
  and\ \bibinfo {author} {\bibfnamefont {F.}~\bibnamefont {Djurabekova}},\
  }\bibfield  {title} {\bibinfo {title} {Machine-learning interatomic potential
  for radiation damage and defects in tungsten},\ }\href@noop {} {\bibfield
  {journal} {\bibinfo  {journal} {Phys. Rev. B}\ }\textbf {\bibinfo {volume}
  {100}},\ \bibinfo {pages} {144105} (\bibinfo {year} {2019})}\BibitemShut
  {NoStop}%
\bibitem [{\citenamefont {Cai}\ \emph {et~al.}(1998)\citenamefont {Cai},
  \citenamefont {Snell}, \citenamefont {Beardmore},\ and\ \citenamefont
  {{Gr{\o}nbech-Jensen}}}]{Cai98}%
  \BibitemOpen
  \bibfield  {author} {\bibinfo {author} {\bibfnamefont {D.}~\bibnamefont
  {Cai}}, \bibinfo {author} {\bibfnamefont {C.~M.}\ \bibnamefont {Snell}},
  \bibinfo {author} {\bibfnamefont {K.~M.}\ \bibnamefont {Beardmore}},\ and\
  \bibinfo {author} {\bibfnamefont {N.}~\bibnamefont {{Gr{\o}nbech-Jensen}}},\
  }\bibfield  {title} {\bibinfo {title} {{Simulation of phosphorus implantation
  into silicon with a single parameter electronic stopping power model}},\
  }\href@noop {} {\bibfield  {journal} {\bibinfo  {journal} {International J.
  Modern Physics C}\ }\textbf {\bibinfo {volume} {9}},\ \bibinfo {pages} {459}
  (\bibinfo {year} {1998})}\BibitemShut {NoStop}%
\bibitem [{g3d()}]{g3data}%
  \BibitemOpen
  \href@noop {} {}\bibinfo {note} {G3data---grab graph data, a program for
  extracting data from graphs.
  \url{https://gitlab.com/pn2200/g3data}}\BibitemShut {NoStop}%
\bibitem [{\citenamefont {Ziegler}()}]{SRIM-2013}%
  \BibitemOpen
  \bibfield  {author} {\bibinfo {author} {\bibfnamefont {J.~F.}\ \bibnamefont
  {Ziegler}},\ }\href@noop {} {}\bibinfo {note} {{SRIM-2013 software package,
  available online at \textit{http://www.srim.org}.}}\BibitemShut {Stop}%
\bibitem [{\citenamefont {Ziegler}\ \emph {et~al.}(2008)\citenamefont
  {Ziegler}, \citenamefont {Biersack},\ and\ \citenamefont
  {Ziegler}}]{SRIMbook}%
  \BibitemOpen
  \bibfield  {author} {\bibinfo {author} {\bibfnamefont {J.~F.}\ \bibnamefont
  {Ziegler}}, \bibinfo {author} {\bibfnamefont {J.~P.}\ \bibnamefont
  {Biersack}},\ and\ \bibinfo {author} {\bibfnamefont {M.~D.}\ \bibnamefont
  {Ziegler}},\ }\href@noop {} {\emph {\bibinfo {title} {{SRIM - The Stopping
  and Range of Ions in Matter}}}}\ (\bibinfo  {publisher} {SRIM Co.},\ \bibinfo
  {address} {Chester, Maryland, USA},\ \bibinfo {year} {2008})\BibitemShut
  {NoStop}%
\bibitem [{\citenamefont {Sillanp{\"a}{\"a}}\ \emph {et~al.}(2000)\citenamefont
  {Sillanp{\"a}{\"a}}, \citenamefont {Nordlund},\ and\ \citenamefont
  {Keinonen}}]{Sil99}%
  \BibitemOpen
  \bibfield  {author} {\bibinfo {author} {\bibfnamefont {J.}~\bibnamefont
  {Sillanp{\"a}{\"a}}}, \bibinfo {author} {\bibfnamefont {K.}~\bibnamefont
  {Nordlund}},\ and\ \bibinfo {author} {\bibfnamefont {J.}~\bibnamefont
  {Keinonen}},\ }\bibfield  {title} {\bibinfo {title} {{Electronic stopping of
  Silicon from a 3D Charge Distribution}},\ }\href@noop {} {\bibfield
  {journal} {\bibinfo  {journal} {Phys. Rev. B}\ }\textbf {\bibinfo {volume}
  {62}},\ \bibinfo {pages} {3109} (\bibinfo {year} {2000})}\BibitemShut
  {NoStop}%
\bibitem [{\citenamefont {Allen}\ and\ \citenamefont
  {Tildesley}(1989)}]{Allen-Tildesley}%
  \BibitemOpen
  \bibfield  {author} {\bibinfo {author} {\bibfnamefont {M.~P.}\ \bibnamefont
  {Allen}}\ and\ \bibinfo {author} {\bibfnamefont {D.~J.}\ \bibnamefont
  {Tildesley}},\ }\href@noop {} {\emph {\bibinfo {title} {{Computer Simulation
  of Liquids}}}}\ (\bibinfo  {publisher} {Oxford University Press},\ \bibinfo
  {address} {Oxford, England},\ \bibinfo {year} {1989})\BibitemShut {NoStop}%
\bibitem [{\citenamefont {Nordlund}\ \emph {et~al.}(1996)\citenamefont
  {Nordlund}, \citenamefont {Keinonen\markthis},\ and\ \citenamefont
  {Mattila}}]{Nor96}%
  \BibitemOpen
  \bibfield  {author} {\bibinfo {author} {\bibfnamefont {K.}~\bibnamefont
  {Nordlund}}, \bibinfo {author} {\bibfnamefont {J.}~\bibnamefont
  {Keinonen\markthis}},\ and\ \bibinfo {author} {\bibfnamefont
  {T.}~\bibnamefont {Mattila}},\ }\bibfield  {title} {\bibinfo {title}
  {{Formation of ion irradiation-induced small-scale defects on graphite
  surfaces}},\ }\href@noop {} {\bibfield  {journal} {\bibinfo  {journal} {Phys.
  Rev. Lett.}\ }\textbf {\bibinfo {volume} {77}},\ \bibinfo {pages} {699}
  (\bibinfo {year} {1996})}\BibitemShut {NoStop}%
\bibitem [{\citenamefont {Juslin}\ \emph {et~al.}(2005)\citenamefont {Juslin},
  \citenamefont {Erhart}, \citenamefont {Tr{\"a}skelin}, \citenamefont {Nord},
  \citenamefont {Henriksson}, \citenamefont {Nordlund}, \citenamefont
  {Salonen},\ and\ \citenamefont {Albe}}]{Jus05}%
  \BibitemOpen
  \bibfield  {author} {\bibinfo {author} {\bibfnamefont {N.}~\bibnamefont
  {Juslin}}, \bibinfo {author} {\bibfnamefont {P.}~\bibnamefont {Erhart}},
  \bibinfo {author} {\bibfnamefont {P.}~\bibnamefont {Tr{\"a}skelin}}, \bibinfo
  {author} {\bibfnamefont {J.}~\bibnamefont {Nord}}, \bibinfo {author}
  {\bibfnamefont {K.~O.~E.}\ \bibnamefont {Henriksson}}, \bibinfo {author}
  {\bibfnamefont {K.}~\bibnamefont {Nordlund}}, \bibinfo {author}
  {\bibfnamefont {E.}~\bibnamefont {Salonen}},\ and\ \bibinfo {author}
  {\bibfnamefont {K.}~\bibnamefont {Albe}},\ }\bibfield  {title} {\bibinfo
  {title} {{Analytical interatomic potential for modelling non-equilibrium
  processes in the W-C-H system}},\ }\href@noop {} {\bibfield  {journal}
  {\bibinfo  {journal} {J. Appl. Phys.}\ }\textbf {\bibinfo {volume} {98}},\
  \bibinfo {pages} {123520} (\bibinfo {year} {2005})}\BibitemShut {NoStop}%
\bibitem [{\citenamefont {Ghaly}\ and\ \citenamefont {Averback}(1994)}]{Gha94}%
  \BibitemOpen
  \bibfield  {author} {\bibinfo {author} {\bibfnamefont {M.}~\bibnamefont
  {Ghaly}}\ and\ \bibinfo {author} {\bibfnamefont {R.~S.}\ \bibnamefont
  {Averback}},\ }\bibfield  {title} {\bibinfo {title} {{Effect of Viscous Flow
  on Ion damage near Solid Surfaces}},\ }\href@noop {} {\bibfield  {journal}
  {\bibinfo  {journal} {Phys. Rev. Lett.}\ }\textbf {\bibinfo {volume} {72}},\
  \bibinfo {pages} {364} (\bibinfo {year} {1994})}\BibitemShut {NoStop}%
\bibitem [{\citenamefont {Nordlund}\ \emph {et~al.}(1998)\citenamefont
  {Nordlund}, \citenamefont {Wei}, \citenamefont {Zhong},\ and\ \citenamefont
  {Averback}}]{Nor98}%
  \BibitemOpen
  \bibfield  {author} {\bibinfo {author} {\bibfnamefont {K.}~\bibnamefont
  {Nordlund}}, \bibinfo {author} {\bibfnamefont {L.}~\bibnamefont {Wei}},
  \bibinfo {author} {\bibfnamefont {Y.}~\bibnamefont {Zhong}},\ and\ \bibinfo
  {author} {\bibfnamefont {R.~S.}\ \bibnamefont {Averback}},\ }\bibfield
  {title} {\bibinfo {title} {{Role of electron-phonon coupling on collision
  cascade development in Ni, Pd and Pt}},\ }\href@noop {} {\bibfield  {journal}
  {\bibinfo  {journal} {Phys. Rev. B (Rapid Comm.)}\ }\textbf {\bibinfo
  {volume} {57}},\ \bibinfo {pages} {13965} (\bibinfo {year}
  {1998})}\BibitemShut {NoStop}%
\bibitem [{\citenamefont {Bj{\"o}rkas}\ and\ \citenamefont
  {Nordlund}(2007)}]{Bjo07a}%
  \BibitemOpen
  \bibfield  {author} {\bibinfo {author} {\bibfnamefont {C.}~\bibnamefont
  {Bj{\"o}rkas}}\ and\ \bibinfo {author} {\bibfnamefont {K.}~\bibnamefont
  {Nordlund}},\ }\bibfield  {title} {\bibinfo {title} {{Comparative study of
  cascade damage in Fe simulated with recent potentials}},\ }\href@noop {}
  {\bibfield  {journal} {\bibinfo  {journal} {Nucl. Instr. Meth. Phys. Res. B}\
  }\textbf {\bibinfo {volume} {259}},\ \bibinfo {pages} {853} (\bibinfo {year}
  {2007})},\ \bibinfo {note} {notes that 'The cutoﬀ radius in the cluster
  analysis was second nearest neighbor and third nearest neighbor for the
  vacancy and interstitial defects, respectively.'}\BibitemShut {NoStop}%
\bibitem [{\citenamefont {Sand}\ \emph {et~al.}(2016)\citenamefont {Sand},
  \citenamefont {Dequeker}, \citenamefont {Becquart}, \citenamefont {Domain},\
  and\ \citenamefont {Nordlund}}]{San15a}%
  \BibitemOpen
  \bibfield  {author} {\bibinfo {author} {\bibfnamefont {A.~E.}\ \bibnamefont
  {Sand}}, \bibinfo {author} {\bibfnamefont {J.}~\bibnamefont {Dequeker}},
  \bibinfo {author} {\bibfnamefont {C.~S.}\ \bibnamefont {Becquart}}, \bibinfo
  {author} {\bibfnamefont {C.}~\bibnamefont {Domain}},\ and\ \bibinfo {author}
  {\bibfnamefont {K.}~\bibnamefont {Nordlund}},\ }\bibfield  {title} {\bibinfo
  {title} {{Non-equilibrium properties of interatomic potentials in cascade
  simulations in tungsten}},\ }\href@noop {} {\bibfield  {journal} {\bibinfo
  {journal} {J. Nucl. Mater.}\ }\textbf {\bibinfo {volume} {470}},\ \bibinfo
  {pages} {119} (\bibinfo {year} {2016})}\BibitemShut {NoStop}%
\bibitem [{\citenamefont {Belko}\ \emph {et~al.}(2002)\citenamefont {Belko},
  \citenamefont {Posselt},\ and\ \citenamefont {Chagarov}}]{Bel02}%
  \BibitemOpen
  \bibfield  {author} {\bibinfo {author} {\bibfnamefont {V.}~\bibnamefont
  {Belko}}, \bibinfo {author} {\bibfnamefont {M.}~\bibnamefont {Posselt}},\
  and\ \bibinfo {author} {\bibfnamefont {E.}~\bibnamefont {Chagarov}},\
  }\bibfield  {title} {\bibinfo {title} {{Improvement of the repulsive part of
  the classical interatomic potential for SiC}},\ }\href@noop {} {\bibfield
  {journal} {\bibinfo  {journal} {Nucl. Instr. Meth. Phys. Res. B}\ }\textbf
  {\bibinfo {volume} {202}},\ \bibinfo {pages} {18} (\bibinfo {year}
  {2002})}\BibitemShut {NoStop}%
\bibitem [{Sci()}]{SciPy}%
  \BibitemOpen
  \href@noop {} {}\bibinfo {note}
  {{h}ttps://www.scipy.org/scipylib/}\BibitemShut {NoStop}%
\bibitem [{\citenamefont {Gombas}(1949)}]{Gom49}%
  \BibitemOpen
  \bibfield  {author} {\bibinfo {author} {\bibfnamefont {P.}~\bibnamefont
  {Gombas}},\ }\href@noop {} {\emph {\bibinfo {title} {{Die Statistische
  Theorie des Atoms und ihre Anwendungen}}}}\ (\bibinfo  {publisher}
  {Springer-Verlag},\ \bibinfo {address} {Vienna, Austria},\ \bibinfo {year}
  {1949})\BibitemShut {NoStop}%
\bibitem [{\citenamefont {Eckstein}(1991)}]{Eck91}%
  \BibitemOpen
  \bibfield  {author} {\bibinfo {author} {\bibfnamefont {W.}~\bibnamefont
  {Eckstein}},\ }\href@noop {} {\emph {\bibinfo {title} {{Computer Simulations
  of Ion-Solid Interactions}}}}\ (\bibinfo  {publisher} {Springer},\ \bibinfo
  {address} {{Berlin}},\ \bibinfo {year} {1991})\ \bibinfo {note} {p.
  40}\BibitemShut {NoStop}%
\end{thebibliography}%


\end{document}